\documentclass[a4paper,10pt]{article}
\usepackage[pdftex]{graphicx,hyperref}
\usepackage{amsmath,amsfonts,amssymb}
\usepackage{fullpage}
\usepackage{authblk}
\usepackage{setspace}
\usepackage{subcaption}
\usepackage{booktabs}
\usepackage{hyperref}
\usepackage{tabularx}

\usepackage[explicit]{titlesec}
\usepackage{sidecap}
\usepackage{pbox}
\usepackage[superscript]{cite}

\usepackage{amsfonts}
\usepackage{amsmath}
\usepackage{multirow}
\usepackage{longtable}
\usepackage{graphicx}
\date{}
\title{\bf{Evidence for a Conserved Quantity \\ in Human Mobility}}

\author[a,b]{Laura Alessandretti}
\author[b]{Piotr Sapiezynski} 
\author[b,d]{Vedran Sekara}
\author[b,c,*]{Sune Lehmann}
\author[a,*]{Andrea Baronchelli}

\affil[a]{City,  University of London, London EC1V 0HB, United Kingdom}
\affil[b]{Technical University of Denmark, DK-2800 Kgs. Lyngby, Denmark}
\affil[c]{{Niels Bohr Institute, University of Copenhagen, DK-2100 K\o benhavn \O , Denmark}}
\affil[d]{Sony Mobile Communications, Mobilv\"agen, 221 88 Lund, Sweden \newline {\small $^*$Corresponding authors:  sune.lehmann@gmail.com, Andrea.Baronchelli.1@city.ac.uk}}
\begin{document}
\maketitle

\textbf{Recent seminal works on human mobility have shown that individuals constantly exploit a small set of repeatedly visited locations \cite{song2010limits, schwanen2008fixed, golledge1997spatial}. A concurrent literature has emphasized the explorative nature of human behavior, showing that the number of visited places grows steadily over time \cite{song2010modelling,gonzalez2008understanding,pappalardo2015returners,alessandretti2017multi}. How to reconcile these seemingly contradicting facts remains an open question. Here, we analyze high-resolution multi-year traces of $\sim$40,000 individuals from 4 datasets and show that this tension vanishes when the long-term evolution of mobility patterns is considered. We reveal that mobility patterns evolve significantly yet smoothly, and that the number of familiar locations an individual visits at any point is a conserved quantity with a typical size of $\sim$25 locations. We use this finding to improve state-of-the-art modeling of human mobility \cite{song2010modelling,jiang2016timegeo}. Furthermore, shifting the attention from aggregated quantities to individual behavior, we show that the size of an individual's set of preferred locations correlates with the number of her social interactions. This result suggests a connection between the conserved quantity we identify, which as we show can not be understood purely on the basis of time constraints, and the `Dunbar number' \cite{dunbar1993coevolution,gonccalves2011modeling} describing a cognitive upper limit to an individual's number of social relations. We anticipate that our work will spark further research linking the study of Human Mobility and the Cognitive and Behavioral Sciences.}

\vspace{1cm}
There is a disagreement between the current scientific understanding of human mobility as highly predictable and stable over time \cite{gonzalez2008understanding, song2010modelling, song2010limits}, and the fact that individual lives are constantly evolving due to changing needs and circumstances \cite{sarason1978assessing}. 
The role of cultural, social and legal constraints on the space-time fixity of daily activities has long been recognized \cite{hagerstraand1970people, burns1980transportation, schwanen2008fixed}.
Recent studies based on the analysis of human digital traces including mobile phone records \cite{csaji2013exploring, sevtsuk2010does}, online location-based social networks \cite{cho2011friendship, cheng2011exploring, Cecilia, noulas2012tale, bapierre2015mobile}, and Global Positioning System (GPS) location data of vehicles \cite{giannotti2011unveiling, scellato2011nextplace, liang2012scaling, riccardo2012towards, bazzani2010statistical, jiang2009characterizing} have shown that individuals universally exhibit a markedly regular pattern characterized by few locations, or points of interest \cite{mulligann2011analyzing,phithakkitnukoon2010activity}, where they return regularly \cite{isaacman2011identifying, pappalardo2015returners} and predictably \cite{song2010modelling}.  
However, the observed regularity mainly concerns human activities taking place at the daily \cite{schneider2013unravelling, bagrow2012mesoscopic,phithakkitnukoon2010activity} or weekly \cite{cheng2011exploring, csaji2013exploring, sevtsuk2010does} time-scales, such as commuting between home and office \cite{csaji2013exploring, sevtsuk2010does, ranjan2012call, zang2011anonymization}, pursuing habitual leisure activities, and socializing with established friends and acquaintances \cite{cho2011friendship}.
Thus, while the role played by slowly occurring changes on the evolution of individuals' social relationships has been widely investigated \cite{kossinets2006empirical, kossinets2009origins, romero2011maintaining, martin2006persistence, miritello2013limited, saramaki2014persistence, burt2000decay, arnaboldi2013dynamics}, their effects on human mobility behavior are not well understood and not included in most available models  \cite{jiang2016timegeo, song2010modelling, isaacman2012human, lee2009slaw, kim2006extracting, jia2012empirical, han2011origin, PAPPALARDO2016934}. 

Here, we investigate individuals' routines across months and years.
We reveal how individuals balance the trade-off between the exploitation of familiar places and the exploration of new opportunities, we point out that predictions of state-of-the-art models can be significantly improved if a finite memory is assigned to individuals, and we show that individuals' exploration-exploitation behaviors in the social and spatial domain are correlated. 

Our study is based on the analysis of $\sim40\,000$ high resolution mobility trajectories of two samples of individuals measured for at least $12$ months (see Suppementary Table~1): the users of the Lifelog mobile application (Lifelog), traced over $19$ months, and the participants in a longitudinal experiment, the Copenhagen Networks Study (CNS) \cite{stopczynski2014measuring}, spanning 24 months. Results were corroborated with data from two other experiments with fixed rate temporal sampling, but lower spatial resolution and sample size (see Supplementary Table~1): the Lausanne Data Collection Campaign (MDC), lasted for $19$ months~\cite{kiukkonen2010towards, laurila2012mobile} and the Reality Mining dataset (RM)~\cite{eagle2006reality, eagle2009inferring}, spanning $10$ months. 
Our datasets rely on different types of location data and collection methods (see section Data Description, Supplementary Note 1.1, and Supplementary Figures~1~to~6), but share the high spatial resolution and temporal sampling necessary to capture mobility patterns beyond highly regular ones such as home-work commuting~\cite{ccolak2015analyzing}. 

All the datasets considered display statistical properties consistent with those reported in previous studies focusing on larger samples but shorter timescales~\cite{gonzalez2008understanding,song2010modelling} (see Supplementary Note 1.2 and Supplementary Figures~7~to~9), and their temporal resolution and duration make them ideal for investigating the evolution of individual geo-spatial behaviors on longer timescales. Moreover, three of the datasets considered (CNS, MDC, RM) include also information on individuals' interactions across multiple social channels (phone call, sms, Facebook), allowing us to connect individuals' spatial and social behaviors across long timescales. Two of the datasets (CNS and RM), describing together $\sim2\%$ of the individuals analyzed in this study, consist of the trajectories of university students (CNS, RM) and faculty members (RM). These subjects are homogeneous with respect to socio-demographic indicators affecting mobility behavior \cite{lenormand2014influence}, and their displacements are constrained by a similar academic schedule. Notwithstanding this possible source of bias, all results presented below hold for the four considered datasets.

\begin{figure*}[h!]
\centering
\includegraphics[width=.9\textwidth]{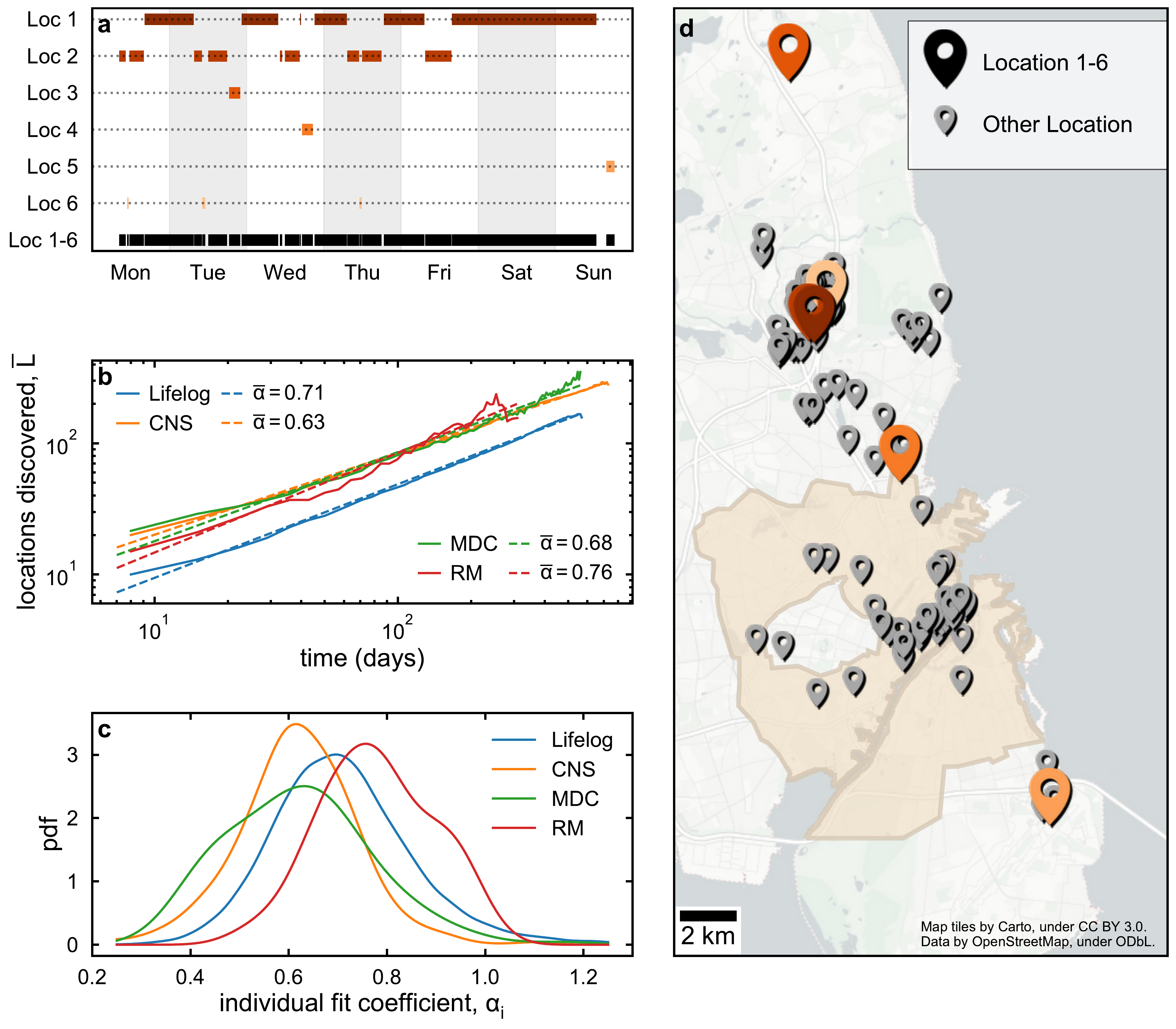}
\caption{\textbf{Activity set and exploration of new locations.} (\textbf{a}) An example of an individual's mobility trace. The visiting temporal pattern of the six most visited locations are shown (Loc1, ..., Loc6) along with the black trace including all visits to these 6 locations (Loc1-6). (\textbf{b}) Total number of discovered locations in time. The figure shows the average across users for each dataset (colored filled lines), and a power-law fitting function (dashed lines) with exponent $\alpha$. (\textbf{c}) The probability density functions of individuals' power-law fit coefficients for different datasets (colored filled lines) are peaked around their average value. (\textbf{d}) Example of an individual's activity set. Locations are represented as pins on a map, after random noise was added to protect the user's privacy. The six most visited locations are displayed as larger pins using the same color scheme of panel \textbf{a}. The light orange area shows the city of Copenhagen. }
\label{Figure1}
\end{figure*}

Our first finding is that individuals' sets of visited locations grows with characteristic sub-linear exponent.
When initiating a transition from a place to another, individuals may either choose to return to a previously visited place, or explore a new location. 
To characterize this exploration-exploitation trade-off, we represent individual geo-spatial trajectories as sequences of locations, where `locations' are defined as places where participants in the study stopped for more than $10$ minutes  (Fig.~\ref{Figure1}a, see also Supplementary Note 1.1). CNS locations' typical extent after pre-processing matches that of places like commercial activities, metro stations, classrooms and other areas within the University campus (see Supplementary Figure~6). Despite the differences in data spatial resolution, the number of unique locations visited weekly is comparable among all 4 datasets (see Supplementary Table~2).

A central question concerning the long term \emph{exploration behavior} of the individuals is whether an individual's set of known locations continuously expands, or saturates over time. 
We find that the total number of unique locations $L_i(t)$ an individual $i$ has discovered up to time $t$ grows as $L_i\propto t^{\alpha_i}$ (Fig.~\ref{Figure1}b), and that individuals' exploration is homogeneous across the populations studied, with $\alpha_i$ peaked around $\overline{\alpha}$ (Lifelog: $\overline{\alpha}=0.71$, CNS: $\overline{\alpha}=0.63$, MDC: $\overline{\alpha}=0.68$, RM: $\overline{\alpha}=0.76$) (Fig.~\ref{Figure1}c). 
This sub-linear growth occurs regardless of how locations are defined, when in time the measurement starts, and individuals' age (see Supplementary Figures~19~to~21). 
This behavior is a characteristic signature of Heaps' law  \cite{heaps1978information}, and consistent with findings from previous studies focusing on shorter time-scales \cite{song2010modelling}. 

While continually exploring new places, individuals allocate most of their time among a small subset of all visited locations (see Supplementary Figure~10), in agreement with previous research on human mobility behavior \cite{isaacman2011identifying, pappalardo2015returners, song2010modelling} and time-geography \cite{hagerstraand1970people, horton1971effects, mazey1981effect, golledge1997spatial, yuan2016analyzing}. 
Hence, at any point in time, each individual is characterized by an \emph{activity set} within which she visits as a result of her daily activities \cite{sherman2005suite, mazey1981effect}. 
This is defined to capture important locations visited multiple times even if for short visits \cite{isaacman2011identifying,zhou2007mining}, and it is closely related to the concept of `activity space' widely used in geography \cite{sherman2005suite}. 
Operationally, we define it as the set $AS_i(t)=\{\ell_1,\ell_2,...,\ell_k,...\ell_C\}$ of locations $\ell_k$ that individual $i$ visited at least twice and where she spent on average more than $10$ minutes/week during a time-window of $20$ consecutive weeks preceding time $t$.
The results presented below are robust with respect to variations of this definition, such as changes of the time-window size or the definition of a location (Supplementary Note 1.3, Supplementary Figures~11,~13,~19 and Supplementary Tables~3,~4,~6).

\begin{figure*}[t]
\centering

\includegraphics[width=.9\textwidth]{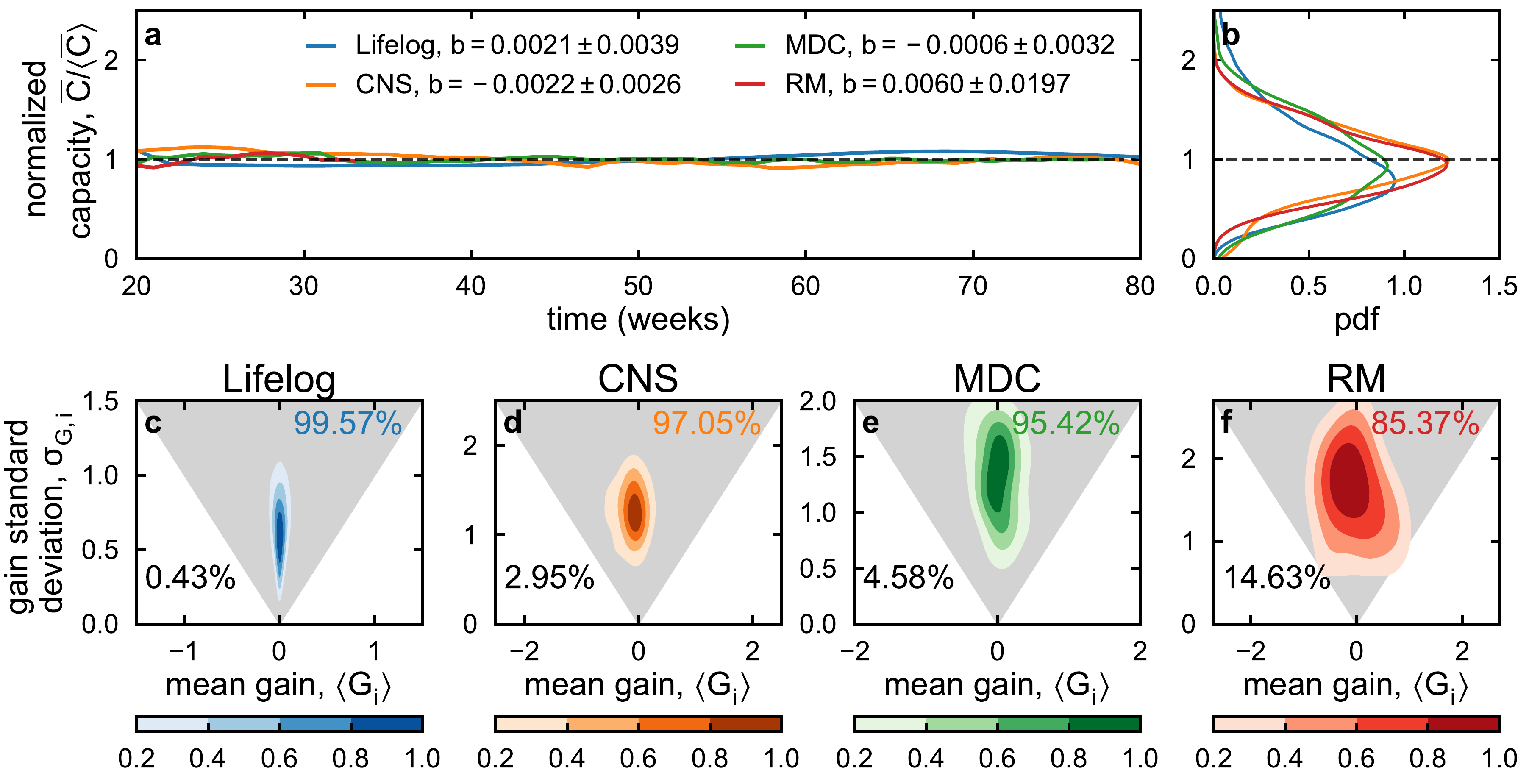}

\caption{\textbf{Conserved size of evolving activity sets.} \textbf{(a)} Evolution of the average normalized capacity for the 4 datasets considered. The dashed black line corresponds to constant capacity. The error on the angular coefficient $b$ of a linear fit, reported in the legend, shows that the fit is compatible with a constant line. \textbf{(b)} Probability density function of individuals' average capacity for the 4 datasets considered.   \textbf{(c, d, e, f)} Gain standard deviation $\sigma_{G,i}$ vs the average gain  $\langle G_i \rangle$ for the Lifelog \textbf{(e)}, CNS \textbf{(f)}, MDC \textbf{(g)} and RM \textbf{(h)} datasets. Lines representing cumulative probabilities are obtained through a kernel density estimation from the data, the grey area corresponds to individuals for which $|\langle G_i \rangle| < \sigma_{G,i}$, i.e. whose average gain is compatible with zero. It contains $99.57\%$ (Lifelog), $97.05\%$ (CNS), $95.42\%$ (MDC) and $85.37\%$ (RM) of the population.}
\label{Figure2}
\end{figure*}

Thus, individuals continually explore new places yet they are loyal to a limited number of familiar ones forming their activity set. 
But how does discovery of new places affect an individual's activity set? 
We find that the average probability $\overline{P}$ that a newly discovered location will become part of the activity set stabilizes at $P^*$ (CNS: $P^*=15\%$, Lifelog: $P^*=7\%$, MDC: $P^*=15\%$, RM: $P^*=20\%$) over the long term, indicating that individuals' activity sets are inherently unstable and new locations are continually added.
However, over time individuals may also cease to visit locations that are part of the activity set. 
The balance between newly added and dismissed familiar locations is captured by the temporal evolution of the activity set, which we characterize by the \emph{location capacity} and \emph{net gain}. 
We define \emph{location capacity} $C_i$ as the number of an individual's familiar locations, i.e.~the activity set size, at any given moment. 
The \emph{net gain} $G_i$ is defined as the difference between the number of locations that are respectively added $(A_i)$ and removed $(D_i)$ at a specific time, hence  $G_i=A_i-D_i$. Fig.~\ref{Figure2}a shows the evolution of the average capacity $\overline{C} $ for the populations considered, normalized to account for the effects due to different data collection methods (see Supplementary Note 1.1).

We find that $\overline{C}$ is constant in time, with a linear fit of the form $\overline{C}=a+b \cdot t$ yielding $b$ not significantly different than 0 (Lifelog: $b = 0.0021 \pm 0.0039 $, CNS: $b=-0.0022 \pm 0.0026$, MDC: $b=-0.0006 \pm 0.0032$, RM: $b=0.0060 \pm 0.0197$). 
Analogously, a power-law fit of the form $\overline {C (t)} \propto t^{\beta}$ yields $\beta$ consistent with 0 (Lifelog: $9\cdot 10^{-4} \pm 3\cdot 10^{-2}$, CNS: $-2 \cdot 10^{-3} \pm 6 \cdot 10^{-2}$, MDC: $-2 \cdot 10^{-4} \pm 5 \cdot 10^{-3}$, RM: $2 \cdot 10^{-3} \pm 3 \cdot 10^{-1}$).
As a further control, we performed a multiple hypothesis test with false discovery rate correction to compare the averages of the capacity distribution at different times (see Supplementary Table~3). We find no evidence for rejecting the hypothesis that the average capacity does not change in time. 
Additionally, we find that, for the CNS and the Lifelog datasets, the radius of gyration\cite{gonzalez2008understanding} of the activity set, a measure of its spatial extent, is on average constant in time (see Supplementary Figure~31) under the two tests above.
Thus, despite individual activity set evolving over time, the average location capacity is a conserved quantity.

The conservation of the average location capacity may result from either (i) each individual maintaining a stable number of familiar locations over time or (ii) a substantial heterogeneity of the populations considered, with certain individuals shrinking their activity sets and other expanding theirs. 
We test the two hypotheses by measuring the individual average net gain across time $\langle G_i \rangle $ and its standard deviation $\sigma_{G,i}$. If a participant's average gain is closer than one standard deviation to 0, hence $|\langle G_i \rangle |/\sigma_{G,i}<1$, then the net gain is consistent with $\langle G_i \rangle=0$. 
If this is true for the majority of individuals, the location capacity is conserved at the individual level and hypothesis (i) holds. If, on the other hand, $|\langle G_i \rangle |/\sigma_{G,i}\geq1$, the individual capacity must either increase or decrease in time, supporting hypothesis (ii). 
We find that hypothesis (i) holds for most individuals (Lifelog: 99.57\%, CNS: 97.05\%, MDC: 95.42\%, RM: 85.37\%) (Fig.~\ref{Figure2}c-f, see also Supplementary Table~4). 
For the large majority of each population, the average net gain of familiar locations added or removed to the activity set at any point is not significantly different from 0, hence their individual capacity is conserved. 
Also, we find that the individual capacity has low variability with the ratio between the average individual capacity and its standard deviation $\langle C_i \rangle / \sigma_{C,i}$ typically limited below 30\%  (Lifelog: $30\%$, CNS: $28\%$, MDC: $27\%$, RM: $14\%$), demonstrating that fluctuations of the capacity are relatively small. Further evidence suggesting the conservation of individual location capacity is provided in Supplementary Note~1.5 and Supplementary Figures~33~to~35. 

These results indicate that each individual is characterized by a fixed-size but evolving activity set of familiar locations. We find that the typical size of the activity set saturates at $\sim 25$ for increasingly larger values of the time-window defining the activity set (see Supplementary Figure~12). This value is consistent across all $4$ samples, prior rescaling to account for the differences in time coverage. Individuals' values are homogeneously distributed around the sample mean (Fig.~\ref{Figure2}b, see also Supplementary Figure~14). Previous analyses identified two distinct classes of individuals, `returners', whose characteristic travelled distance is dominated by movements between few important locations, and `explorers', characterized by a larger number of places \cite{pappalardo2015returners}. We observe that `explorers' typically have higher location capacity than `returners' (see Supplementary Figures~8,~9~and~32).

To interpret the information contained in the measured value of the location capacity, we randomize the temporal sequences of locations in two ways, preserving individual routines only up to the daily level. 
After breaking individual time series into modules of 1 day length, (a) we randomize individual timeseries preserving the module/day units (local randomization) or (b) we create new sequences by assembling together modules extracted randomly by the whole set of individual traces (global randomization, see Supplementary Figure~22). 
Due to the absence of temporal correlations, the capacity is constant in time also for the randomized datasets. However, the capacity of the random sets is significantly higher than in the real time series for both randomizations under the Kolmogorov-Smirnov test (see Supplementary Table~5), implying that the observed value in real data is not a simple consequence of time constraints. Instead, the fixed capacity is an inherent property of human behavior.

The time evolution of the activity set supports this finding. We measure the turnover of familiar locations using the Jaccard similarity $J_i(t,\gamma)$ between the weekly activity set at $t$ and at $t+\gamma$ (see Fig.~\ref{Figure4}a-d). 
Despite seasonality effects (see Supplementary Figures~15~and~16), which imply fluctuations around a typical behavior, $J_i$ does not depend on the initial point but only on the waiting time $\gamma$, and we can consider $J_i(\gamma)$ independently of $t$ (see Supplementary Figure~17). 
We find that the average similarity decreases as a power law $\overline{J} \propto \gamma^{\lambda}$ with coefficient $\lambda$ significantly different than 0 (Lifelog: $\lambda=-0.15$, CNS: $\lambda=-0.31$,  MDC:  $\lambda=-0.49$,  RM: $\lambda=-3.00$, see also Supplementary Figure~18). Furthermore, the center of mass of the activity set changes position across time (see Supplementary Figure~30). On the other hand, for the randomized sequences, the Jaccard similarity is constant in time as familiar locations are never abandoned ($\overline{J} \propto \gamma^{0}$).  
This confirms that individual activity sets change continually and individual routines evolve gradually in time.

\begin{figure*}[h]
\centering
\includegraphics[width=.9\textwidth]{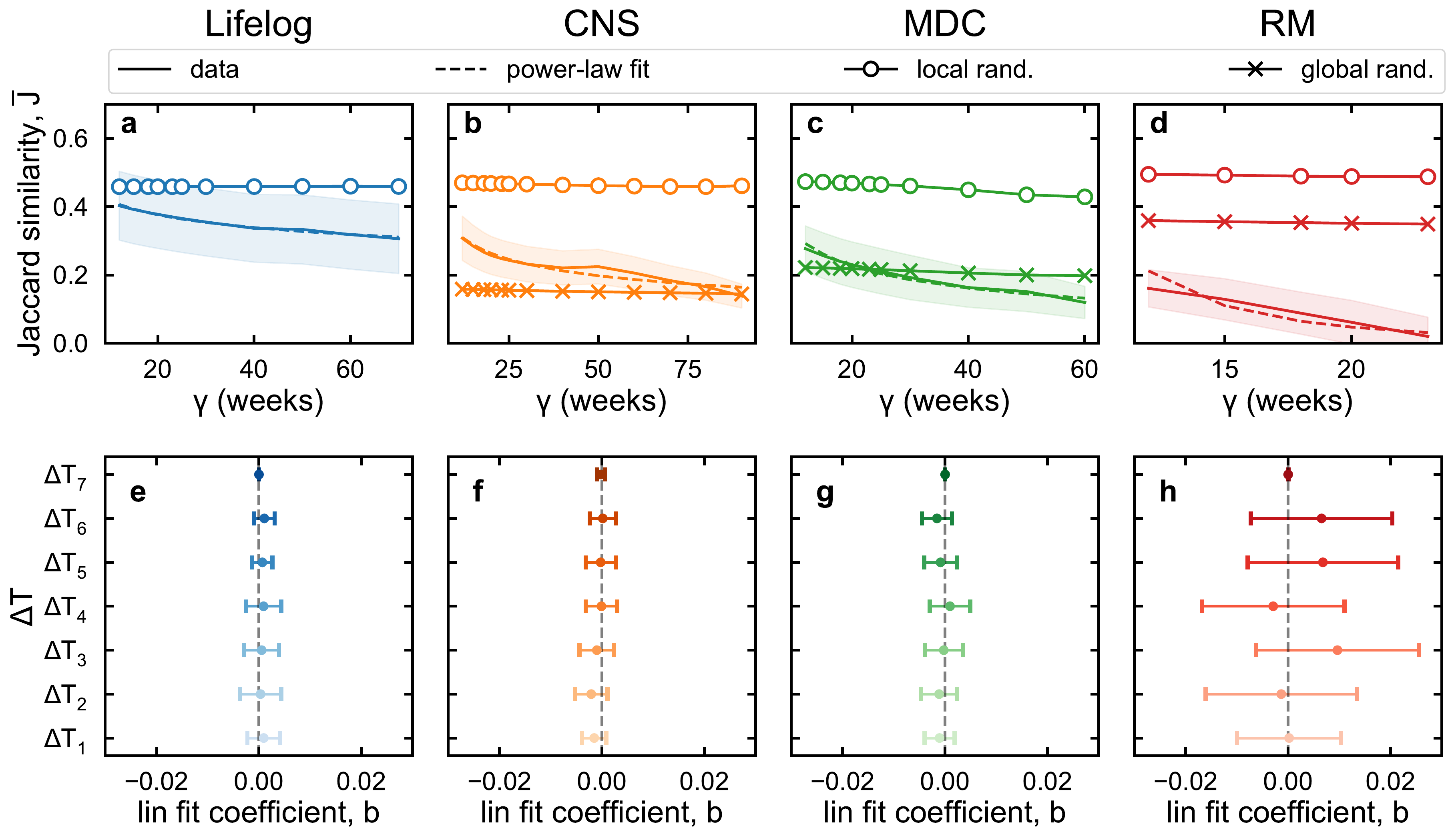}
\caption{\textbf{Evolution of activity sets and conservation of time allocation} \textbf{(a-d)} The average Jaccard similarity $\overline{J}$ between the weekly activity sets measured at $t$ and $t+\gamma$ as a function of $\gamma$ for data (filled lines), the globally randomized series (lines with crosses) and the locally randomized series (lines with dots). Filled areas correspond to the 50\% interquartile range. Dashed lines correspond to power-law fits $\overline{J} \sim \gamma^{\lambda}$, with $\lambda=-0.15$ for the Lifelog data \textbf{(a)}, $\lambda=-0.31$ for CNS \textbf{(b)}, $\lambda=-0.49$ for MDC \textbf{(c)} and $\lambda=-3.00$ for RM \textbf{(d)}. Results are obtained with $w=10$ weeks. The anonymization procedure applied by SONY Mobile before supplying the data makes it impossible to perform the global randomization on the Lifelog trajectories. \textbf{(e-h)}  Linear fit coefficients of the average capacity vs time for several categories of locations $\Delta T$ are consistent with $0$ within errors. The intervals considered are $\Delta T_1=10-30$ min/w, $\Delta T_2=30-60$ min/w, $\Delta T_3=1-6$ h/w, $\Delta T_4=6-12$ h/w, $\Delta T_5=12-24$ h/w, $\Delta T_6=24-48$ h/w, $\Delta T_7=48-168$ h/w. Results are shown for the Lifelog \textbf{(e)}, CNS \textbf{(f)}, MDC \textbf{(g)} and RM \textbf{(h)} datasets. }
\label{Figure4}
\end{figure*}

In order to characterize the structure of the activity set, we investigate how individuals allocate time among different location classes defined on the basis of their average visit duration. We consider intervals $\Delta T$, with $\Delta T$ ranging from $10$ to $30$ minutes per week (the time it takes to visit a bus stop or grocery shop) up to 48 to 168 hours per week (such as for home locations). For each of these locations classes, we compute the evolution of the \emph{capacity} $c^{\Delta T}_i$ and the \emph{gain} $G^{\Delta T}_i$, and test the hypothesis $G^{\Delta T}_i=0$, as above. We find that, although the activity set subsets are continuously evolving (see Supplementary Table~7),  $c^{\Delta T}_i$ is conserved for each $\Delta T$ (Fig.~\ref{Figure4}e-h, see also Supplementary Figures~24 to 27 and Supplementary Table~6), indicating that the number of places where individuals spend a range of time $\Delta T$ does not change over time. This result holds independently of the choice of specific $\Delta T$ and implies that the individual capacity $C_i=\sum c_i^{\Delta T}$, where both $C_i$ and each $c_i^{\Delta T}$ are conserved across time. Thus, both location capacity and time allocation are conserved quantities.

Our results have consequences for the modeling of human mobility.
The renown exploration and preferential return model \cite{song2010modelling,jiang2016timegeo} describe agents that, when not exploring a new location, return to a previously visited place selected with a probability proportional to the number of former visits. Another state of the art model introduces a mechanism assigning higher return probability to recently visited locations \cite{barbosa2015effect}.
These models reproduce some of the empirical observations described above, including the conservation of the \emph{location capacity} (Fig.~\ref{Figure2}), but fail to describe the time evolution of the activity set (Fig.~\ref{Figure4}). To overcome this limitation, we start from the observation that the exploitation probability for a location is time-dependent \cite{barbosa2015effect, szell2012understanding} and endow the agents with a finite memory $M$ so that the probability of returning to a location is based on the number of visits occurred in the last $M$ days. The model including this simple modification qualitatively reproduces all the observations, including the long-term evolution of the activity set (see Supplementary Note 1.4 and Supplementary Figure~28).

Finally, we analyse the connection between the social and spatial domain. 
Empirical observations suggest that there are upper limits to the size of an individual's social circle, the so-called Dunbar number \cite{dunbar1993coevolution, miritello2013limited, saramaki2014persistence, gonccalves2011modeling}, due to cognitive constraints \cite{dunbar1993coevolution}, and it has been hypothesized that the geography of one's activity set is proportional to one's social network geography \cite{axhausen2007activity}.
Motivated by these observations, we test the hypothesis of a correlation between individuals' $\emph{location capacity}$ and the size of their social circle, as measured by the people contacted by phone (see Fig.~\ref{Figure8}), and Facebook (see Supplementary Figure~36) over a period of $20$ weeks. We find that a significant positive correlation exists (see caption of Fig.~\ref{Figure8}). Furthermore, for the CNS dataset, we are able to show that both quantities correlate with
the individual personality trait of extraversion \cite{costa1992four}, which tend to be manifested in outgoing, talkative and energetic behavior \cite{kalish2006psychological}
(see Supplementary Figure~29; Pearson correlation $\rho = 0.22$, 2-tailed $p<10^{-9}$  for location capacity vs extraversion; $\rho = 0.40$, 2-tailed $p<10^{-28}$  for size of social network vs extraversion\cite{pollet2011extraverts}). We consider that these observations call for further analyses on the  connections between human social and spatial behavior.

\begin{figure*}[h]
\centering
\includegraphics[width=.9\textwidth]{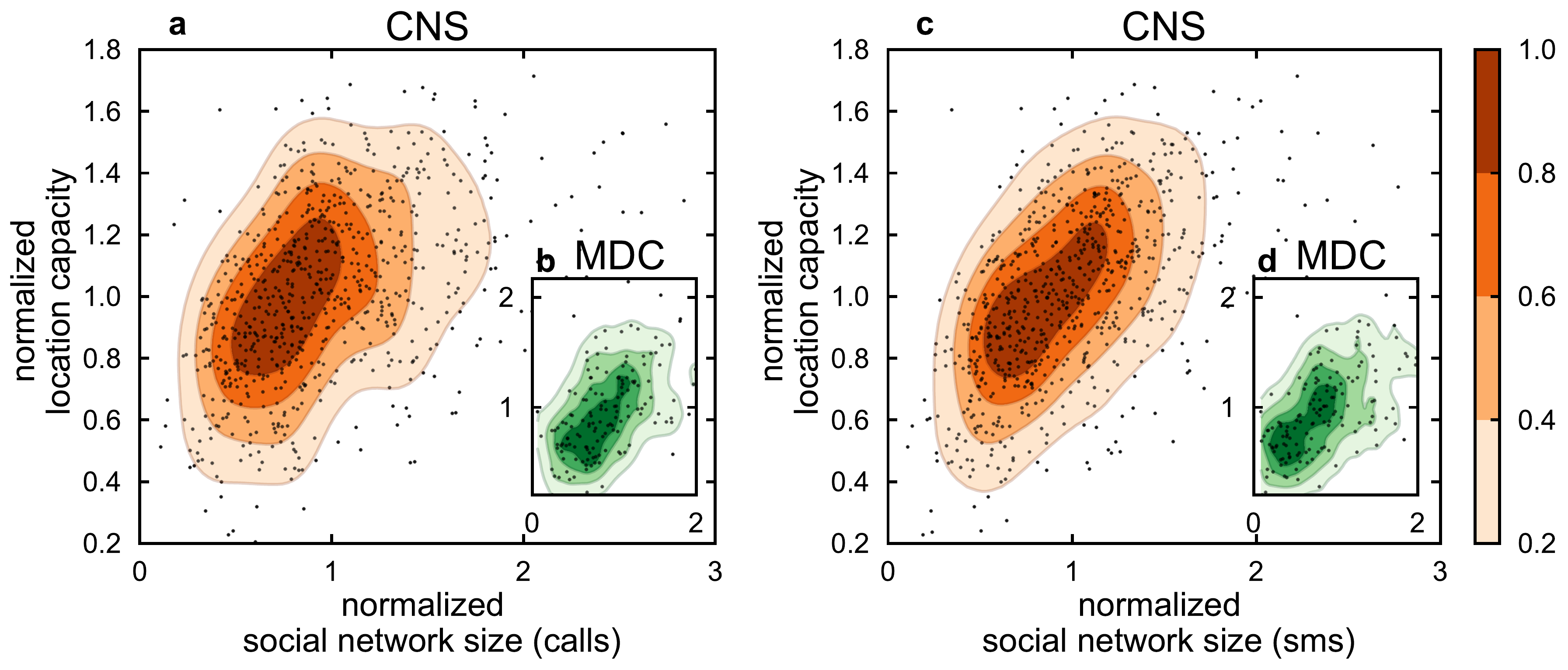}
\caption{\textbf{Correlation between location capacity and social network size.} Values of individuals' average normalized location capacity vs their normalized social network size computed from phone call interactions \textbf{(a, c)} and sms interactions \textbf{(b, d)} (black dots, see Supplementary Figure~36 for results obtained using Facebook interactions). Colored filled areas correspond to cumulative probabilities estimated via Gaussian Kernel Density estimations for visualization purposes. Results are shown for the CNS \textbf{(a, b)} and MDC \textbf{(c, d)} datasets. The values of the Pearson correlation coefficient are 0.31 \textbf{(a)}, 0.48 \textbf{(b)}, 0.52 \textbf{(c)}, 0.54 \textbf{(d)} (2-tailed $p<10^{-13}$ in all cases). Social network size is normalized to the population average value.}
\label{Figure8}
\end{figure*}

In summary, we have shown that the number of locations an individual visits regularly is conserved over time, even while individual routines are unstable in the long term because of the continual exploration of new locations. 
This individual \emph{location capacity} is peaked around a typical value of $\sim 25$ locations across the population, and significantly (typically, at least $30\%$) smaller than what would be expected if only time-constraints were at play (see Supplementary Table~5, and Supplementary Figure~23).

The \emph{location capacity} is hierarchically structured, indicating that individual time allocation for categories of places is also conserved. These results have allowed us to improve existing models of human mobility which are unable to fully account for long-term instabilities and fixed-capacity effects. 

Taken together, these findings shed new light on the underlying dynamics shaping human mobility, with potential impact for a better understanding of  phenomena such as urban development and epidemic spreading. 

Extending our scope beyond mobility, we have shown that individuals' \emph{location capacity} is  correlated with the size of their social circles. 
In this respect, it is interesting to note that fixed-size effects in the social domain~\cite{dunbar1993coevolution, miritello2013limited, saramaki2014persistence,
gonccalves2011modeling} have been put in direct relation with human cognitive abilities \cite{dunbar1993coevolution}. We anticipate that our results will stimulate new research exploring this connection.

\section*{Methods}\label{dataDescr}

\subsection*{Data Description}

\textbf{RM dataset:} The Reality Mining project was conducted from 2004-2005 at the MIT Media Laboratory. It measured 94 subjects using mobile phones over the course of nine months. Of these 94 subjects, 68
were colleagues working in the same building on campus (90\% graduate students, 10\% staff) while the remaining 26 subjects were incoming students at the university's business school \cite{eagle2006reality}. An application installed on users' phones continuously logs location data from cell tower ids at fixed rate sampling. The study was approved by the MIT Committee on the Use of Humans as Experimental Subjects (COUHES). Subjects were provided with detailed information about the type of information captured and provided informed consent \cite{eagle2010reality}. 

\textbf{CNS dataset:} The Copenhagen Networks Study experiment took place between September 2013 and September 2015 \cite{stopczynski2014measuring} and involved 851 Technical University of Denmark students ($\sim 22\%$ female, $\sim 78\%$ male) typically aged between 19 and 21 years old. Participants' position over time was estimated combining their smart-phones WiFi and GPS data using the method described in~\cite{sapiezynski2015opportunities} (see also Supplementary Note 1.1, and Supplementary Figure~6). The location estimation error is below 50 meters in 95\% of the cases. Participants calls and sms activity was also collected as part of the experiment. Individuals' background information were obtained through a 310 questions survey including the Big Five Inventory \cite{john1999big}, measuring five broad domains of human personality traits (openness, conscientiousness, extraversion, agreeableness, neuroticism). Data collection was approved by the Danish Data Protection Agency. All participants provided informed consent.

\textbf{Lifelog dataset:} The dataset consists of anonymized GPS location data for $\sim 36000$ users of the Lifelog app between 2015 and 2016. Lifelog users are geo-localized across the world (see Supplementary Note 1.1, and Supplementary Figure~4), and are aged between 18 and 65 years old, with average at 36 years old. About 1/3 of users are female. Data is not collected with a fixed time interval. Instead, the app gets updates when there is a change is the motion-state of the device (if the accelerometer registers a change, see Supplementary Figure~5). Location estimation error is below $100$ meters for 93\% of data points. To preserve privacy, GPS traces were pre-processed (internally at SONY Mobile) to infer stop-locations using the method described in \cite{cuttone2014inferring}. The method is built on the idea that a stop corresponds to a temporal sequence of locations within a maximal distance $d_{max}$ from each other. The results presented are for $d_{max}=30 m$. Data collection for the Sony dataset has been approved by the Sony Mobile Logging Board and informed consent has been obtained for all study participants according to the Sony Mobile Application Terms of Service and the Sony Mobile Privacy Policy. 

\textbf{MDC dataset:} Data was collected by the Lausanne data Collection Campaign between October 2009 and March 2011. The campaign involved an heterogeneous sample of $\sim185$ volunteers with mixed backgrounds from the Lake Geneva region (Switzerland), who were allocated smart-phones \cite{laurila2012mobile}. In this work we used GSM data since the GSM data has higher sampling frequency than the GPS data collected from the same experiment. Following Nokia's privacy policy, individuals participating in the study provided informed consent \cite{laurila2012mobile}. The Lausanne Mobile Data Challenge experiment involves 62\% male and 38\% female participants, where the age range 22-33 year-old accounts for roughly 2/3 of the population \cite{laurila2013big}.

\section*{Data Availability Statement}
\textbf{CNS dataset:} Data from the Copenhagen Networks study are not publicly available due to privacy considerations including European Union regulations and Danish Data Protection Agency rules. Due to the data security of participants, data cannot be shared freely, but are available to researchers who meet the criteria for access to confidential data, sign a confidentiality agreement, and agree to work under supervision in Copenhagen. Please direct your queries to Sune Lehmann, the Principal Investigator of the study, at sljo@dtu.dk. \\
\textbf{MDC dataset:} The Lausanne Mobile Data Challenge data are available from Idiap Research Institute but restrictions apply to the availability of these data, which were used under license for the current study, and so are not publicly available. Data are however available from Idiap Research Institute to eligible institutions upon reasonable request (\url{https://www.idiap.ch/dataset/mdc/download}). \\
\textbf{Lifelog dataset:} Raw data are not publicly available to preserve users' privacy under Sony Mobile Privacy Policy. Derived data supporting the findings of this study are available from the corresponding authors upon request. \\
\textbf{RM data:}  The Reality Mining Dataset is available from MIT Human Dynamics Lab (\url{http://realitycommons.media.mit.edu/realitymining4.html}).
\section*{Code Availability Statement}
The code used to generate results is available from the corresponding authors upon request.

\section*{Correspondence}
Correspondence and requests for materials should be addressed to Sune Lehmann (sljo@dtu.dk) or Andrea Baronchelli (a.baronchelli.work@gmail.com).

\section*{Acknowledgements}\label{aknow}

This work was partially supported by the Villum Foundation (``High resolution Networks" project, SL is PI), the  UCPH-2016 grant (``Social Fabric", SL is Co-PI), and the Danish Council for Independent Research (``Microdynamics of influence in social systems", grant id. 4184-00556, SL is PI). Portions of the research in this paper used the MDC Database made available by Idiap Research Institute, Switzerland and owned by Nokia. VS was supported by Sony Mobile Communications. The funders had no role in study design, data collection and analysis, decision to publish, or preparation of the manuscript

\section*{Author Contributions}\label{contributions}

LA, SL and AB designed the research. LA, PS and VS pre-processed the data. LA performed the data analysis. LA, SL and AB analysed the results and wrote the paper.

\bibliographystyle{unsrt}
\bibliography{biblio}

\section*{Competing Interests Statement}\label{competing}
The authors declare no competing interests.

\clearpage
\renewcommand{\figurename}{\textbf{Supplementary Figure}}

\onehalfspacing
\begin{Large}{\begin{center}
{\bf{Evidence for a Conserved Quantity \\
 in Human Mobility~-~Supplementary Information}}
\end{center} 
 }
\end{Large}
\thispagestyle{empty}
\tableofcontents
\clearpage

\section{Supplementary Notes}
\subsection{Data pre-processing}
\label{preprocessing} 
The four datasets considered collect different types of location data. For each of them we obtained sequences of intervals describing individuals' pauses at a given location:

\begin{tabular}{c c c c}
 \textit{User} & \textit{Interval Start} & \textit{Interval End} & \textit{Location} \\
\end{tabular}

In this section, we describe data collection and the pre-processing applied to obtain such records. A summary of the datasets' characteristics is presented in \ref{Table_data}. Other properties are shown in Supplementary Figures~\ref{duration_experiment}, \ref{temporal_resolution}, \ref{temporal_resolution2}, and Table~\ref{Table0}. For all datasets, we consider only intervals longer than $10$ minutes.

\subsubsection*{Lifelog dataset}
\label{Sony} 

\begin{description}

\item[Data Collection]{Data was collected by the Lifelog Sony app \cite{Lifelog}. The app is opportunistic in collecting location data. (i.e. if another app requests location data for the device, Lifelog will get a copy of the location). The app does not collect locations with a fixed time interval. Instead, the heuristic is to get updates when there is a change is the motion-state of the device (if the accelerometer registers a change), or if the app uploads/downloads data to/from the servers, which by default is set to at least once per day. Communication with servers can be more frequent as the app will connect to the servers every time it is opened. If two data-points are close together in time (less than 15 minutes) and space the backend aggregates them. The spatial distribution of data points is shown in Supplementary Figure~ \ref{SonySpatial}. }

\item[Selection of users]{
 We have selected users who have data for at least 365 days ($\sim 36.000 users$).}

\item[Definition of Locations]{
GPS data is pre-processed to infer stop-locations using the distance grouping method described in \cite{cuttone2014inferring}. The method is built on the idea that a stop corresponds to a temporal sequence of locations within a maximal distance $d_{max}$ from each other. In the main text, results are presented for $d_{max}=50 m$. Below, we show that the same results hold for $d_{max}=30 m$, $d_{max}=40 m$ and $d_{max} = 500 m$ (see section \emph{Robustness Tests} )}

\item[Data Cleaning]{
During the data collection period, the app settings changed causing a considerable change in time coverage for a subset of users (see Supplementary Figure~\ref{temporal_resolution2}). We propose two methods to solve this issue (see Supplementary Figure~\ref{sony_preprocessing}):
\begin{itemize}
\item[(a)]{Users selection: }{We consider only the subset of users for which there is no change in time-coverage over time ($\sim 6 \% $ of all users)}
\item[(b)]{Temporal down-sampling: }{We down-sample data to achieve constant time-coverage across time. The method used relies on:
\begin{itemize}
\item{Find for each user $i$ the week $w_m$ with lowest weekly time-coverage  $tC(w_m)$.}
\item{Down-sample weeks with weekly time-coverage higher than $w_m$ by selecting a random sample of total duration $tC(w_m)*60$ minutes.}
\end{itemize}}
\end{itemize}
Results presented in the main text are produced with method (a). We show below (see section \emph{Robustness Tests}) that results hold also under method (b). 
}

\end{description}

\subsubsection*{CNS mobility dataset}
\label{CNS} 

Location data is obtained combining Wi-Fi data (sampled every $\sim 15 s$) with GPS data (high spatial resolution). The following methodology was implemented to estimate the sequences of individuals stop-locations:\\

\begin{description}

\item[Estimation of Wi-Fi Access Points (AP) position]{Access Points (AP) positions were estimated using participants' sequences of GPS scans. We discarded \textit{mobile APs}, that are located on buses or trains, and \textit{moved APs} that were displaced during the experiment (for example by residents of Copenhagen changing apartment, taking their APs with them). Then, we considered all WiFi scans happening within the same second as a GPS scan to estimate APs location. The APs location estimation error is below 50 meters in 95\% cases. Most of the APs are located in the Copenhagen area (see \cite{sapiezynski2015opportunities} for a detailed description of the methodology). }

\item[Definition of Locations]{

We find locations by clustering APs based on the distance between them. First, we built the indirect graph of APs simultaneous detection $G=(V,E)$. $V$ is the set of geo-localized APs, links $e(j,k)$ exist between pairs of access points that have ever been scanned in the same $1$ min bin by at least one user. Then, we compute the physical distances $dist(j,k)$ for all pairs of $(j,k) \in E$. 
and we consider the set of links $E_D \subset E$ such that $dist(j,k)<d$, where $d$ is a threshold value, to define a new graph $G_d=(V, E_d)$. 
Finally, we define a \emph{location} as a connected component in the graph $G_d$. For $d=5m$ the maximal distance between two APs in the same location is smaller than $10m$ for most locations and at most $\sim 200m$ (see Supplementary Figure~\ref{CNS_locations}-A). The number of APs in the same location is lower that 10 for most locations, but reaches $\sim 1000$ for dense areas such as the University Campuses  (see Supplementary Figure~\ref{CNS_locations}-B).  An example of APs clustering for $d=5m$ and $d=10m$ is shown in Supplementary Figures~\ref{CNS_locations}-C and \ref{CNS_locations}-D. We show below that our findings do not depend on the choice of the threshold (see section \emph{Robustness Tests}).}

\item[Temporal aggregation]{Data was aggregated in bins of length $1$ min, where for each bin we selected the most likely location.}

\end{description}

\subsubsection*{MDC mobility dataset}
\label{MDC} 

Data collection is described in \cite{laurila2012mobile} and \cite{kiukkonen2010towards}. We used the GSM data, sampled every 60 seconds.

\subsubsection*{RM mobility dataset}
\label{RM} 

Data collection is described in \cite{eagle2010reality} and \cite{eagle2006reality}. We used the GSM data.

\subsection{Comparison with previous research}
\label{comparison} 

Our datasets displays statistical properties consistent with previously analyzed data on human mobility. 
\begin{itemize}

\item{\textbf{Rank-frequency distribution of locations:} The visitation frequency of a location, defined as the fraction of visits to that location, goes with the location rank $r$ as $r^{-\zeta}$, with $\zeta \sim 1$.(Supplementary Figure~\ref{agreement_with_previous_research}-A). Our result is consistent with \cite{gonzalez2008understanding}, where the authors found $f(r)\propto 1/r$ , and \cite{song2010modelling}, where it was found  $f(r)\propto r^{-1.2}$.}

\item{\textbf{Distribution of displacements:} The distribution between consecutive jumps $P(\Delta r)$ has a power law tail (Supplementary Figure~\ref{agreement_with_previous_research}-B), with exponent $\beta=1.81$. Gonzalez et. al \cite{gonzalez2008understanding} found $\beta=1.77$ for the truncated power-law distribution, Song et. al \cite{song2010modelling} found a power-law tail with exponent $\beta=1.55$.}

\item{\textbf{Growth of the radius of gyration:}  Individuals' total radius of gyration (see \cite{gonzalez2008understanding}, SI for definition) growth across time is consistent with the logarithmic growth described in \cite{song2010modelling}(Supplementary Figure~\ref{agreement_with_previous_research}-C).}

\item{\textbf{Distribution of the radius of gyration: }Individuals are distributed heterogeneously with respect to their total radius of gyration measured at the end of the experiment, with the probability distribution $P(r_g)$ (Supplementary Figure~\ref{agreement_with_previous_research}-D) decaying as a power-law with coefficient $\beta=-1.47$.  This is comparable with the results  found in \cite{gonzalez2008understanding},  $\beta=-1.65$ and \cite{song2010modelling} $\beta=-1.55$, where both studies relied on CDRs.}

\item{\textbf{Returners and explorers: } In accordance with Pappalardo et al. \cite{pappalardo2015returners}, the distribution of $r^3_{g}/r_{g}$ is bimodal (Figs~\ref{returners_and_explorers} and \ref{returners_and_explorers_2}, A and B), where $r_g$ is the radius of gyration computed across a window of $20$ weeks and $r^3_g$ is the radius of gyration computed within the same window including only the top $2$ locations (see \cite{pappalardo2015returners} for the definition). Hence, within each window, an individual can be categorized as either a \emph{returner} (if $r^3_{g}/r_{g}<0.5$) or as an \emph{explorer} (if $r^3_{g}/r_{g}>0.5$). We find that this categorization is stable in time for  $\sim 50\%$ of individuals (Figs~\ref{returners_and_explorers} and \ref{returners_and_explorers_2}, D).}

\end{itemize}

\subsection{Robustness Tests} 
\label{Robustness} 

The results presented in the main text do not depend on how locations are defined, nor on the time-window used to investigate the long-term behavior. In this section, we show how the results are derived and we demonstrate their statistical robustness. To avoid confusion, we will indicate with $\overline{x}$ the average value of a quantity $x$ across the population, and $\langle x \rangle$ the average across time.

\subsubsection*{Conservation of the location capacity} 

The \emph{activity set} is defined here as the set $AS_i(t)=\{\ell_1,\ell_2,...,\ell_k,...\ell_C\}$ of locations $\ell_k$ that individual $i$ visited at least twice and where she spent on average more than $10$ minutes/week during a time-window of $W$ consecutive weeks preceding time $t$. In Supplementary Figure~\ref{activity_space}, we show that for $W=10$ weeks, the set contains on average a small fraction of all locations seen during the same 10 weeks. Yet, the time spent in these locations is on average close to the total time (Supplementary Figure~\ref{activity_space}). 
Given this definition, the number of locations an individual $i$ visits regularly is equivalent to the set size $C_i(t)=|AS_i (t)|$. We call this quantity \emph{location capacity}.

\textbf{Evidence 1} The average individual location capacity $\overline{C}$ is constant in time regardless of the definition of location or the choice of the window size $W$ (Table \ref{conservation_cap_1} and Table \ref{conservation_cap_2}). This result is tested in several ways:

\begin{description}

\item[1]{Linear Fit Test: We perform a linear fit of the form $\overline{C(t)}= a + b \cdot t$, computed with the least squares method. We test the hypothesis $H_0: b=0$, under independent 2-samples t-tests.}

\item[2]{Power Law Fit Test: We perform a power-law fit of the form $\overline {C(t)} \propto t^{\beta}$, computed with the least squares method. We test the hypothesis $H_1:\beta=0$, under independent 2-samples t-tests.}

\item[3]{Multiple intervals test: We compare the value of  $\overline{C}$ across different time-intervals  $\delta t_k$. We divide the total time range into time-intervals  $\delta t_k$ spanning $w$ weeks. We compute the average capacity $\overline{C (\delta t_k)}$ and its standard deviation $\sigma_C(\delta t_k) $ for each time-interval $\delta t_k$. We test the hypotheses $H_{j,k}: \overline{C (\delta t_j)}=\overline{C (\delta t_k)}$ for all pairs $\delta t_k, \delta t_j$. }

\end{description}

For all the datasets considered, all choices of $W$, and  definitions of locations the hypotheses $H_0$, $H_1$ and $H_{j,k}$ (for all intervals $j$ and $k$) can not be rejected at $\alpha=0.05$ with p-value$>\alpha$ under 2-tailed tests. Results are reported in Table \ref{conservation_cap_1}. \\

\textbf{Evidence 2} The individual weekly \emph{net gain} of locations is equal to zero. The \emph{net gain} defined as $G_i(t)=A_i(t)-D_i(t)$, where  $A_i(t)=|AS_i(t) \setminus AS_i (t-dt)|$ is the number of location added and  $D_i(t)=|AS_i(t-dt)\setminus AS_i(t)|$ (the difference between the sets) is the number of location removed from the set during $dt$, where $dt=1$ week. This is verified by testing for all individuals $i$ if the ratio $\sigma_{G,i}/ \langle G _i \rangle>1$, where $\sigma_{G,i}$ is the standard deviation of the average individual net gain across time (see main text).
We find that $\sigma_{G,i}/ \langle G _i \rangle>1$ hold for a large majority of individuals, under different definitions of locations and choices of $W$, for all datasets considered. Results are reported in Table \ref{conservation_cap_2} and Supplementary Figure~ \ref{gain_window}.

\textbf{Evidence 3} The average value of location capacity saturates for increasing values of the time-window $W$. We find that for all datasets the average time coverage $\overline{\langle C} \rangle \sim 25$. This result is obtained after accounting for the differences in data collection by considering the normalized location capacity $C_i/TC_i$, where $TC_i$ is the weekly time coverage of individual $i$ (see Supplementary Figures~ \ref{25}, \ref{capacity_window}). Individuals' capacity values are distributed homogeneously around the mean (Supplementary Figure~ \ref{individual_capacity}). \\

\subsubsection*{Evolution of the activity set:  Invariance under time translation} 
We verified that the evolution of the activity set is not influenced by the particular time at which the data collection started or by the time elapsed from that moment. We borrow the concept of \textit{aging} from the physics of glassy systems \cite{struik1977physical,mukherjee2011aging}. A system is said to be in equilibrium when it shows invariance under time translations; if this holds, any observable comparing the system at time $t$ with the system at time $t+\gamma$ is independent of the starting time $t$. In contrast, a system undergoing aging is not invariant under time translation. This property can be revealed by measuring correlations of the system at different times.

We measure the evolution of the activity set, starting at different initial times $t$, to verify if the system undergoes aging effects. The evolution is quantified measuring the Jaccard similarity $J_i(t,\gamma)=|AS_i(t)\cap AS_i(t+\gamma)|/|AS_i(t)\cup AS_i(t+\gamma)|$ (see MS). The average similarity $\overline{ J (t,\gamma)}$ decreases in time: power-law fits of the form  $\overline{ J(t,\gamma)}=\sim \gamma^{\lambda(t)}$ yield $\lambda<0$ for all $t$. The fit coefficient $\lambda(t)$ fluctuates around a typical value, probably due to seasonality effects (see Supplementary Figures~\ref{seasonality1} and \ref{seasonality2}). However, it does not changes substantially as a function of the starting time $t$ (Supplementary Figure~\ref{evolution_aging}), hence $\overline{J(t, \gamma)}=\overline{J(\gamma)}$. This implies that the rate at which the activity set evolves does not substantially depends on when the measure is initiated. We conclude that our data reflect the `equilibrium' behavior of the monitored individuals. The fact that our dataset allow us to replicate measures performed on other datasets obtained with different methods (see above) further confirms this finding. 

Note that the evolution of the activity set can be measured as the cosine similarity between vectors constructed from the sets $AS_i(t)$ and $AS_i(t+\gamma)$. The vector components are the probability of visiting locations (i.e. the fraction of time spent in that location). Results (see Supplementary Figure~\ref{cosine_similarity}) confirm that the activity set evolves in time.

\subsubsection*{Sub-linear growth of number of locations} 
We quantify exploration behavior, measuring the number of locations $L_{i}(t)$ discovered up to day $t$. In the main text, we show that $L_{i}(t)$ grows sub-linearly in time. Here, we show that this holds also changing the definition of locations (See Supplementary Figure~ \ref{definitions_of_locations}). This property of exploration behavior is not affected by the waiting time before starting the measure as we verify by repeating the same measures starting $M$ months after the participant received the phone, for several values of $M$ (See Supplementary Figure~ \ref{exploration_aging}).

For the MDC dataset, we find that the growth of locations is sublinear independently of users' age. We find a positive relation between the coefficient ${\alpha_i}$ describing the growth of location $L_i\propto t^{\alpha_i}$ and the age of individual $i$ (Pearson correlation $\rho = 0.2$, $p-value=0.008$, see Supplementary Figure~\ref{age}).

\subsubsection*{Discrepancy relative to the randomized cases} Individual capacity is lower than it could be if individuals were only subject to time constraints. We showed this by randomizing individual temporal sequences of stop-locations for 100 times, and then comparing the average randomized capacity   $\langle C_{rand,i} \rangle$  with the real capacity   $\langle C _i \rangle$. We perform two types of randomizations (see Supplementary Figure \ref{randomization}): \\
\begin{description}
\item[1]{Local randomization: For each individual $i$, we split her digital traces in segments of length 1 day. We shuffle days of each individual. }
\item[2]{Global randomization: For each individual $i$, we split her digital traces in segments of length 1 day. We shuffle days of different individuals.}
\end{description}

The individual randomized capacity $\langle C_{rand,i} \rangle$ averaged across time, (see Supplementary Figure~\ref{randomization2}),  is higher than in the real case both for the global and the local randomization cases. We compute the Kolmogorov–Smirnov test-statistics (Table \ref{ks_statistics}) to compare the real sample with the randomized samples. We reject the hypothesis that the two samples are extracted from the same distribution since $p<\alpha$ with $\alpha=0.05$.

\subsubsection*{Conservation of time allocation} 

Individuals allocate time heterogeneously among locations, due to their different functions (homes, work-places, shops, universities, leisure places...).  We study time allocation between different classes of locations considering subsets of the activity set defined on the basis of the total visitation time. The subsets $AS_i(t)^{\Delta T} \in AS_i(t)$ include all locations  seen in the $W$ weeks preceding $t$ at least twice and such that $W*\Delta t(0)<T_{i,\ell}(t)<W*\Delta t(1)$  where $T_{i,\ell}(t)$ is the time of observation of location $\ell$ during the $W$ weeks preceding $t$.
We test several choices of intervals $\Delta T$. We find that when $\Delta T$ increases, the subsets are empty for many individuals, since no locations satisfy the above-mentioned criteria. In Supplementary Figures~\ref{time_alloc_sony}, \ref{time_alloc_cns}, \ref{time_alloc_mdc}, \ref{time_alloc_rm} we show the distribution of average individual sub-capacities $\langle C^{\Delta T} _i \rangle$.  Only subsets with small enough $\Delta T$ are significant for more than 50\% of the population, and typically each individual has 1 location where he/she spend more than 48 hours per week. 
The average sub-capacities $\overline{ C}^{\Delta T}(t)$ are constant in time for several choices of $\Delta T$ and different definitions of location. This is verified with the linear fit test as detailed in a previous section (see table \ref{table_time_alloc}).  
The Jaccard similarity between the subsets $AS_i(t)^{\Delta T}$ and $AS_i(t+w)^{\Delta T}$ increases, on average, with $\Delta T$ (see Table \ref{subset_evolution}), suggesting highly visited locations are replaced less frequently.

\subsection{The EPR model with memory} 

The state-of-the-art exploration and preferential return model (EPR) \cite{song2010modelling}, and its modifications d-EPR \cite{pappalardo2015returners}, r-EPR \cite{jiang2016timegeo}, and recency-based EPR \cite{barbosa2015effect}, reproduce the conserved size of the individual capacity (Figure~5A), but do not account for the evolution of the activity set (Figure~5C). According to the EPR model, at a given transition $n$, an individual explores a new location with probability $P_{new} =\rho S - \gamma$, or returns to a previously visited location with probability $1 - P_{new}$, with $S$ the number of previously visited locations, and $\rho$ and $\gamma$ parameters of the model. If the individual returns to a previously visited location, she chooses location $i$ with probability $\Pi_{i} = m_i/\sum_{i}m_i(n)$ where $m_i$ is the total number of visits to location $i$ occurring before transition $n$. In the EPR model, time scales with the number of transitions as  $\sim n/\beta$, with $\beta$ a parameter of the model. \\

We introduce the limited-memory exploration and preferential return model. Agents obey the same exploration strategy as in the EPR model, but dispose of a limited memory $M$. Hence, the return probability to a given location $i$ is $\Pi_i= m_i/\sum_{i}m_i(n)$, where $m_i$ is the total number of visits to location $i$ occurring at most $M$ days before transition $n$. \\

In Supplementary Figure~\ref{Figure9} the EPR model with memory is compared to the EPR model with parameters chosen in \cite{song2010modelling}: $\beta = 0.8$, $\rho =0.6$, $\gamma = 0.2$. Time is mapped as $1u = 1 min$, where $u$ is the simulation time unit, and memory is set to $M = 200$ days. The model is also compared with the recency-based model \cite{barbosa2015effect}, with parameters from  \cite{barbosa2015effect}: $\alpha = 0.1$, and $\eta = 1.6$.

\subsection{Additional measures}

\subsubsection*{Spatial properties of the activity set}
Two spatial properties of the activity set $AS$ (see section 3) are its center of mass and its radius of gyration (see also \cite{gonzalez2008understanding, pappalardo2015returners}). The center of mass is computed as:

\[
\vec{r_{cm}} = \frac{1}{T}\sum_{j \in AS} t_j \vec{r_j}
\]
where $T$ is the total time spent in the activity set, $t_j$ is the time spent in location $j$ and $\vec{r_j}$ is the spatial position of location $j$. The radius of gyration is computed as: 

\[
r_{g,i} (t) = \sqrt{\frac{1}{T} \sum_{j \in AS} t_j (\vec{r_j} - \vec{r_{cm}}})^2
\]

We compute the aforementioned quantities for the CNS and Lifelog data (for which spatial coordinates of locations are available). We find that the center of mass of the activity set changes position, on average (see Supplementary Figure~\ref{rcm}): The average distance between $r_{cm}(t)$ and $r_{cm}(t+\gamma)$ increases as a function of $\gamma$, suggesting that individuals displace their important locations gradually in time.
Instead, the median radius of gyration is constant in time, with a linear fit $r_{cm} = a + b \cdot t$ yelding $a = 11 \pm 8$ $Km$  and $b= -0.02 \pm 0.15$ $Km/week$ (see Supplementary Figure~\ref{rg}). 

Finally, we find that the location capacity is different between two classes of individuals defined in \cite{pappalardo2015returners} based on the spatial distribution of their important locations: the so-called `\emph{returners}' and `\emph{explorers}' (see also section `\emph{Comparison with previous research}'). Individuals defined as explorers (see Figs~\ref{returners_and_explorers} and\ref{returners_and_explorers_2}) have higher capacity under the Kolmogorov-Smirnov test-statistics (see Supplementary Figure~\ref{returners_explorers_capacity}). 

\subsubsection*{Additional evidence for the conservation of location capacity}

The evolution of mobility behavior can be also analyzed by measuring the aggregated number $n_{L,i}(T)$; the total number of locations added $n_{A,i}(T)$ and removed $n_{D,i}(T)$ from the activity set after $T$. To ensure large enough samples, in the following we consider $T=12$ months.

First, we find that, for all datasets, the number of locations added,$n_{A,i}$, and dismissed, $n_{D,i}$, constitute a substantial proportion of all important locations $n_{L,i}$ (i.e., they are larger than $n_{L,i}/3$, Supplementary Figure~\ref{n_l,n_a,n_d}). This result confirms that there is turnover of locations in the activity set. 
Secondly, we find that for most users in our database we get $n_{A,i} \sim n_{D,i}$ (see details of Supplementary Figure~\ref{n_a vs n_d}), confirming that the number of newly adopted locations equals the number of dismissed locations. 
Finally, we find that for most users $n_{A,i}\sim b*C{i}$, where $C_{i}$ is the location capacity (Lifelog: $b=1.85$, CNS $b=2.00$, MDC $b=1.61$, RM $b=1.25$), see Supplementary Figure~\ref{n_a vs n_l}). This result implies that the number of newly added and removed locations tends to be proportional to the location capacity, highlighting that individuals with larger capacity have more unstable routines. 
Interestingly, the results above are comparable to those obtained by Miritello et al. \cite{miritello2013limited}, who studied the long-term dynamics of social interactions extracted from mobile phone communication data. Their analysis shows that the number of deactivated ties in a given time window equals the number of activated ties, implying that the number of active ties (defined as \emph{communication capacity}) is conserved in time.

\clearpage

\section{Supplementary Figures}

\begin{figure*}[h]
\centering
\includegraphics[width=.9\textwidth]{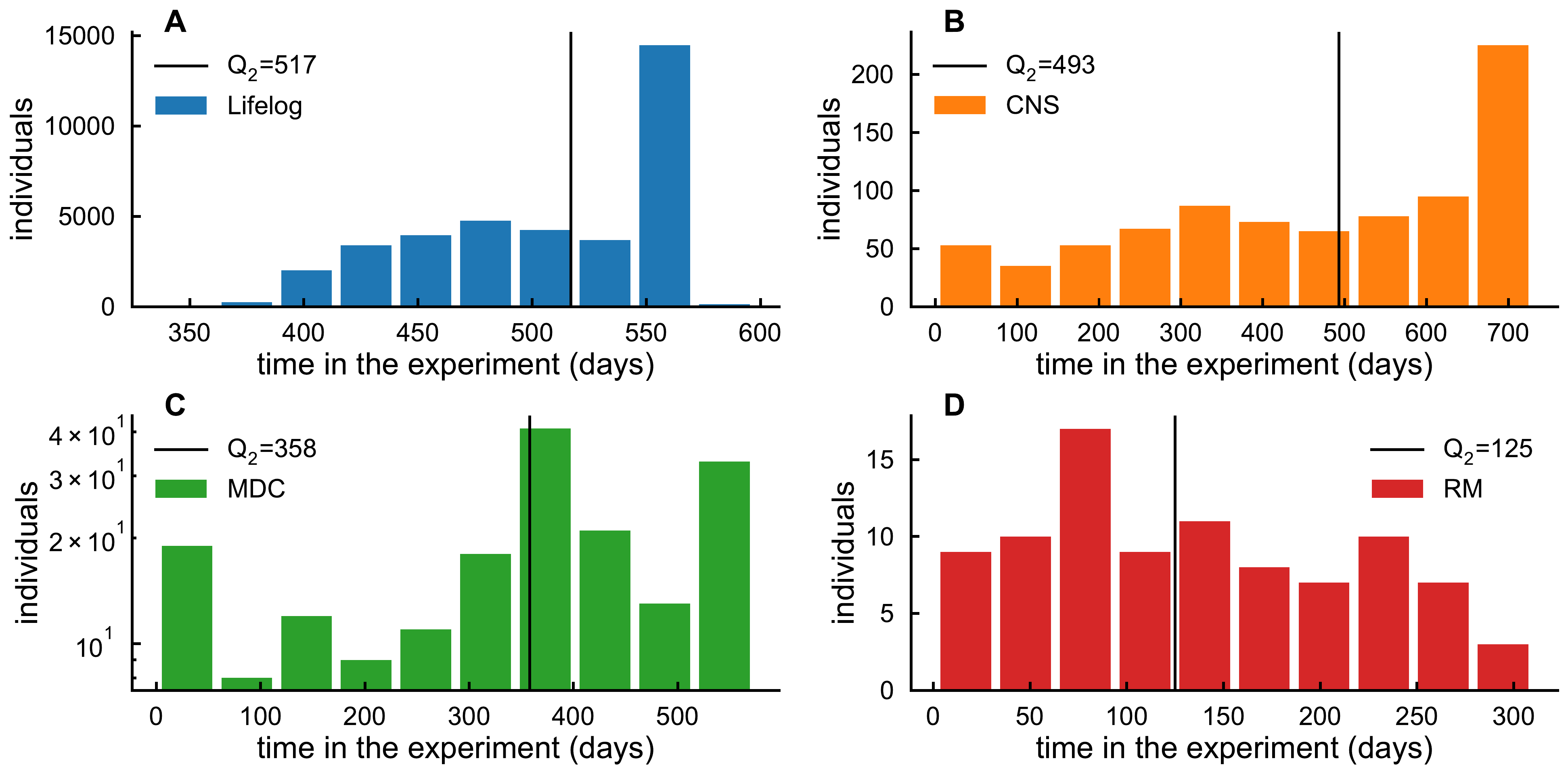}
\caption{\textbf{Long duration of the datasets considered} Frequency histogram of individuals' data collection duration for the Lifelog \textbf{(A)}, CNS \textbf{(B)}, MDC \textbf{(C)} and RM \textbf{(D)} datasets. The black line is the median value across the population.}
\label{duration_experiment}
\end{figure*}
\begin{figure*}[h]
\centering
\includegraphics[width=.9\textwidth]{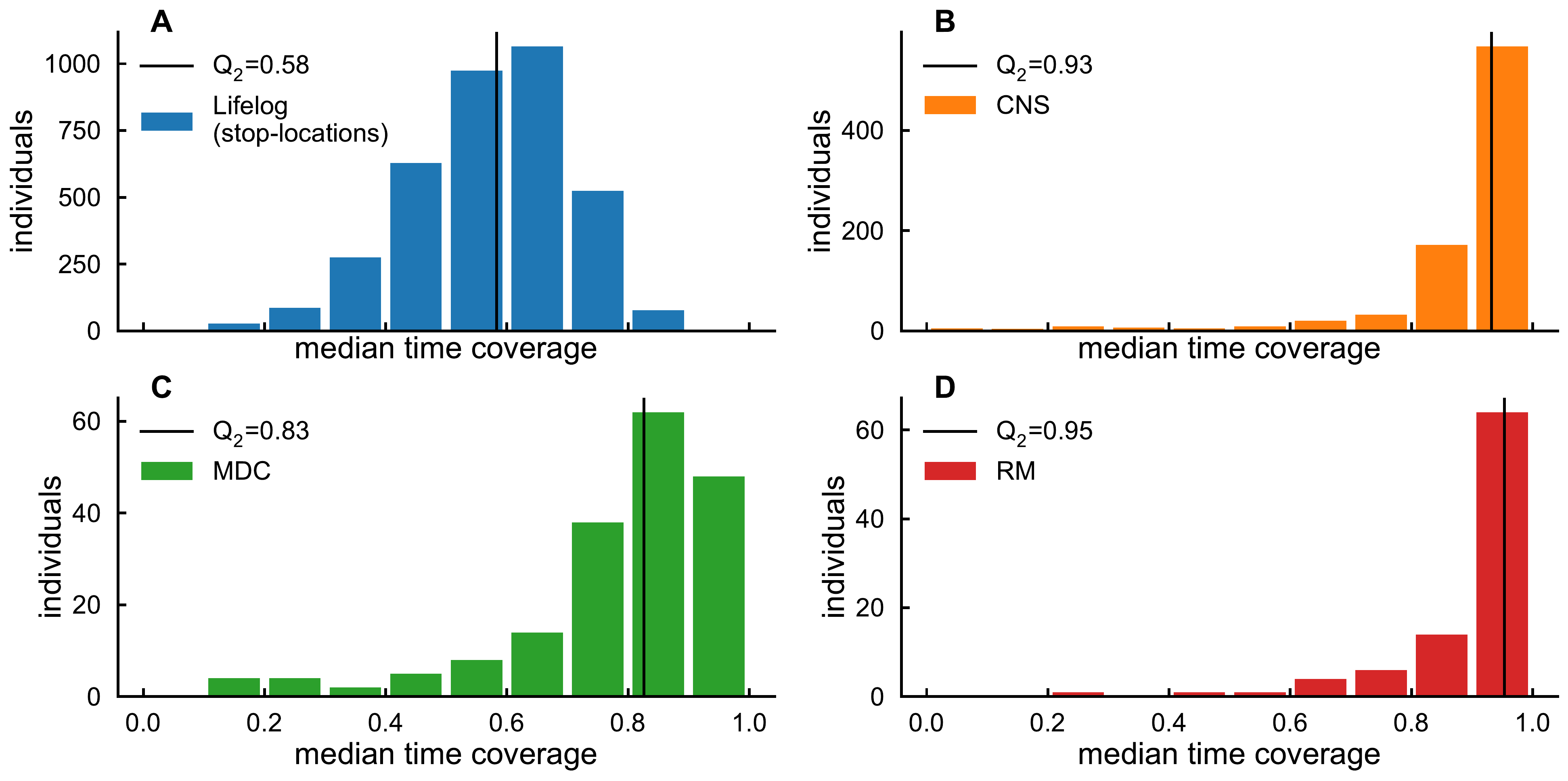}
\caption{\textbf{High individual temporal resolution} Frequency histogram of individuals' median weekly time coverage for the Lifelog \textbf{(A)}, CNS \textbf{(B)}, MDC \textbf{(C)} and RM \textbf{(D)} datasets. The black line is the median value across the population.}
\label{temporal_resolution}
\end{figure*}
\begin{figure*}[h]
\centering
\includegraphics[width=.9\textwidth]{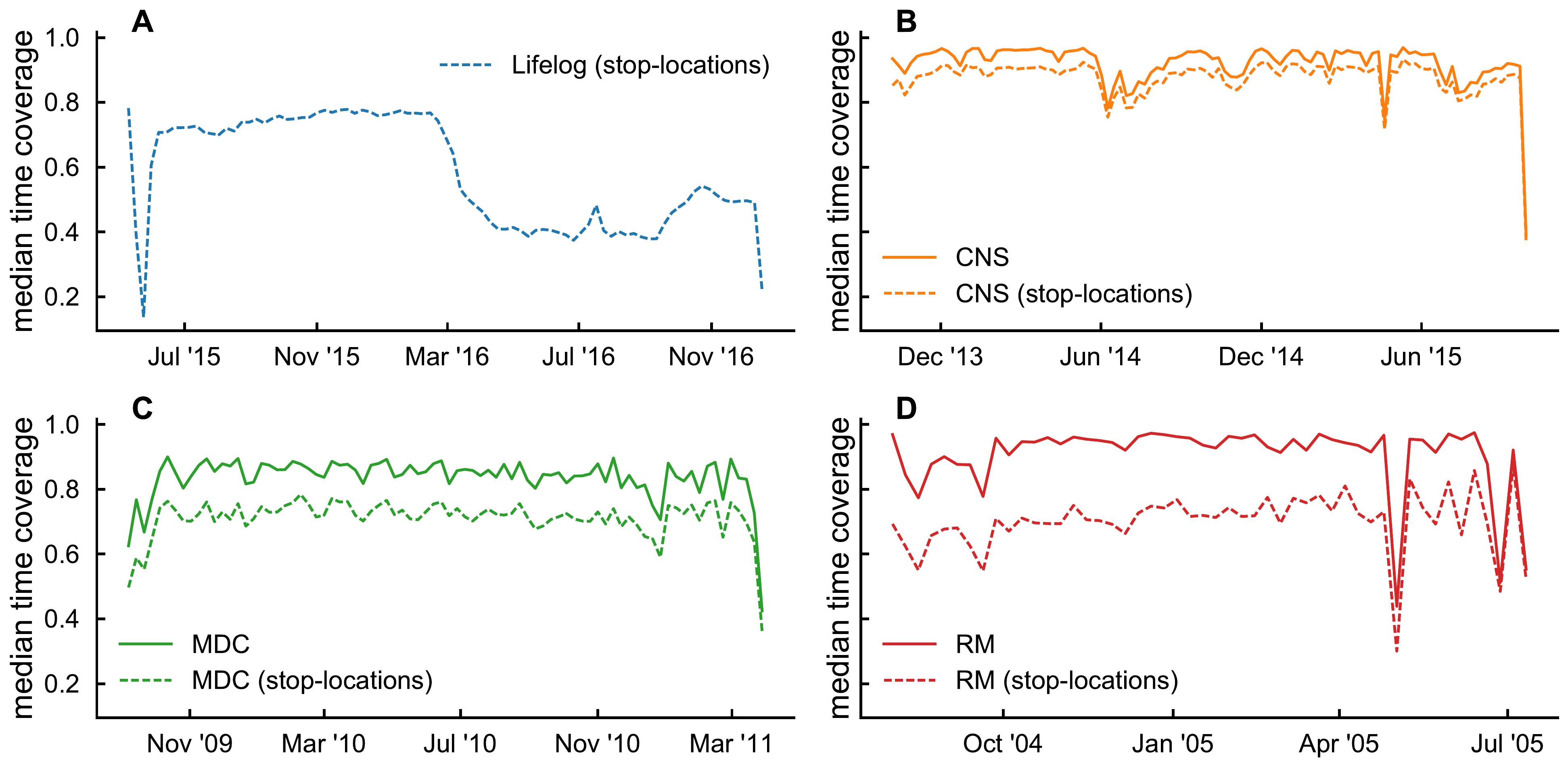}
\caption{\textbf{High temporal resolution} Median weekly time coverage across the population as a function of time for the Lifelog \textbf{(A)}, CNS \textbf{(B)}, MDC \textbf{(C)} and RM \textbf{(D)} datasets. Filled lines are computed considering all locations, dashed lines are computed considering only stop-locations (where individuals spend more than 10 consecutive minutes).}
\label{temporal_resolution2}
\end{figure*}
\begin{figure*}[h]
\centering
\includegraphics[width=.9\textwidth]{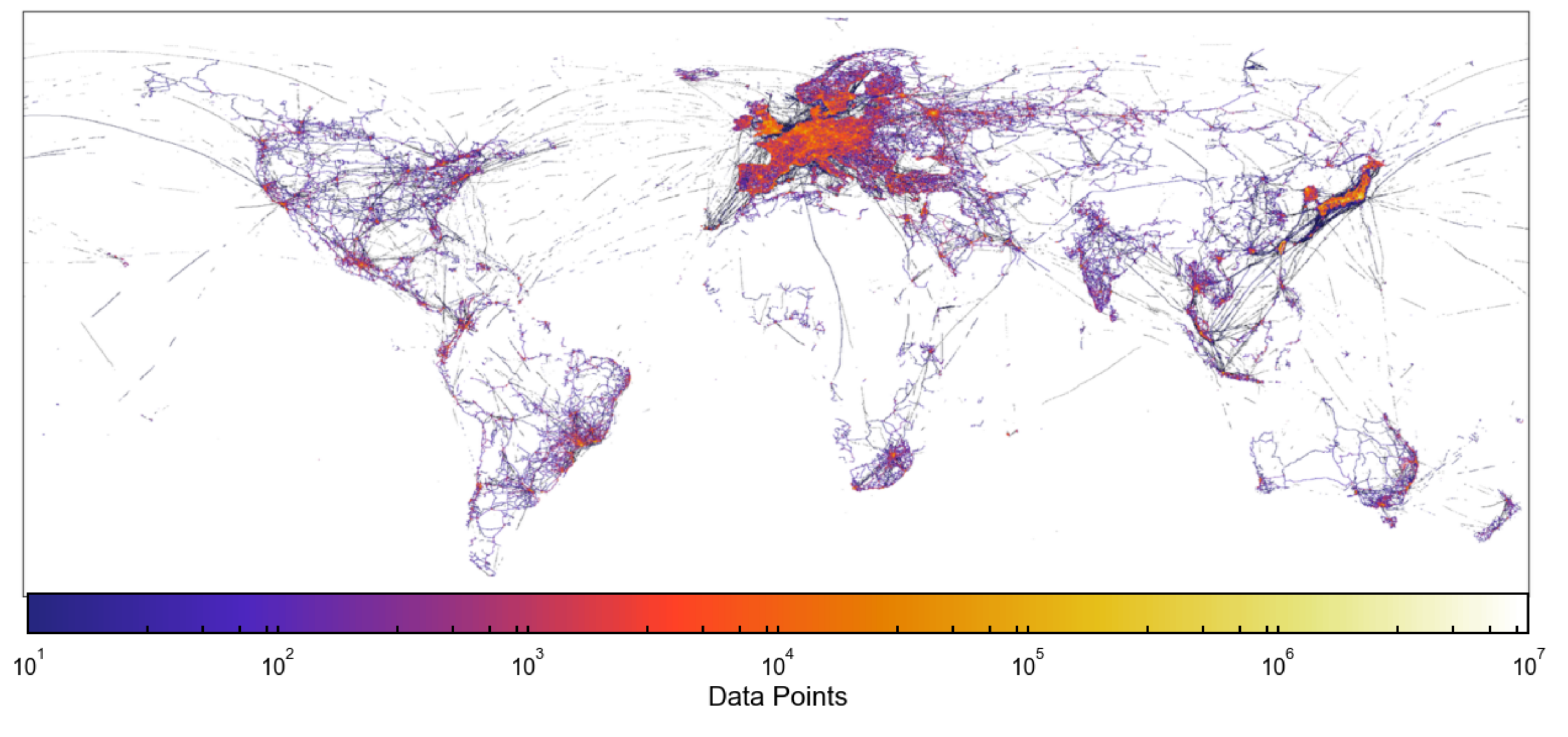}
\caption{\textbf{Broad spatial coverage of the Lifelog dataset.} Heatmap showing the spatial distribution of data points in the Lifelog dataset. }
\label{SonySpatial}
\end{figure*}
\begin{figure*}[h]
\centering
\includegraphics[width=.9\textwidth]{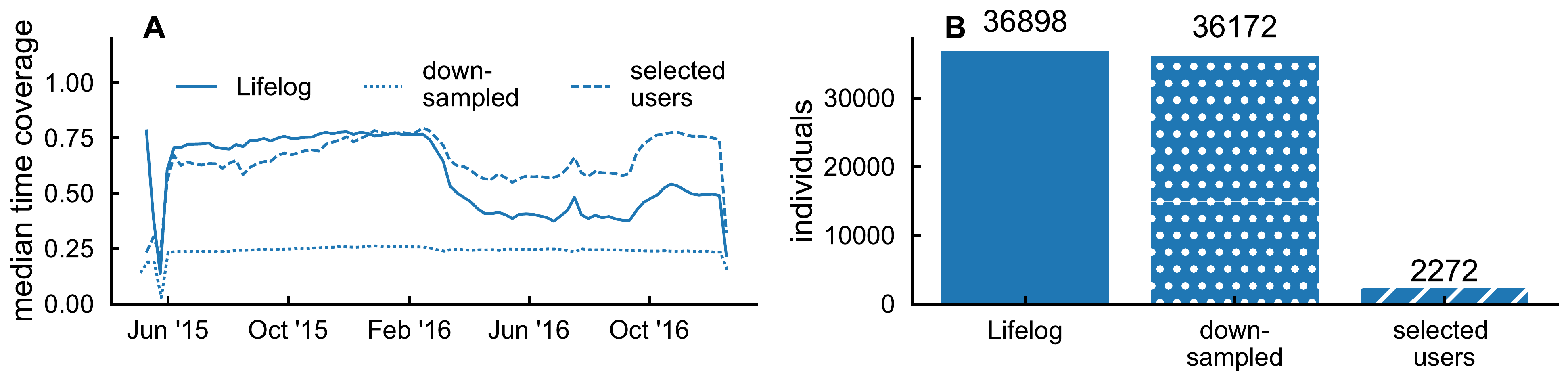}
\caption{\textbf{Lifelog dataset: pre-processing} \textbf{(A)} Median weekly time coverage across the population as a function of time for the raw Lifelog dataset (filled line), after downsampling (dotted line) and after user selection (dashed line). \textbf{(B)} Number of individuals in the dataset (filled bar), after downsampling (dotted bar), and after user selection (dashed bar).}
\label{sony_preprocessing}
\end{figure*}
\begin{figure*}[h]
\centering
\includegraphics[width=.9\textwidth]{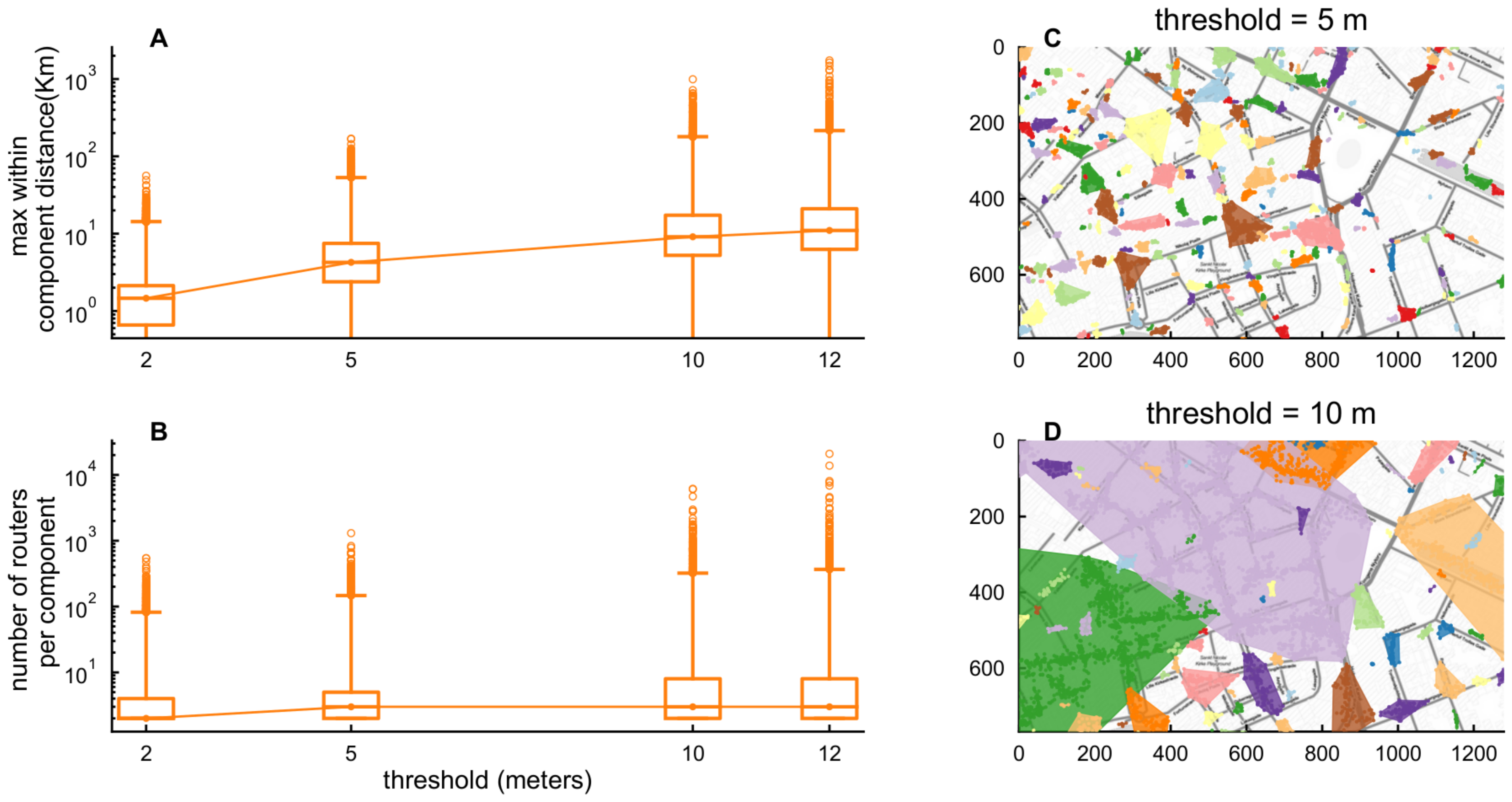}
\includegraphics[width=.9\textwidth]{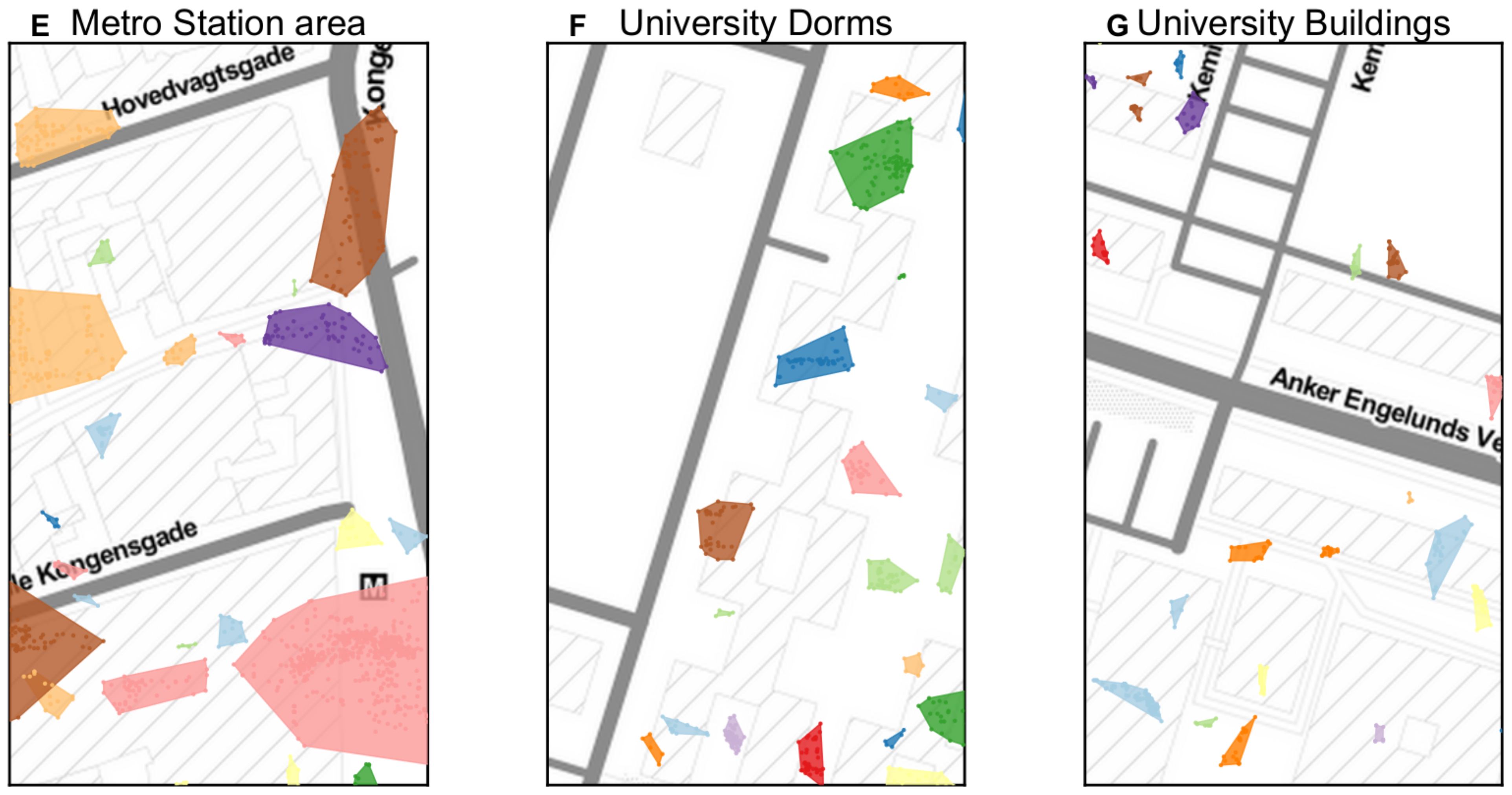}

\caption{\textbf{CNS dataset: Different definitions of locations}
\textbf{(A)} The boxplots of the maximal distance between pairs of geo-localized APs forming a location, as a function of the threshold $d$ used to merge APs. Boxes are set at the 1st and 3rd quantile, while whiskers at 2.5\% and 97.5\%. \textbf{(B)} The boxplots of the locations size (number of APs) as a function of the threshold $d$. 
\textbf{C-D}) An example of the clustering of APs located within Copenhagen city for thresholds $d=5m$\textbf{(C)} and $d=10m$\textbf{(D)}. Dots corresponds to geo-localized APs, colored according to the location they belong to. Note that APs are typically geo-localized outdoor due to poor GPS signal inside buildings \cite{sapiezynski2015tracking}. Colored regions are the convex hulls of the set of APs in a same location. Grey lines are streets.  \textbf{E-G}) Three examples of APs clustering for thresholds $d=5m$.  }
\label{CNS_locations}
\end{figure*}
\begin{figure*}[h]
\centering
\includegraphics[width=.9\textwidth]{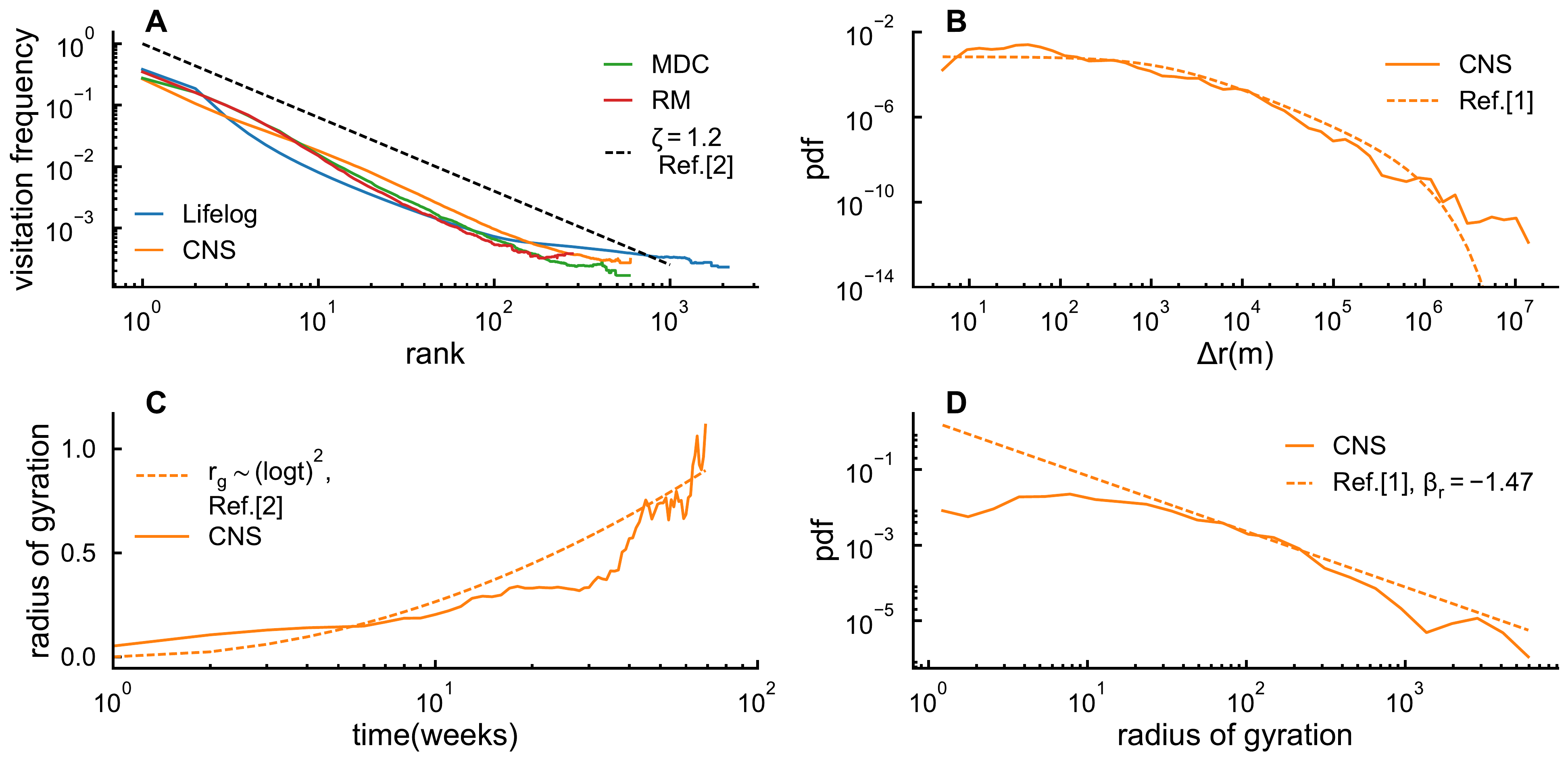}
\caption{\textbf{Agreement with previous research.} 
\textbf{A}) The average visitation frequency $f_k$ as a function of a location rank for the four datasets (filled lines) and the power law fit $f_k \propto k^{\zeta}$, with $\zeta=1.2$ found in \cite{song2010modelling} (dashed line). \textbf{(B)} CNS dataset: The probability density distribution of jump lengths (in m) between consecutive stop-locations (filled line), and the truncated power-law found as the best fit in \cite{gonzalez2008understanding}. \textbf{(C)} CNS dataset: Evolution of the average radius of gyration $r_g$ as a function of time (filled line) and a logarithmic curve $r_g\sim(log t)^2$ found in \cite{song2010modelling} as the best fit. \textbf{(D)} CNS dataset: The probability density function of individuals final radius of gyration (filled line) and the power-law fit (dashed line) found in \cite{gonzalez2008understanding}.}
\label{agreement_with_previous_research}
\end{figure*}
\begin{figure*}[h]
\centering
\includegraphics[width=\textwidth]{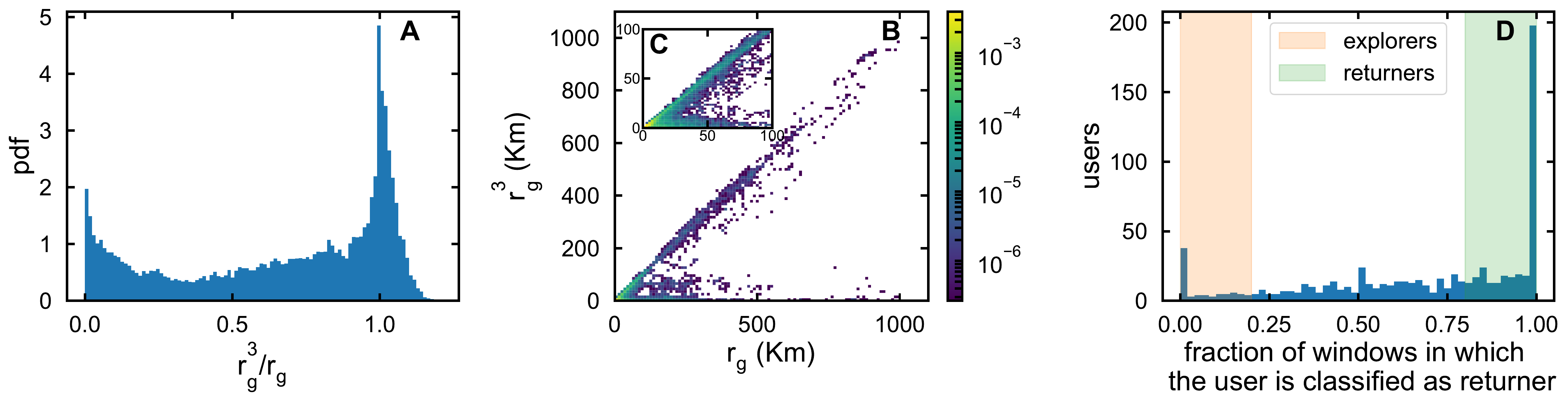}
\caption{\textbf{Returners and Explorers Dichotomy in the CNS dataset.} \textbf{(A)} Distribution of $r^3_g/r_g$, where $r_g$ is the total radius of gyration and $r^3_g$ is the radius of gyration computed considering only the top 3 locations. These quantities are computed across windows of length $20$ weeks for all individuals. \textbf{(B)} Heatmap displaying the joint probability density $p(r_g, r^3_g)$. \textbf{(C)} The joint probability density $p(r_g, r^3_g)$, for $r_g$ and $r^3_g$ $<100 Km$. \textbf{(D)} Based on the definition in \cite{pappalardo2015returners}, \emph{returners} have $r^3_g/r_g>0.5$, while \emph{explorers} have $r^3_g/r_g<0.5$. The figure is the histogram of individuals based on the fraction of times they are assigned to the \emph{returner} category in a window of $20$ weeks (blue bars). For our study, (see Supplementary Figure~\ref{returners_explorers_capacity}), we consider as  \emph{returners} (green shaded area) and \emph{explorers} (red shaded area) only individuals falling in the same category in at least $75\%$ of time-windows (about $64\%$ of all CNS participants).}
\label{returners_and_explorers}
\end{figure*}
\begin{figure*}[h]
\centering
\includegraphics[width=\textwidth]{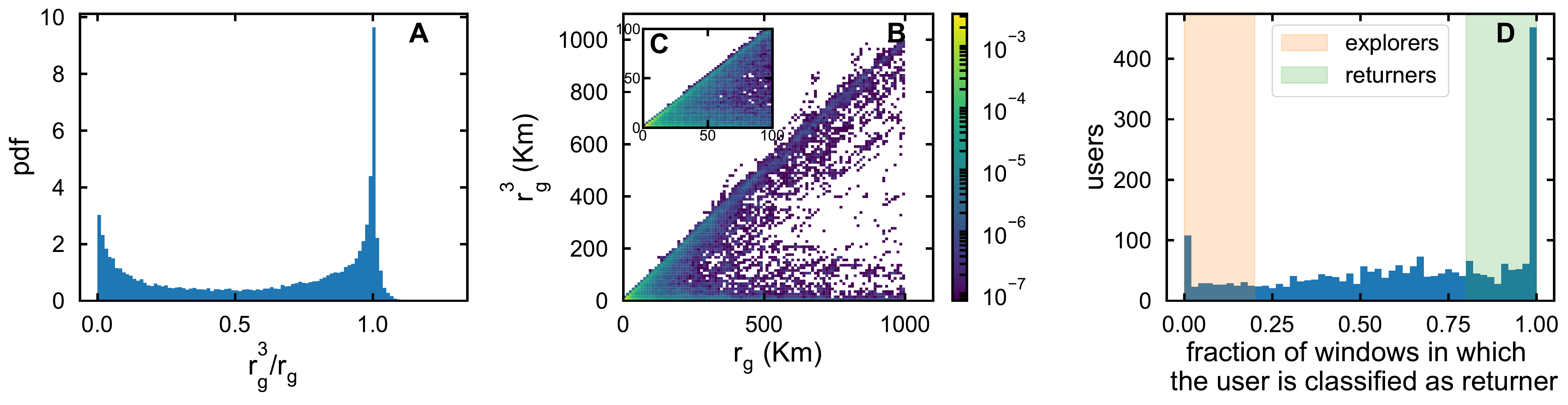}
\caption{\textbf{Returners and Explorers Dichotomy in the Lifelog dataset.} \textbf{(A)} Distribution of $r^3_g/r_g$, where $r_g$ is the total radius of gyration and $r^3_g$ is the radius of gyration computed considering only the top 3 locations. These quantities are computed across windows of length $20$ weeks for all individuals. \textbf{(B)} Heatmap displaying the joint probability density $p(r_g, r^3_g)$. \textbf{(C)} The joint probability density $p(r_g, r^3_g)$, for $r_g$ and $r^3_g$ $<100 Km$. \textbf{(D)} Based on the definition in \cite{pappalardo2015returners}, \emph{returners} have $r^3_g/r_g>0.5$, while \emph{explorers} have $r^3_g/r_g<0.5$. The figure is the histogram of individuals based on the fraction of times they are assigned to the \emph{returner} category in a window of $20$ weeks (blue bars). For our study, (see Supplementary Figure~ \ref{returners_explorers_capacity}), we consider as  \emph{returners} (green shaded area) and \emph{explorers} (red shaded area) only individuals falling in the same category in at least $75\%$ of time-windows (about $56\%$ of all Lifelog users).}
\label{returners_and_explorers_2}
\end{figure*}

\begin{figure*}[h]
\centering
\includegraphics[width=.9\textwidth]{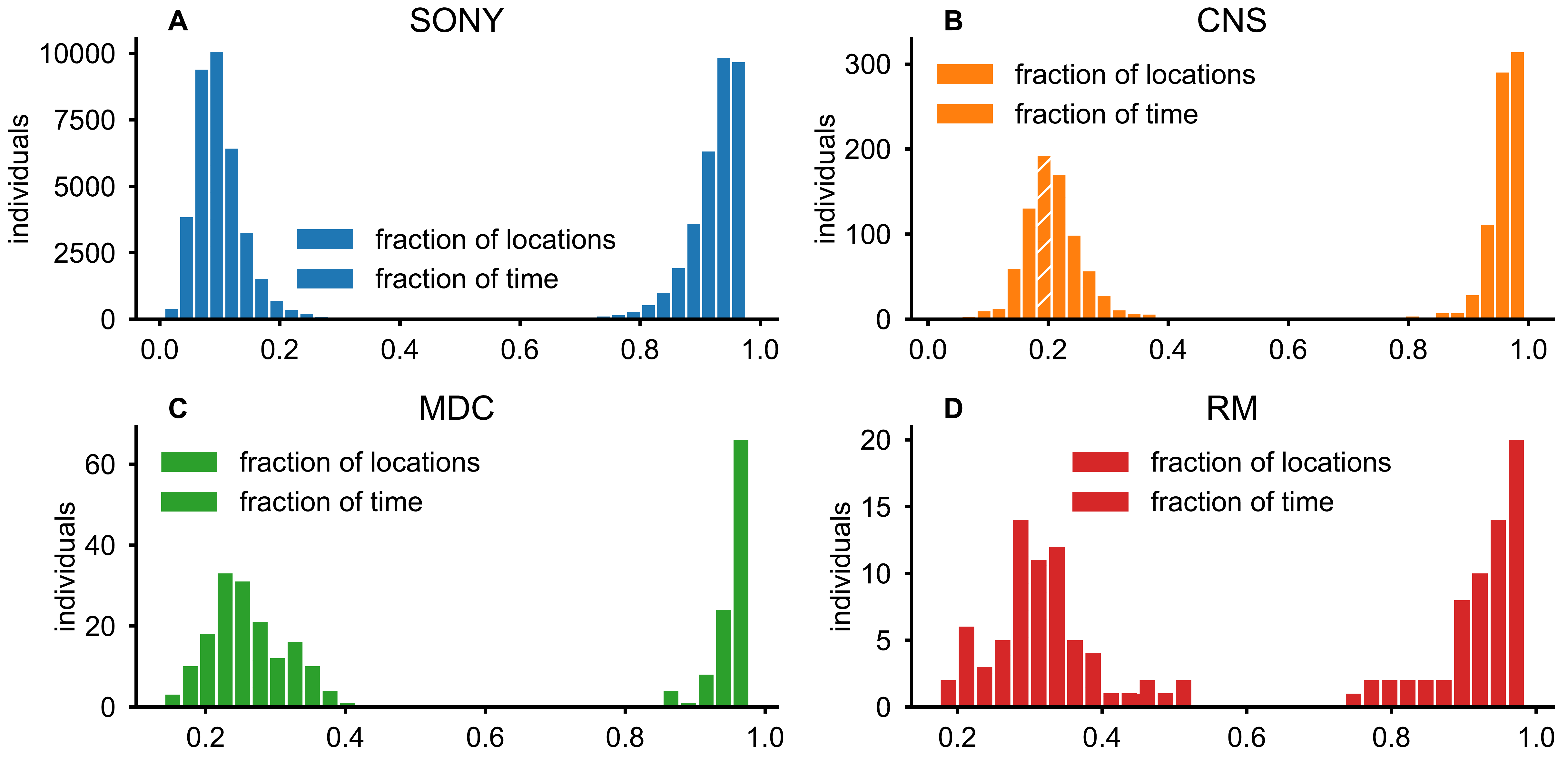}
\caption{\textbf{Establishment of the AS.} Frequency histograms of individuals based on the fraction of all locations seen in a week that are part of the activity set (dashed bars), and on the fraction of time of the week spent in the activity set (full bars). The set is computed for $W=10$ weeks. Results are shown for the Lifelog \textbf{(A)}, CNS \textbf{(B)}, MDC \textbf{(C)} and RM \textbf{(D)}}
\label{activity_space}
\end{figure*}
\begin{figure*}[h]
\centering
\includegraphics[width=.9\textwidth]{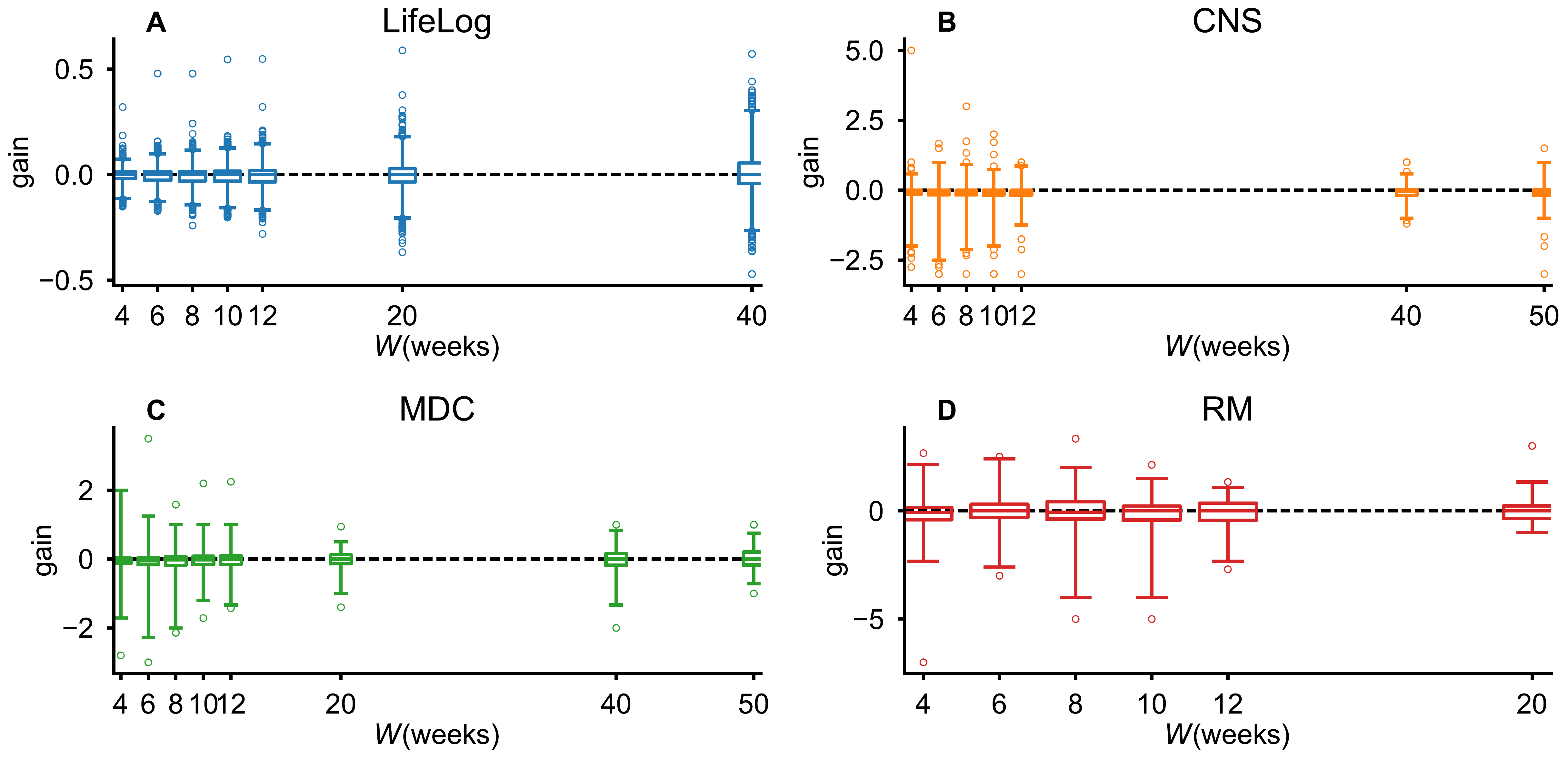}
\caption{\textbf{Gain: window size dependency}
The boxplots of the individual average gain, as a function of the sliding window size for the Lifelog \textbf{(A)}, CNS \textbf{(B)}, MDC \textbf{(C)} and RM \textbf{(D)} datasets. Boxes contains the population interquartile (25 to 75 percentiles) and whiskers contain the 95\% of the population (2.5 to 97.5 percentiles).
}
\label{gain_window}
\end{figure*}
\begin{figure*}[h]
\centering
\includegraphics[width=.5\textwidth]{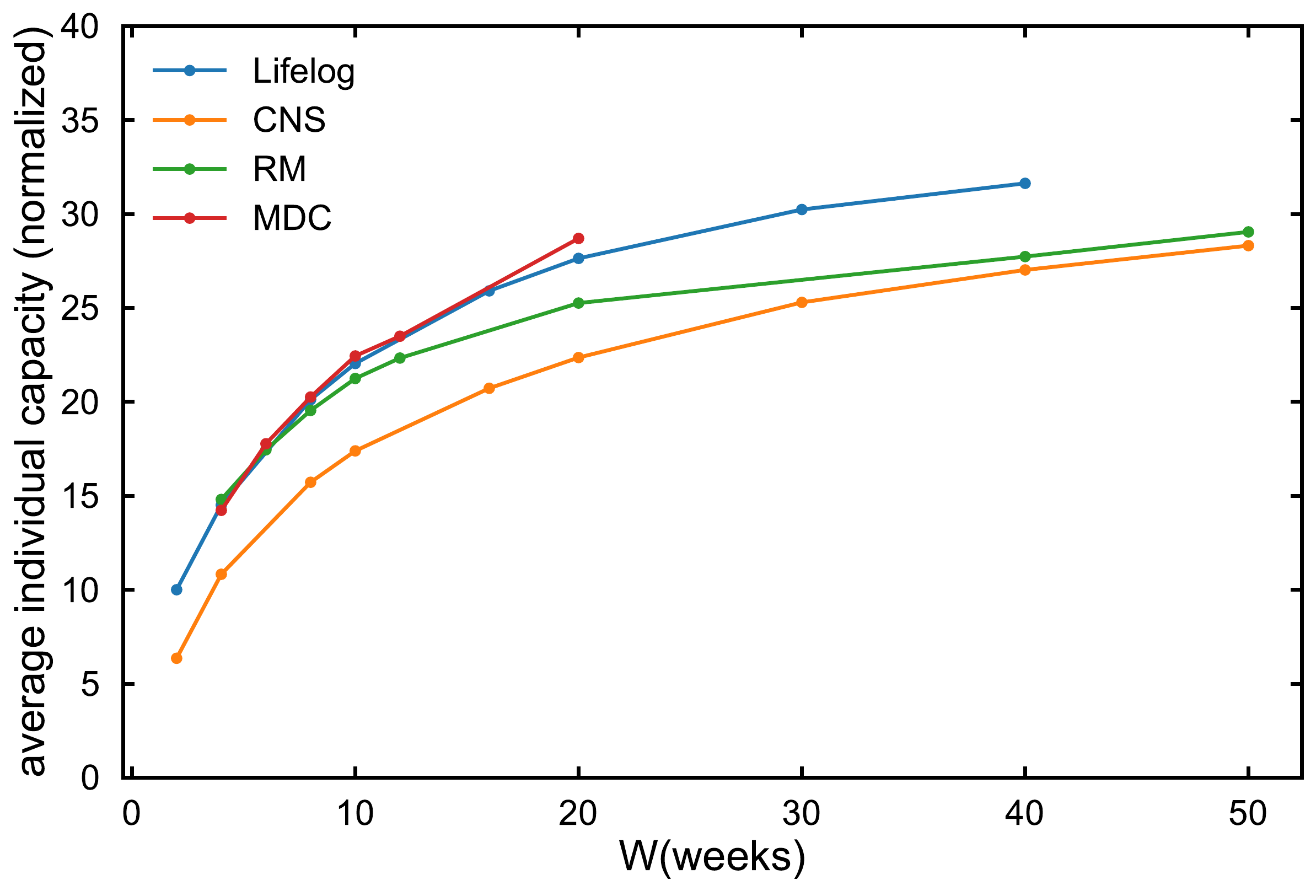}
\caption{\textbf{Saturation of the average normalized capacity.} The average value of the normalized capacity computed for increasing values of the time-window $W$. This result is obtained after accounting for the differences in data collection by computing the normalized location capacity $C_i/TC_i$, where $TC_i$ is the weekly time coverage of individual $i$. \\ }
\label{25}
\end{figure*}
\begin{figure*}[h]
\centering
\includegraphics[width=.9\textwidth]{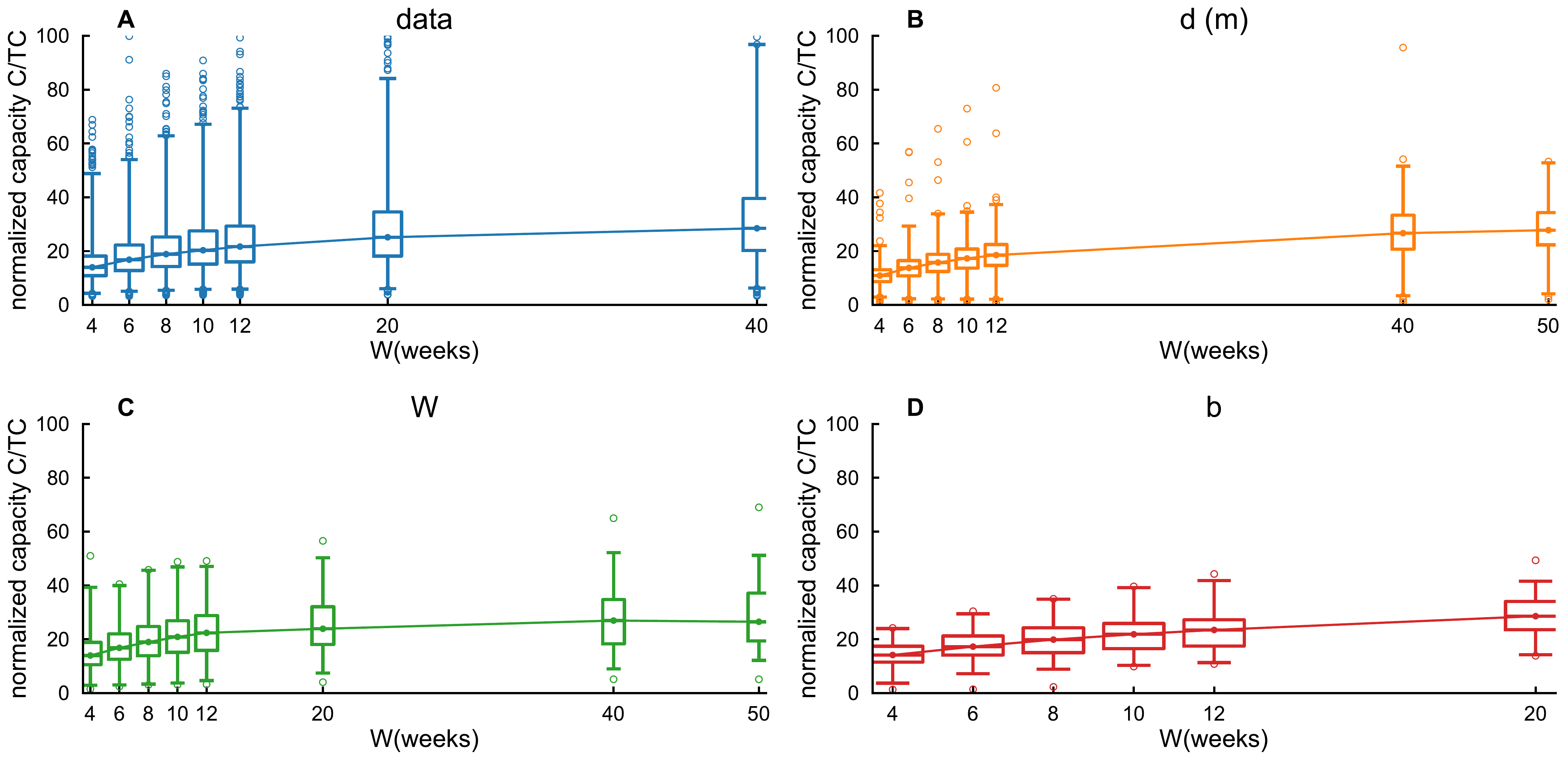}
\caption{\textbf{Capacity: window size dependency}
The boxplots of the individual average capacity, as a function of the sliding window size for the Lifelog \textbf{(A)}, CNS \textbf{(B)}, MDC \textbf{(C)} and RM \textbf{(D)} datasets. Boxes contains the population interquartile (25 to 75 percentiles) and whiskers contain the 95\% of the population (2.5 to 97.5 percentiles). }
\label{capacity_window}
\end{figure*}
\begin{figure*}[h]
\centering
\includegraphics[width=.9\textwidth]{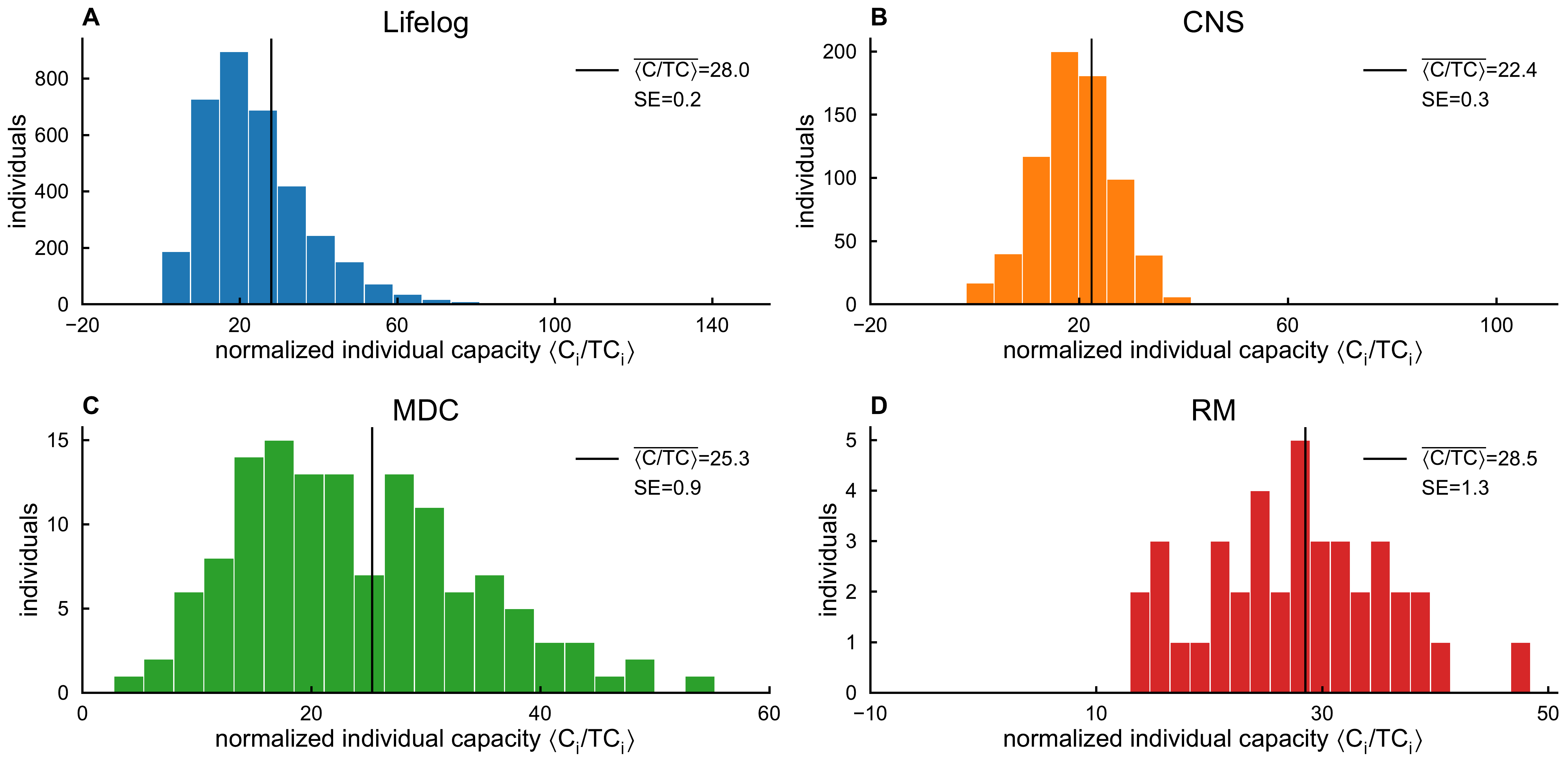}
\caption{\textbf{Individual capacity: population homogeneity} The frequency histogram of the normalized individual capacity $\langle C _i/TC_i \rangle$, where $C_i$ and $TC_i$ are respectively the location capacity and the time coverage of individual $i$. The average value $\overline{ \langle C/TC \rangle}$ (black line) has standard error SE. 
Results are shown for the Lifelog \textbf{(A)}, CNS \textbf{(B)}, MDC \textbf{(C)} and RM \textbf{(D)}, computed with $W=20$.}
\label{individual_capacity}
\end{figure*}

\begin{figure*}[h]
\centering
\includegraphics[width=.9\textwidth]{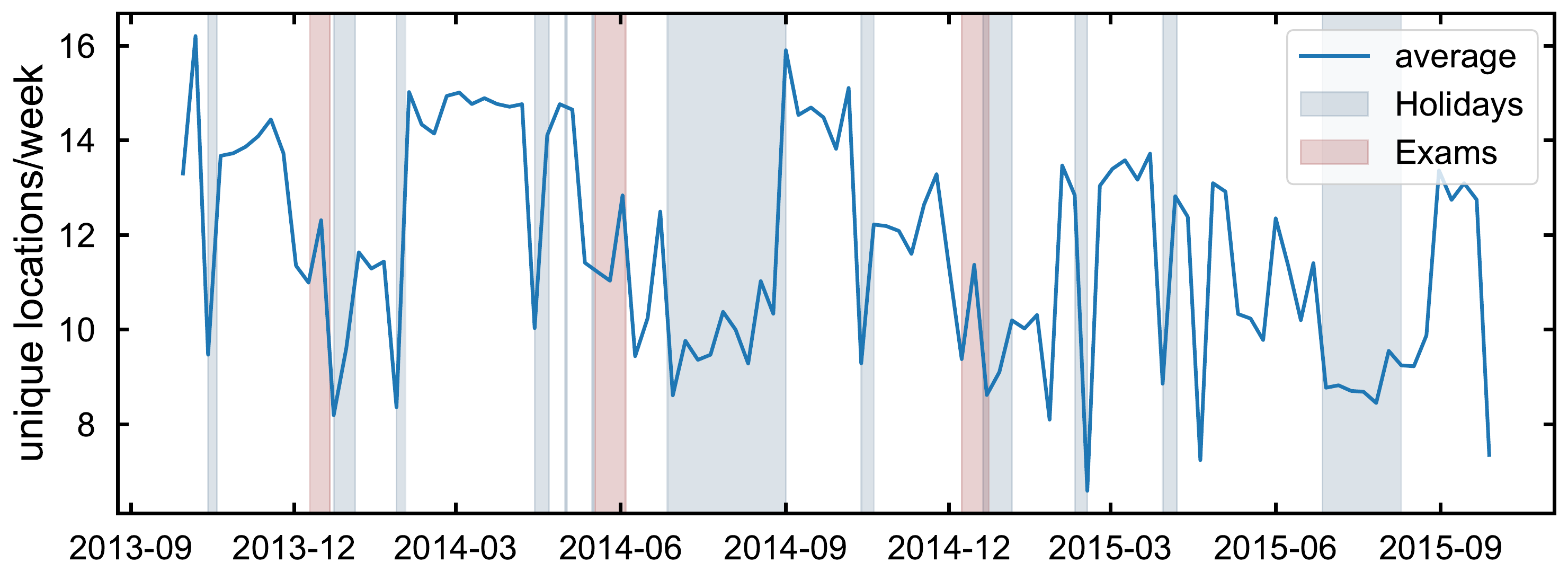}
\caption{\textbf{Seasonality in the CNS data.} The average number of unique locations visited in a week over time (blue line). Blue areas correspond to periods of holidays in the academic schedule; Red areas correspond to exam periods. }
\label{seasonality1}
\end{figure*}

\begin{figure*}[h]
\centering
\includegraphics[width=.9\textwidth]{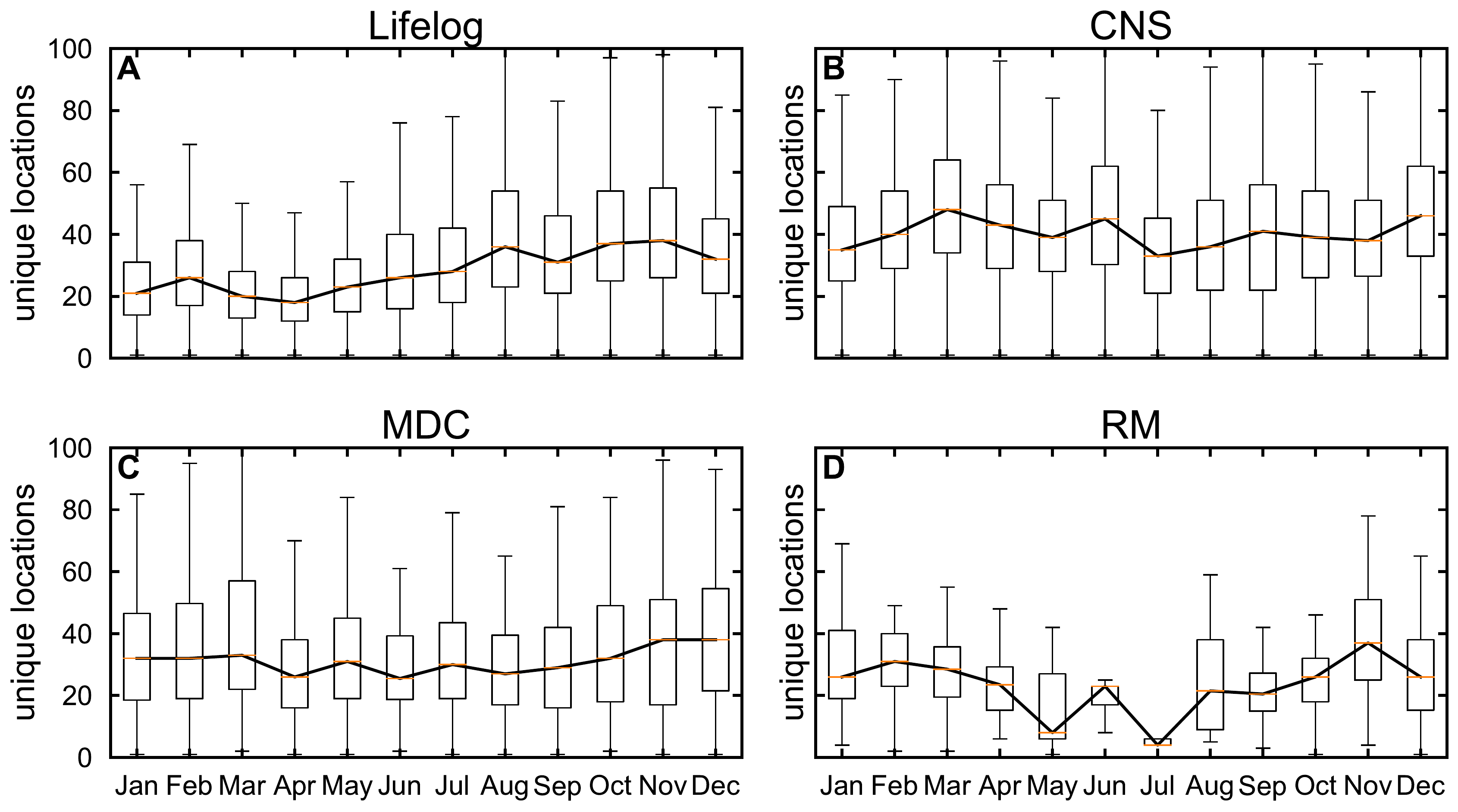}
\caption{\textbf{Seasonality effects.} Boxplot showing the number of unique locations visited in a month for the Lifelog \textbf{(A)}, CNS \textbf{(B)}, MDC \textbf{(C)}, and RM \textbf{(D)} datasets. }
\label{seasonality2}
\end{figure*}
\begin{figure*}[h]
\centering
\includegraphics[width=.45\textwidth]{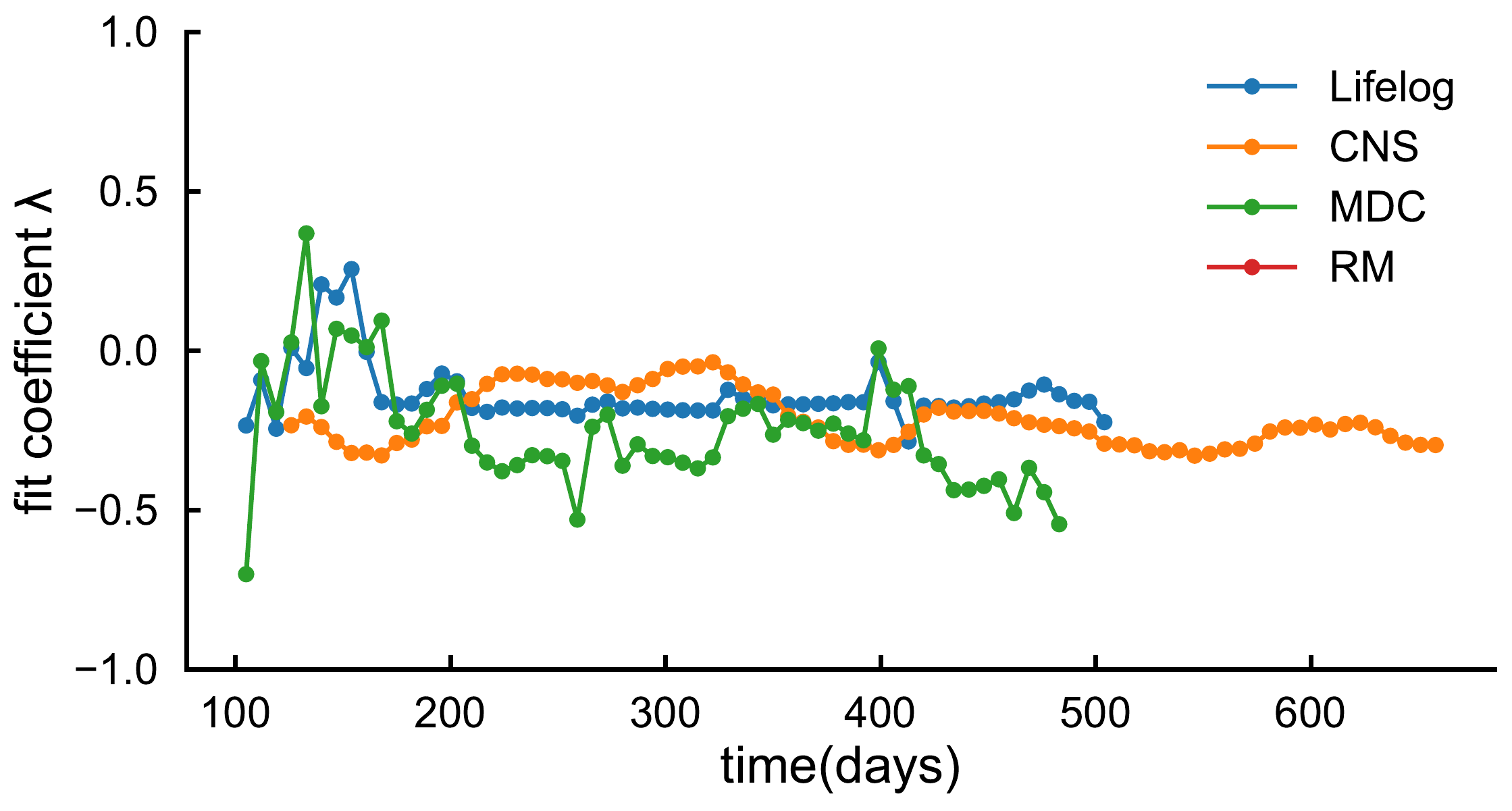}
\caption{\textbf{Evolution of the activity set: invariance under time translation.}  The PL fit coefficients $\lambda$ describing the evolution of the activity set as a function of the starting time of the measurement, for different datasets. }
\label{evolution_aging}
\end{figure*}
\begin{figure*}[h]
\centering
\includegraphics[width=.9\textwidth]{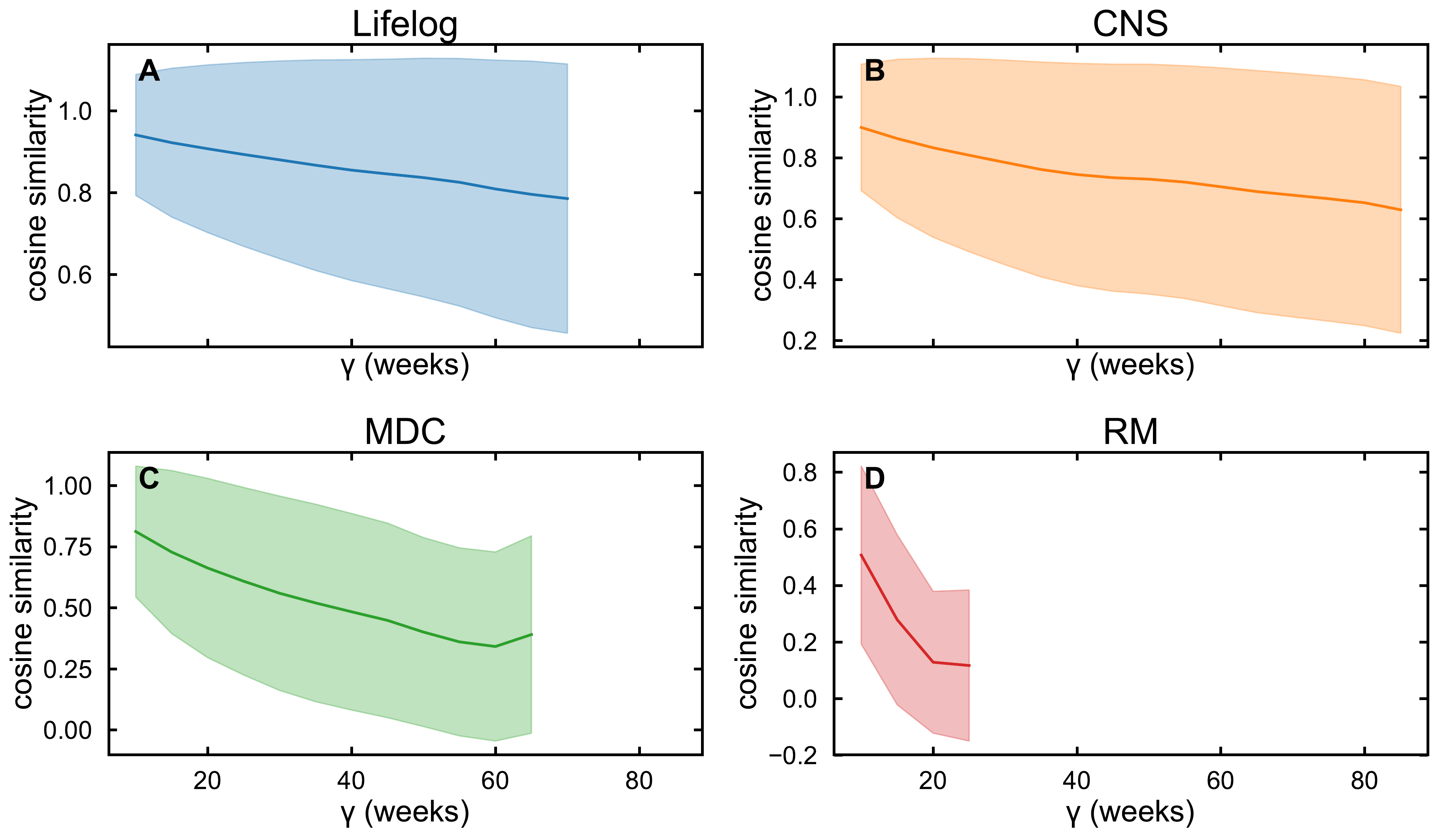}
\caption{\textbf{Evolution of the activity set: cosine similarity. } The average cosine similarity between vectors constructed from the sets $AS_i(t)$ and $AS_i(t+\gamma)$ (full lines), and the corresponding standard deviation (shaded areas) as a function of $\gamma$, in weeks. We consider locations visiting probability (or the fraction of time spent in that location) as vector components. Results are shown for the Lifelog (A), CNS \textbf{(B)}, MDC \textbf{(C)} and RM \textbf{(D)} datasets.}
\label{cosine_similarity}
\end{figure*}

\begin{figure*}[h]
\centering
\includegraphics[width=.9\textwidth]{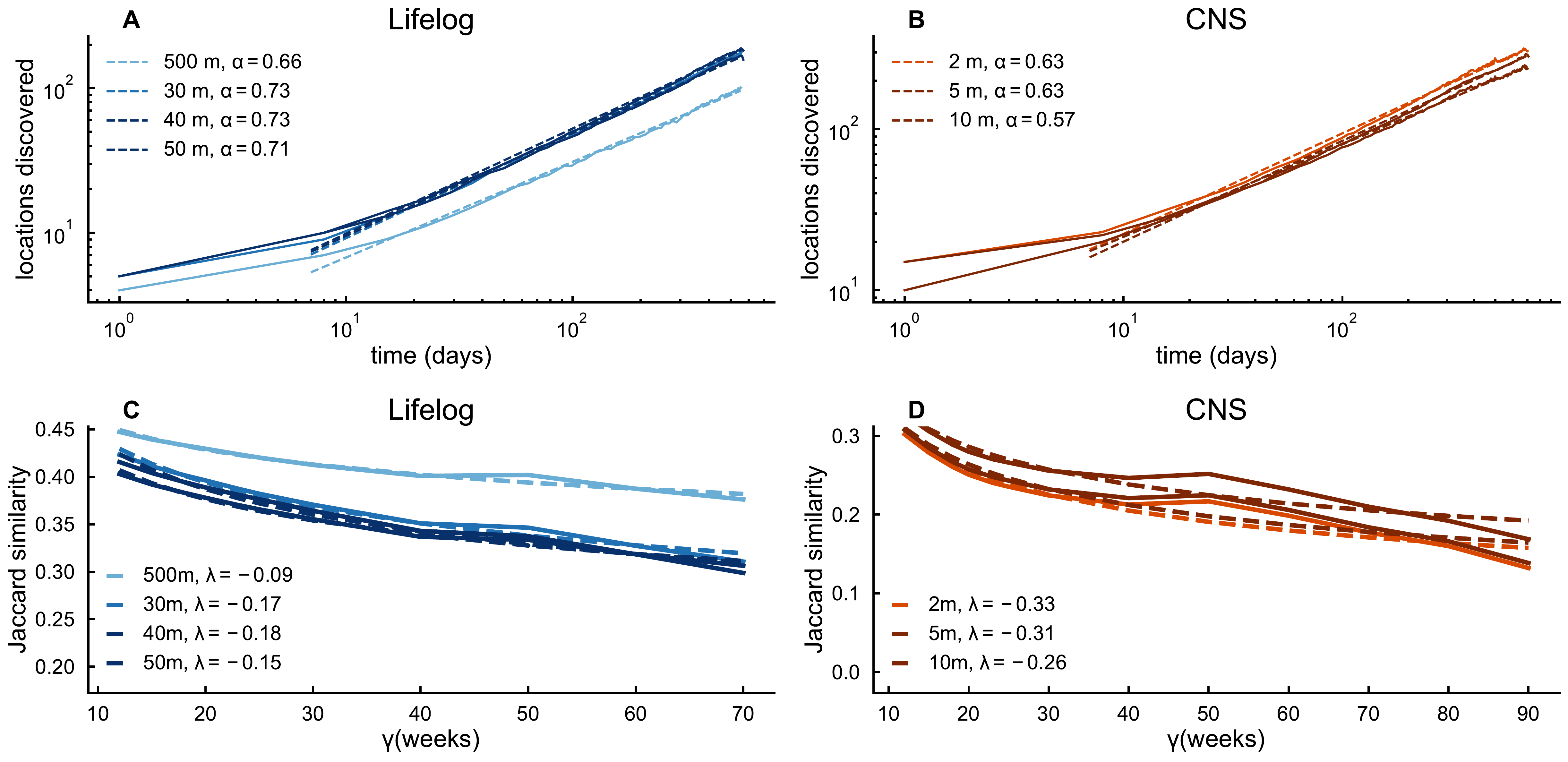}
\caption{\textbf{Effects of different definitions of locations.} The average number of locations discovered up to a given day for different definitions of location, and the corresponding power-law fits (dashed line) with coefficient $\alpha$, for the Lifelog \textbf{(A)} and CNS \textbf{(B)} datasets. \textbf{(C,D)} The average overlap (Jaccard similarity) between the activity set at week $t$ and week $t+\gamma$  (full line), and the corresponding power law fit $\overline{ J(\gamma)} \sim \gamma^{\lambda}$ (dashed lines) (dashed line) for different definitions of location. Results are shown for the Lifelog \textbf{(C)} and CNS \textbf{(D)} datasets. }
\label{definitions_of_locations}
\end{figure*}

\begin{figure*}[h]
\centering
\includegraphics[width=.9\textwidth]{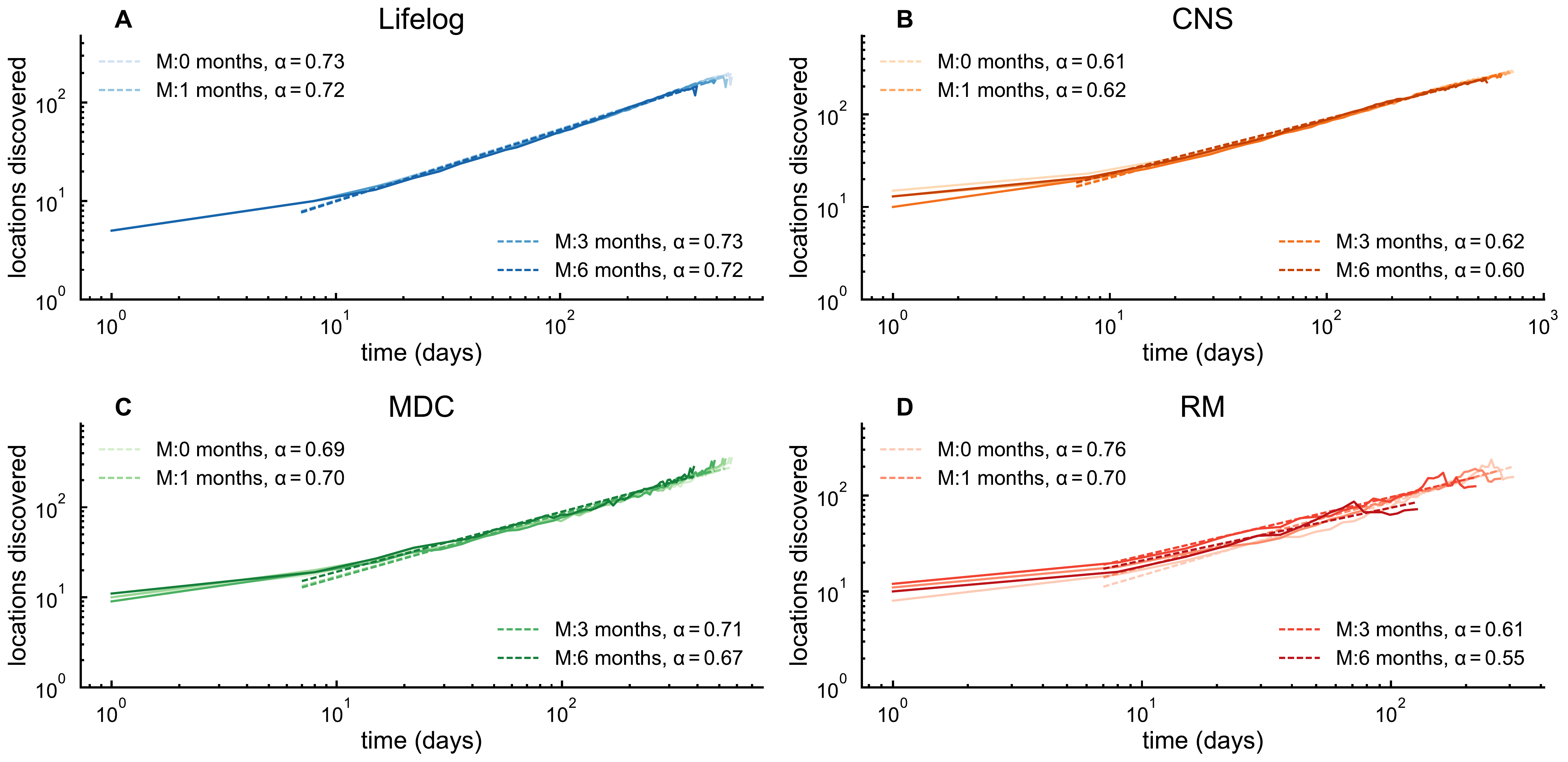}
\caption{\textbf{Exploration behavior: invariance under time translation} The average number of locations individually discovered in time, measured after waiting $M$ months, and the corresponding  power-law function fit with coefficients $\alpha$ (dashed lines) for different values of $M$. Results are shown for the Lifelog \textbf{(A)}, CNS \textbf{(B)}, MDC \textbf{(C)} and RM \textbf{(D)} datasets. }
\label{exploration_aging}
\end{figure*}

\begin{figure*}[h]
\centering
\includegraphics[width=.9\textwidth]{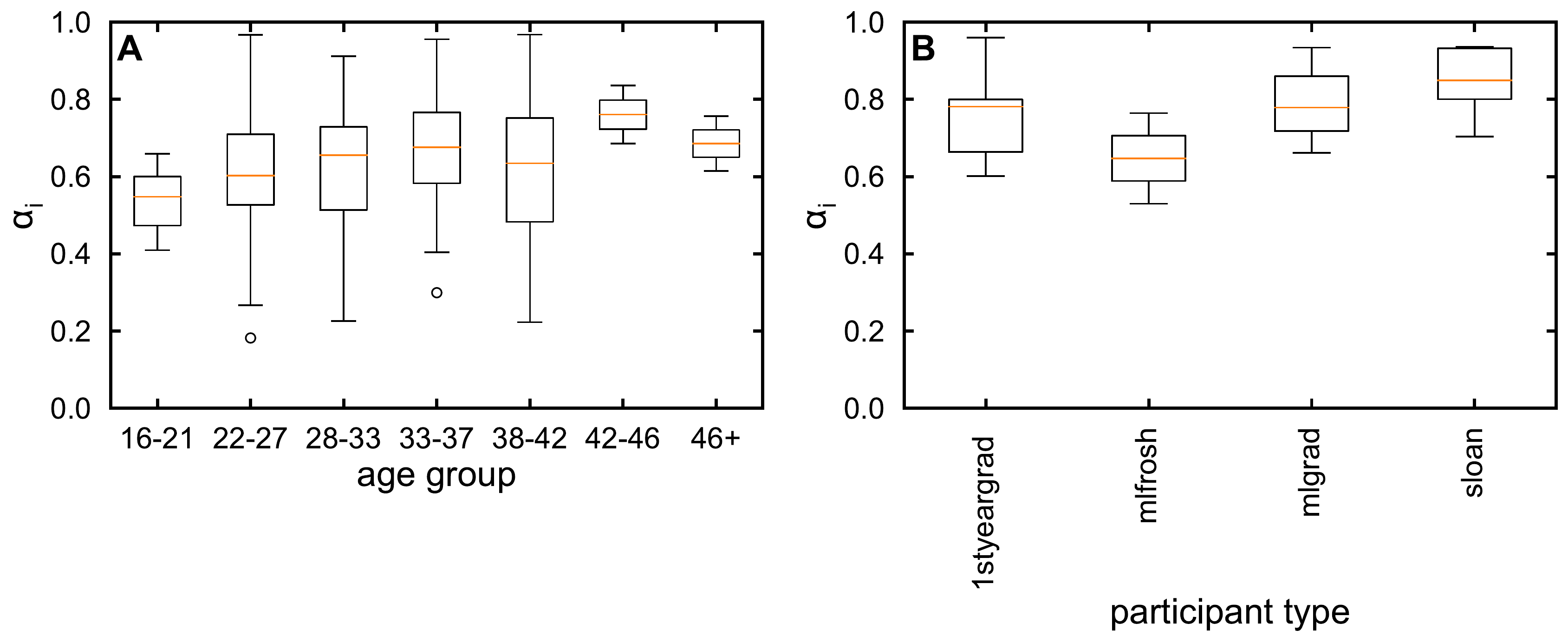}
\caption{\textbf{Relation between individual exploration and age/academic course.}\textbf{(A)} MDC dataset: Boxplot showing the distribution of the coefficient $\alpha_i$ describing the growth of locations $L_i \sim t^{\alpha_i}$ for individuals within different age groups. The correlation between the two is positive and significant (Pearson coefficient $\rho=0.2$, p-value = $0.008$).\textbf{(B)} RM dataset: Boxplot of $\alpha_i$ for different categories of individuals (`mlgrad': Media Lab Graduate Student (not a first year);
`1styeargrad': Media Lab First Year Graduate Student;
`mlfrosh': Media Lab First Year Undergraduate Student; 
`sloan': Sloan Business School)  }
\label{age}
\end{figure*}

\begin{figure*}[h]
\centering
\includegraphics[width=.9\textwidth]{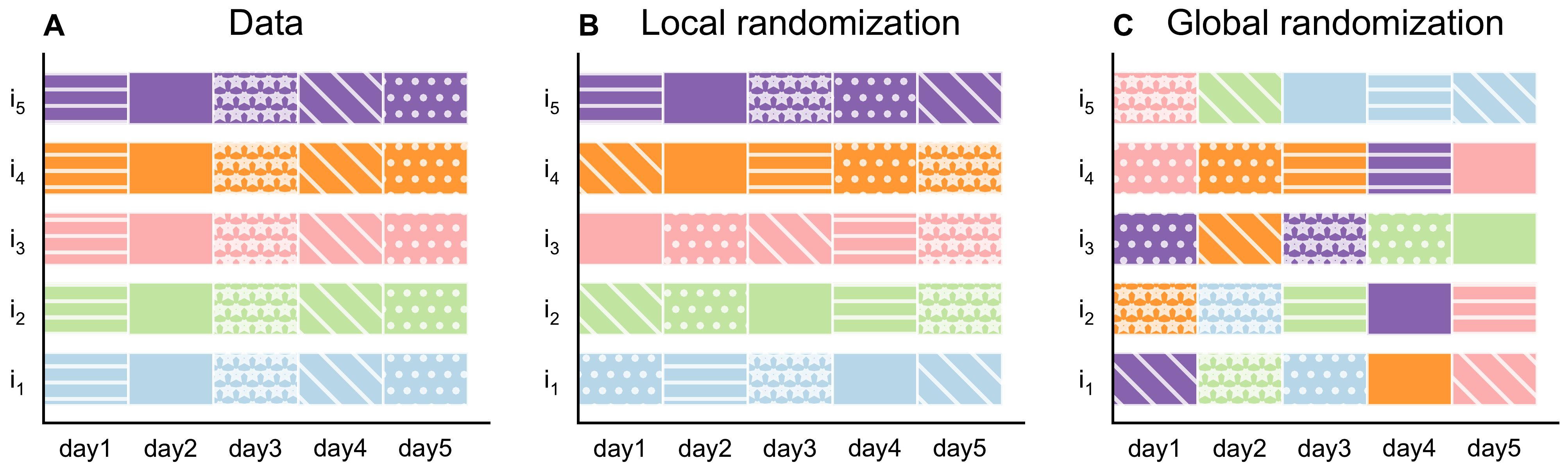}
\caption{\textbf{Data randomization schema.} A schematic representation of local and global randomization. \textbf{(A)} Individual time series for 5 individuals are divided into modules of 1 day length (each day has a specific color pattern). \textbf{(B)} In the \emph{local randomization} individual timeseries are shuffled preserving the module units.  \textbf{(C)} In the \emph{global randomization} new sequences are created assembling together modules extracted randomly from the whole set of individual traces. }
\label{randomization}
\end{figure*}

\begin{figure*}[h]
\centering
\includegraphics[width=.9\textwidth]{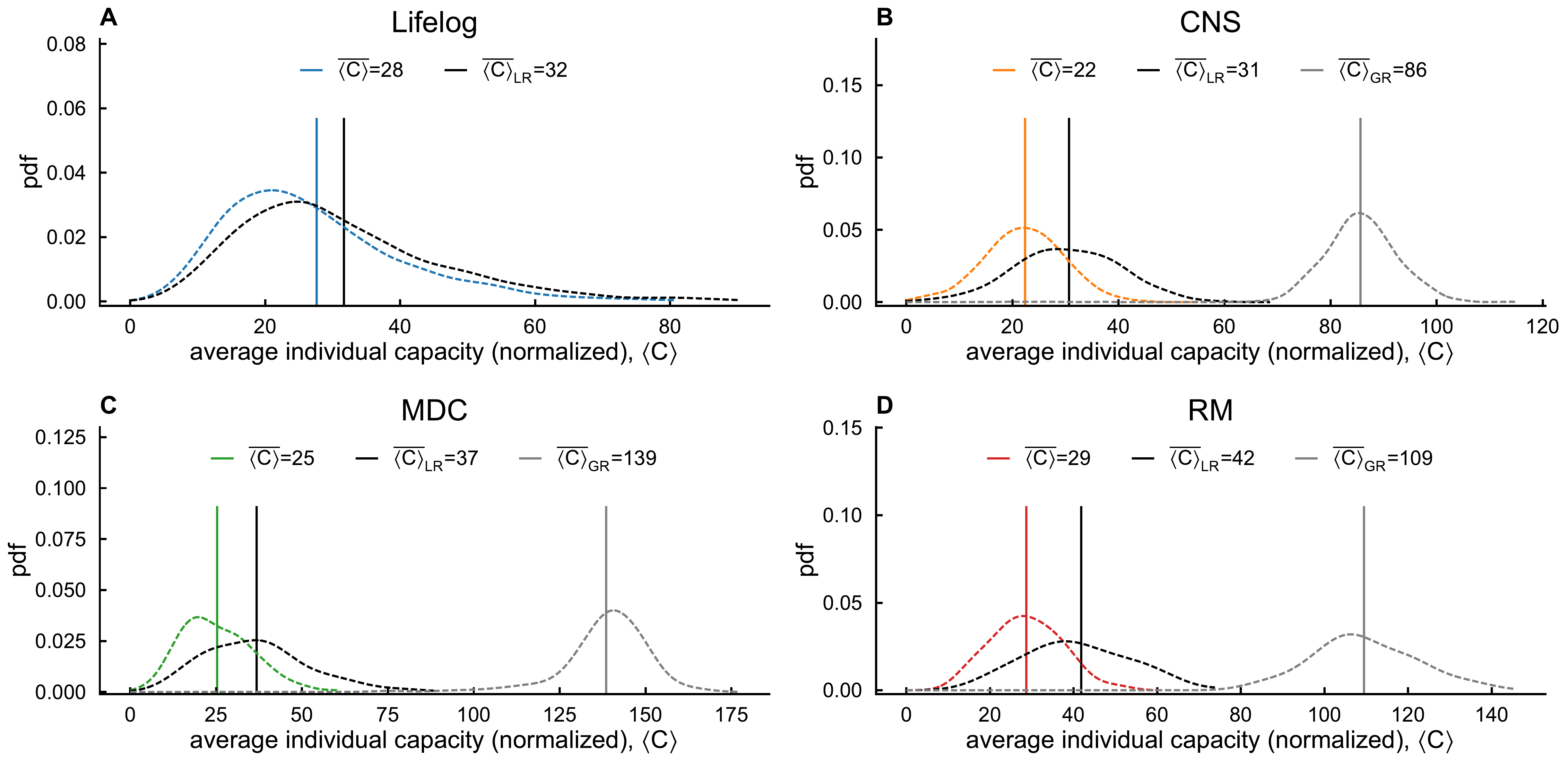}
\caption{\textbf{Discrepancy with the randomized cases.}
The Kernel Density of the average individual capacity (normalized to account for the differences in time coverage) for data ($\langle C \rangle$), local ($\langle C \rangle _{LR}$) and global ($\langle C \rangle_{GR}$) randomizations (dashed lines), and the corresponding average values (full lines) computed across the population. The Kolmogorov–Smirnov test-statistics (Table \ref{ks_statistics}) rejects the hypothesis  that the three samples are  extracted from the same distribution.}
\label{randomization2}
\end{figure*}

\clearpage
\begin{figure*}[h]
\centering
\includegraphics[width=.9\textwidth]{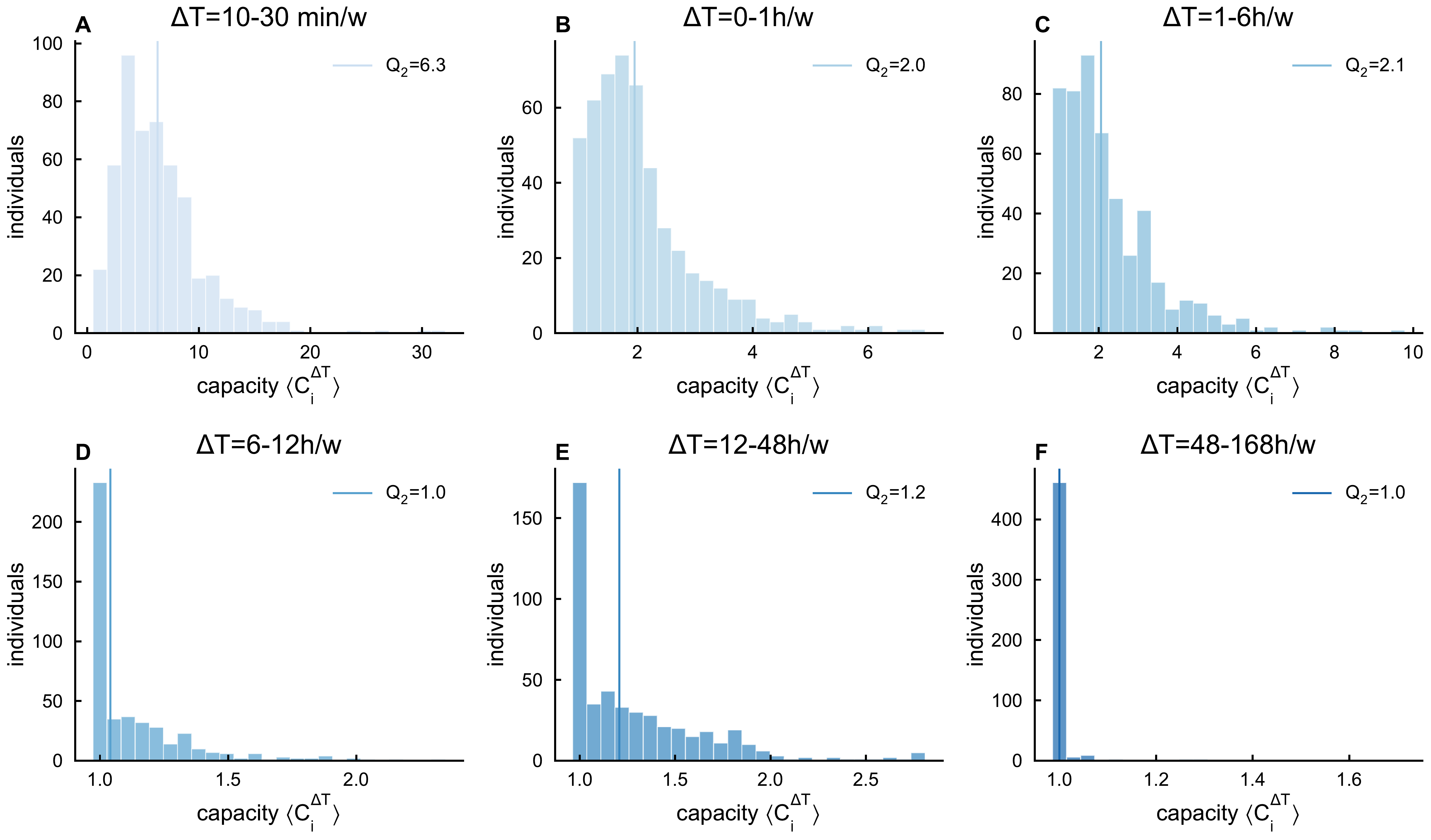}
\caption{\textbf{Lifelog dataset: Composition of the AS.} 
 \textbf{A-F}) The distribution of the average individual capacity $\langle C _i \rangle^{\Delta T} \rangle$, considering locations seen for a time included in $\Delta T$.}
\label{time_alloc_sony}
\end{figure*}
\begin{figure*}[h]
\centering
\includegraphics[width=.9\textwidth]{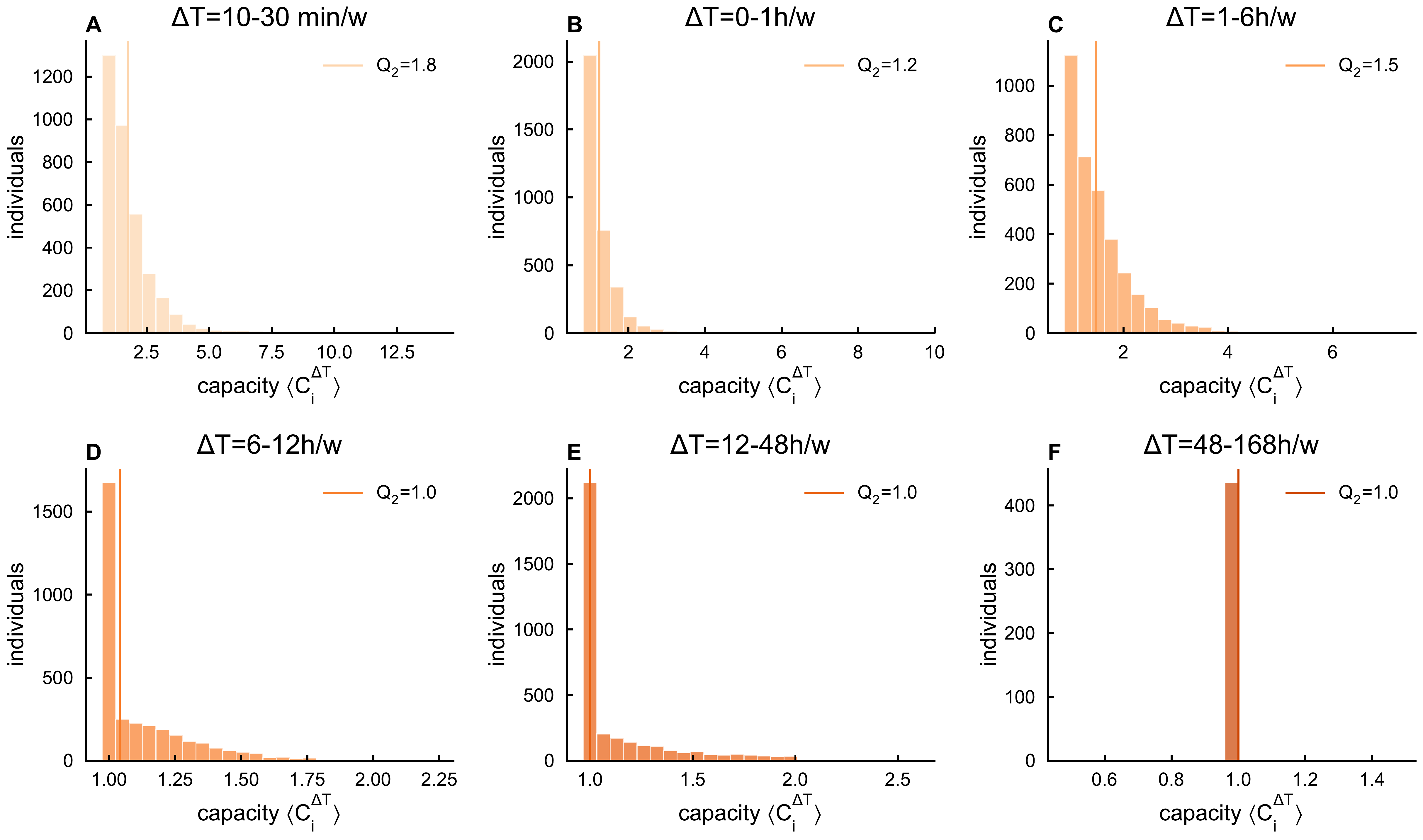}
\caption{ \textbf{CNS dataset: Composition of the AS.}
 \textbf{A-F}) The distribution of the average individual capacity $\langle C _i \rangle^{\Delta T} \rangle$, considering locations seen for a time included in $\Delta T$.  }
\label{time_alloc_cns}
\end{figure*}
\begin{figure*}[h]
\centering
\includegraphics[width=.9\textwidth]{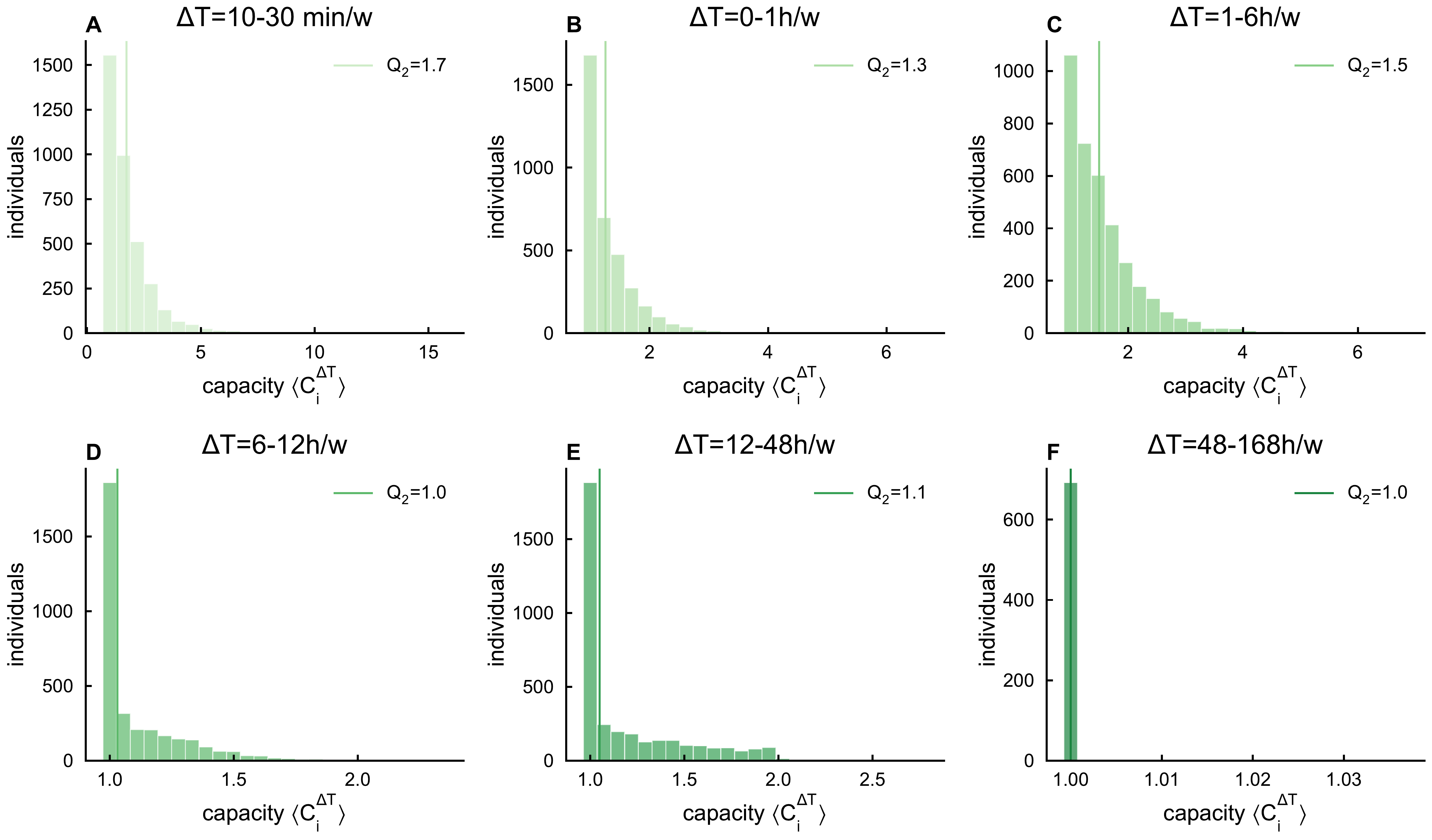}
\caption{\textbf{MDC dataset: Composition of the AS.} \textbf{A-F})  The distribution of the average individual capacity $\langle C _i \rangle^{\Delta T} \rangle$, considering locations seen for a time included in $\Delta T$. }
\label{time_alloc_mdc}
\end{figure*}
\begin{figure*}[h]
\centering
\includegraphics[width=.9\textwidth]{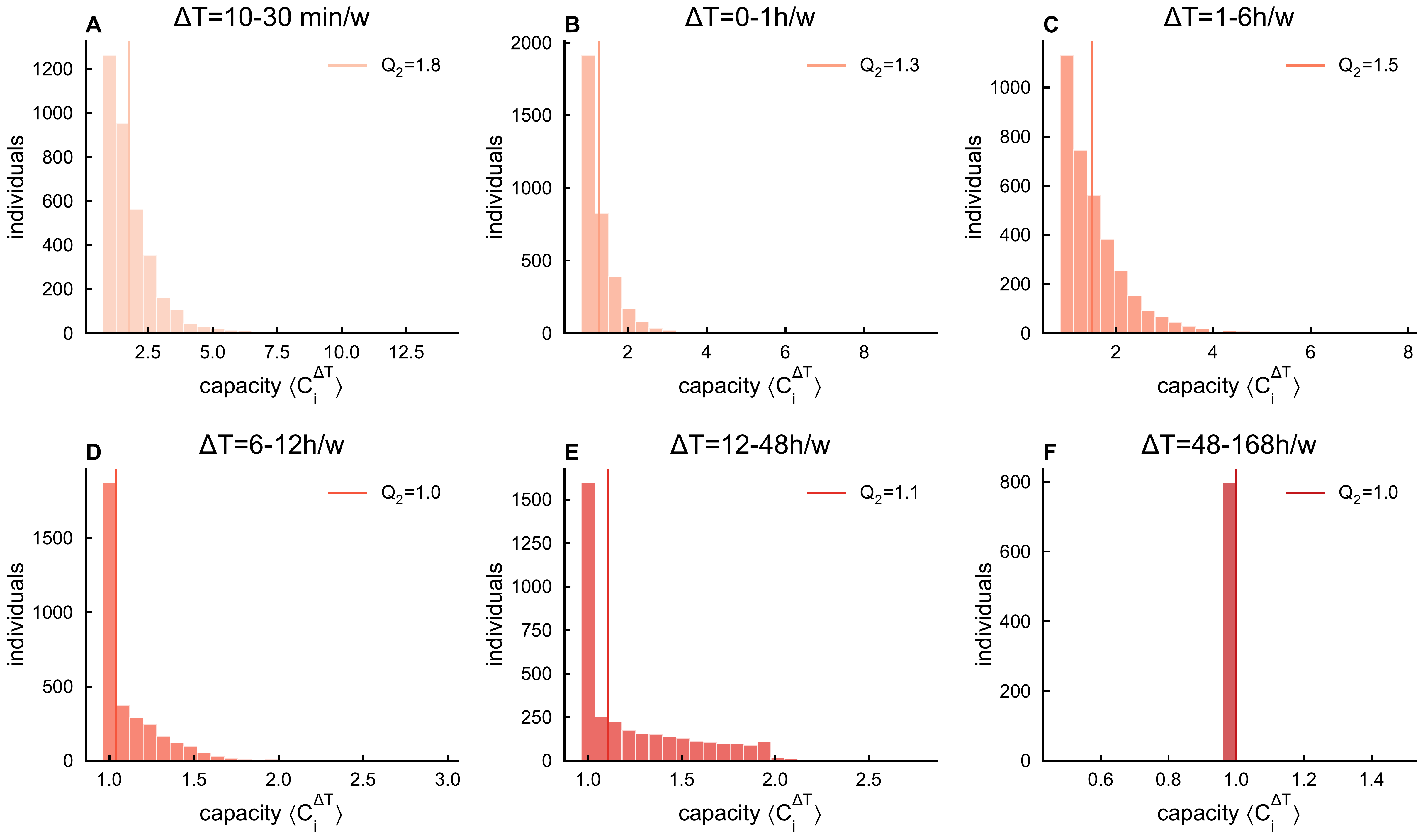}
\caption{\textbf{RM dataset: Composition of the AS.} \textbf{A-F}) The distribution of the average individual capacity $\langle C _i \rangle^{\Delta T} \rangle$, considering locations seen for a time included in $\Delta T$.  }
\label{time_alloc_rm}
\end{figure*}

\begin{figure*}[h]
\centering
\includegraphics[width=.9\textwidth]{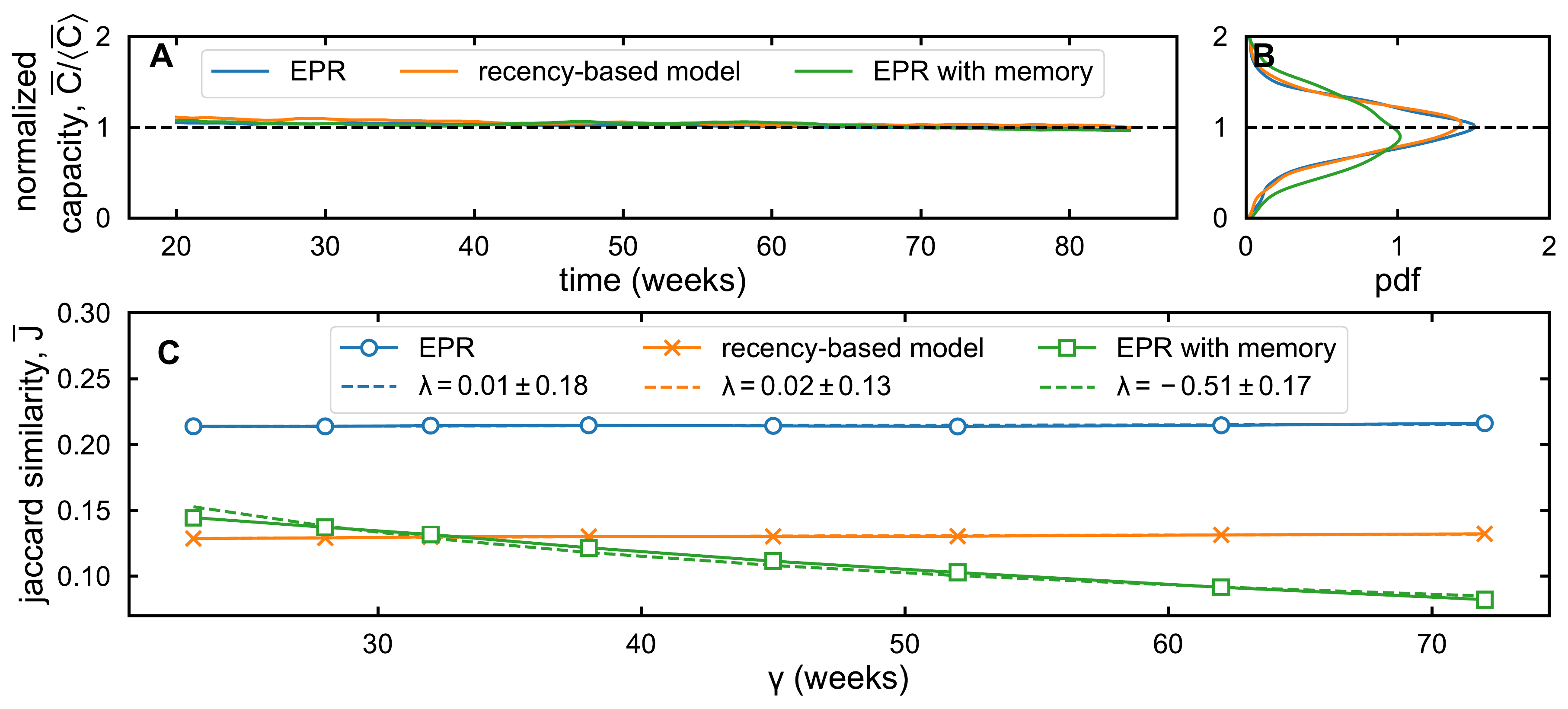}
\caption{ \textbf{Including finite memory improves modeling.}  \textbf{(A)} Average normalized capacity for the EPR model \cite{song2010modelling} (blue line), the recency-based EPR model \cite{barbosa2015effect} (orange line) and the EPR model with finite memory (green line).  \textbf{(B)} Probability density of the average normalized capacity across the population for the three models. \textbf{(C)} The average Jaccard similarity $\overline{J}$ between the set measured at $t$ and $t+\gamma$ as a function of $\gamma$ for the three models. Dashed lines correspond to power-law fits $\overline{J} \sim \gamma^{\lambda}$. Simulations are ran for $10^3$ individuals. Parameters are taken from \cite{song2010modelling} and \cite{barbosa2015effect} $\rho = 0.6$, $\gamma = 0.2 $ and $\beta = 0.8$ (for the three models), $\alpha = 0.1$ and $\eta = 1.6$ (for the recency-based EPR model), $M=200$ days (for the EPR model with memory). We consider that $1$ time unit in the simulation (the shortest duration extracted from the distribution of waiting times) corresponds to 1 minute (the time unit considered to analyse our data). All measures are computed after waiting for a period corresponding to $7$ months. }
\label{Figure9}
\end{figure*}

\begin{figure*}[h]
\centering
\includegraphics[width=.9\textwidth]{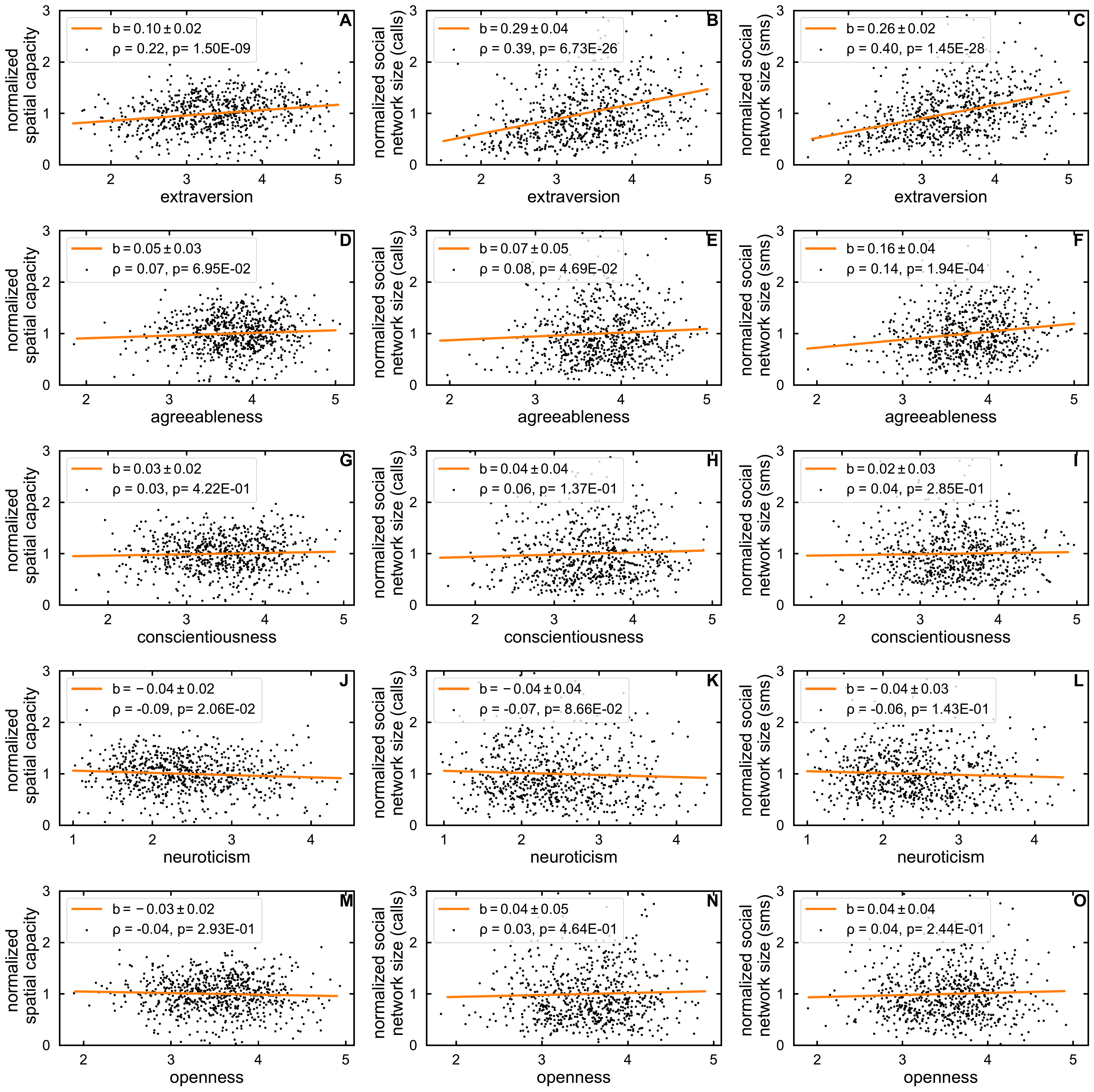}
\caption{\textbf{Social network size and location capacity correlate with extraversion.} The average normalized location capacity (left column), social network size computed from calls interactions(center column) and sms interactions (right column) as a function of each of the BigFive personality traits, measured on a scale from 0 to 5. The personality traits are: extraversion (first row), agreableness (second row), conscientiousness (third row), neuroticism (fourth row) and openness (fifth row). The legend report the value of the slope $b$ of a linear regression line, the Pearson correlation coefficient $\rho$, with associated p-value $p$. Results are shown for $W=20$ weeks. }
\label{model2}
\end{figure*}

\begin{figure*}[h]
\centering
\includegraphics[width=.9\textwidth]{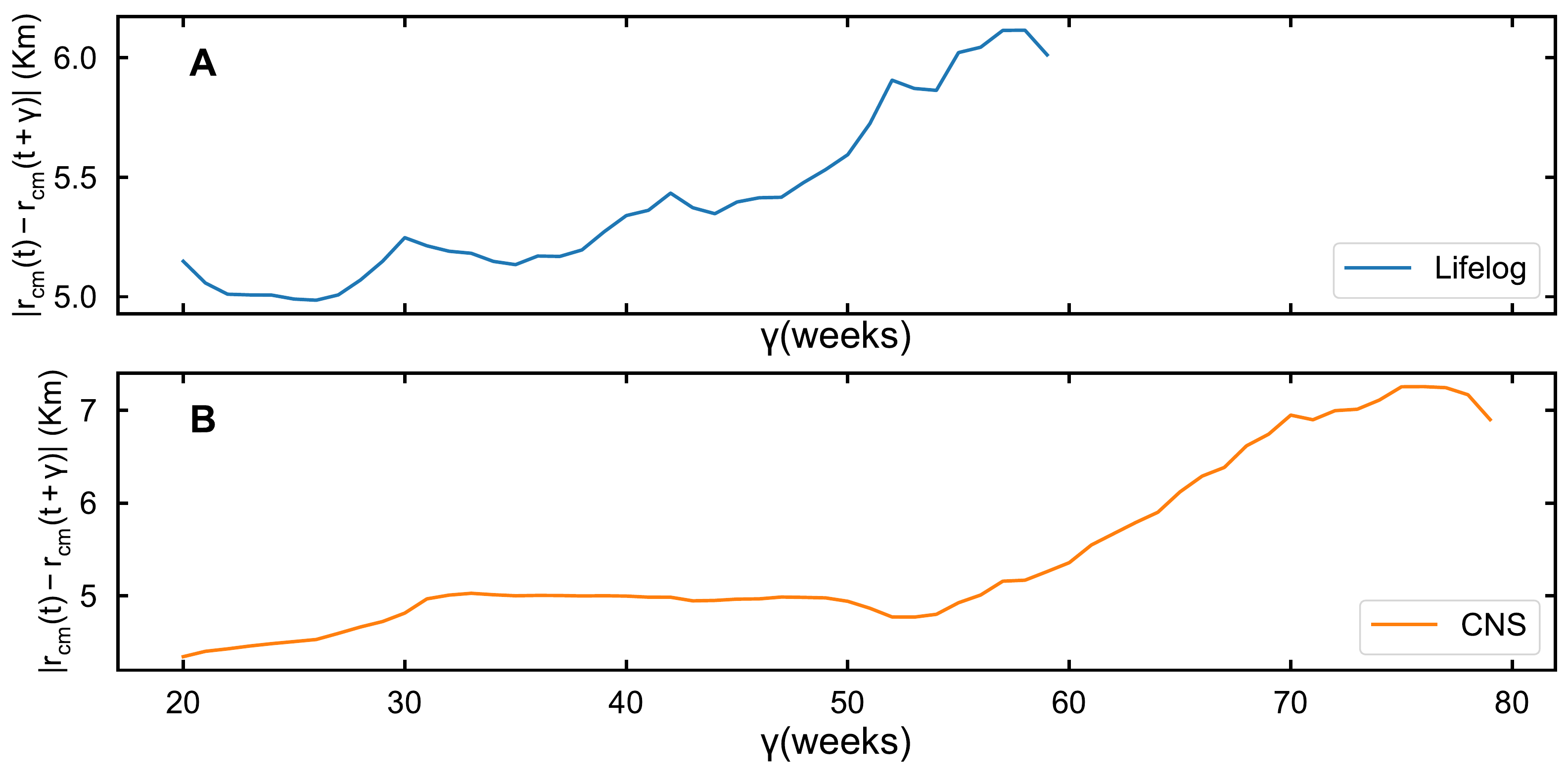}
\caption{\textbf{Displacement of the activity set center of mass.} The average distance between the center of mass of the activity set $r_{cm}(t)$ computed at time $t$, and the same quantity computed at time $(t + \gamma)$. The distance is averaged across values of $t$ and plotted as a function of the delay $\gamma$. Results are shown for sets computed using a sliding window of size $W=20$ weeks, for the Lifelog \textbf{(A)} and CNS \textbf{(B)} datasets. For the CNS dataset, the displacement of the center of mass occurs mainly in the first $7$ months and after $1$ year. This could be explained knowing that many of the CNS participants moved home location from the University campus to the city center after the first year at University. For the Lifelog dataset, we observe an overall growth.}
\label{rcm}
\end{figure*}

\begin{figure*}[h]
\centering
\includegraphics[width=.9\textwidth]{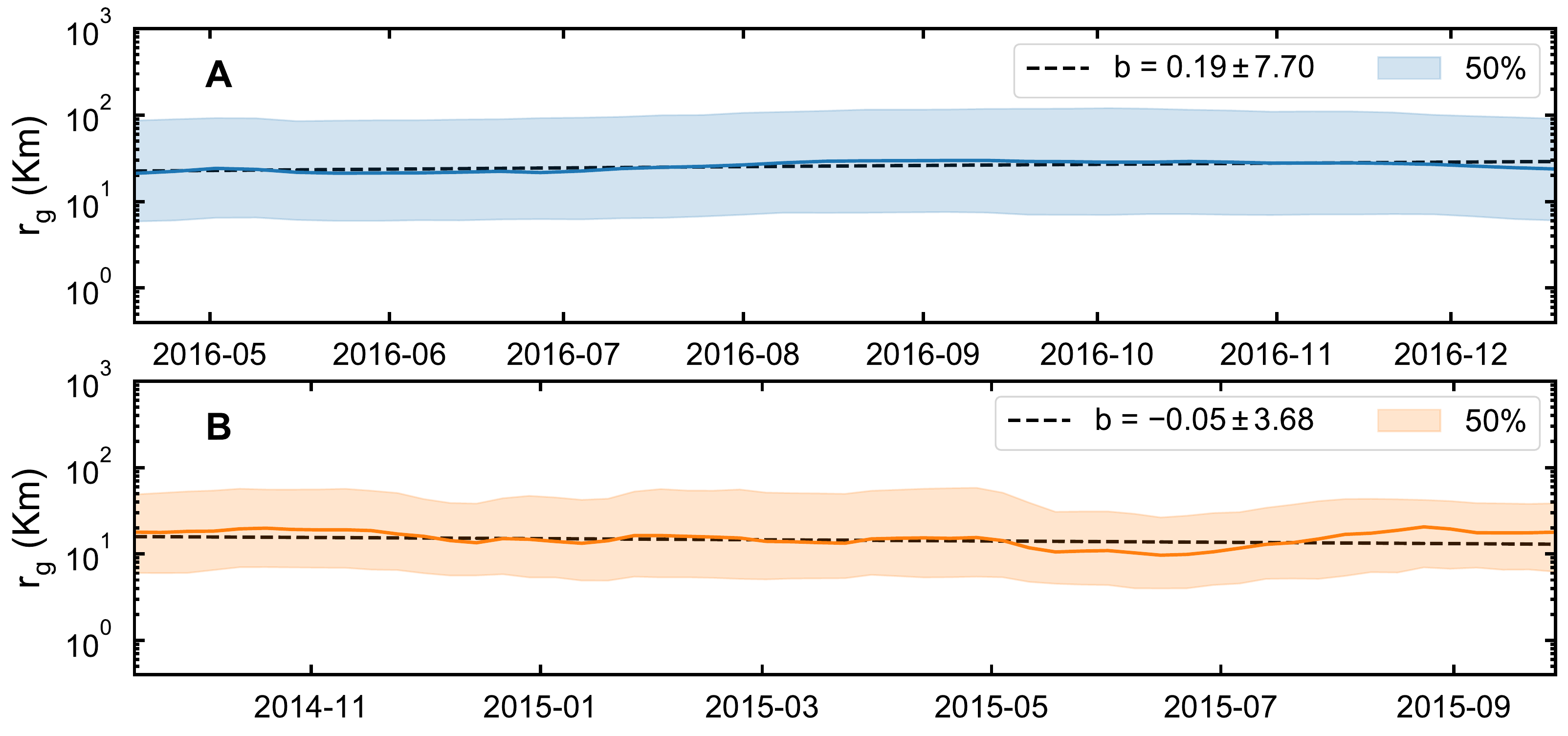}
\caption{\textbf{Constant radius of gyration of the activity set.} Median value of the radius of gyration  $r_g(t)$ of the activity set as a function of time (blue line). The light blue shaded area is the $50\%$ of the sample around the median. The dashed line is a linear fit with coefficient $b = -0.02 \pm 0.15$. Results are shown for the Lifelog \textbf{(A)} and CNS \textbf{(B)} datasets.  }
\label{rg}
\end{figure*}

\begin{figure*}[h]
\centering
\includegraphics[width=.9\textwidth]{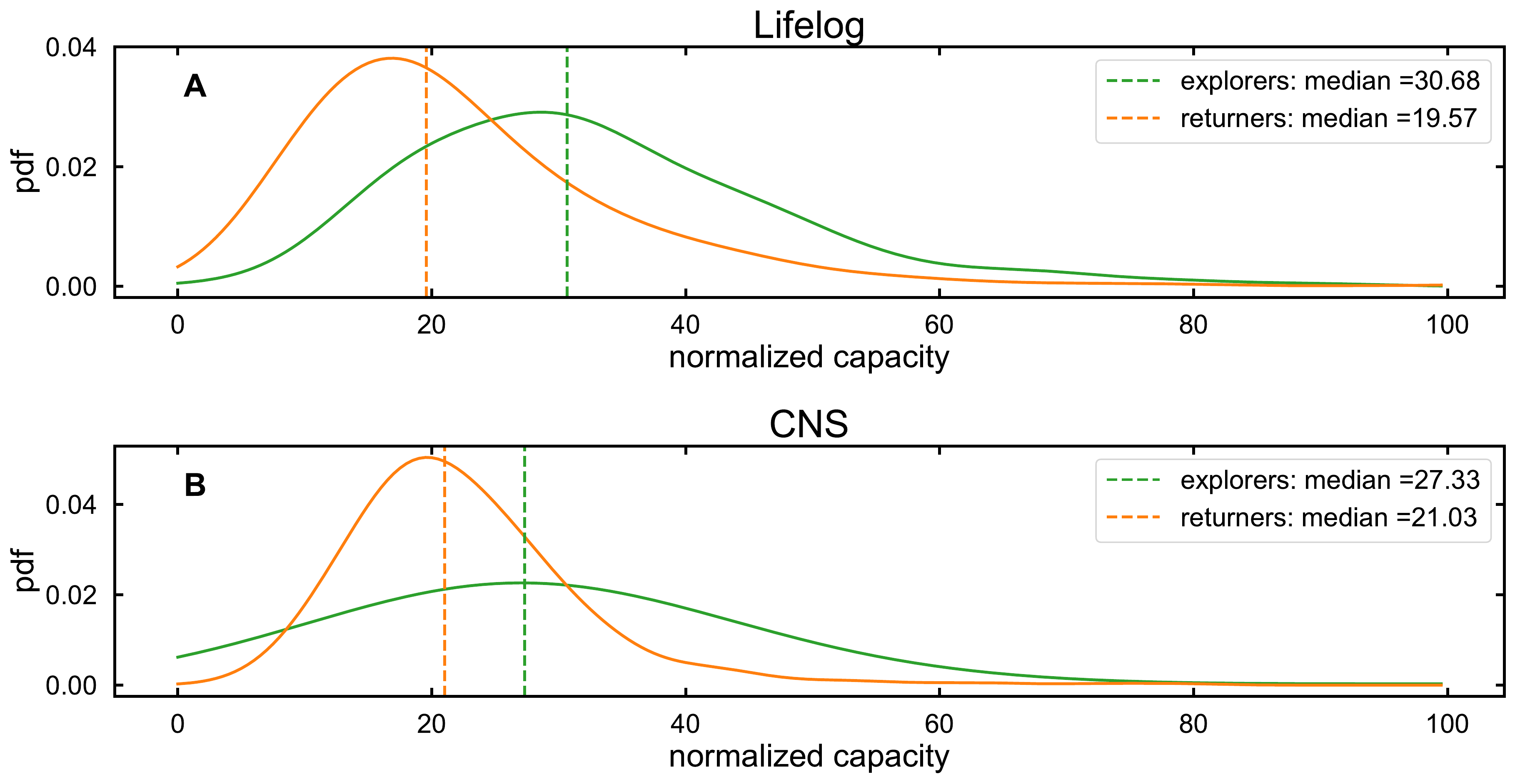}
\caption{\textbf{Different location capacity between returners and explorers.} Probability distribution of the average normalized location capacity for \emph{returners} (orange line) and \emph{explorers} (green line), according to the definition in \cite{pappalardo2015returners} (see also Supplementary Figure~\ref{returners_and_explorers_2}). Results are shown for the Lifelog \textbf{(A)} and CNS \textbf{(B)} datasets. Dashed lines show the median across users.  }
\label{returners_explorers_capacity}
\end{figure*}

\begin{figure*}[h]
\centering
\includegraphics[width=.9\textwidth]{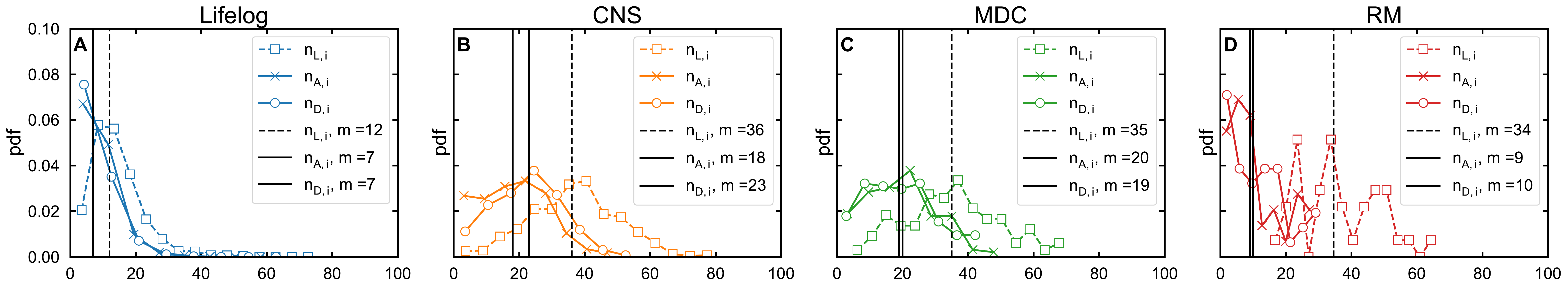}
\caption{\textbf{The aggregated number of locations, locations added and removed in the activity set.} The distribution of the aggregated number of locations $n_{L,i}$ in the activity set during the first $12$ months (dashed line, squared markers), the aggregated number of locations added $n_{A,i}$ (full line, cross markers), and removed $n_{D,i}$ (full line, circles) from it, for the Lifelog \textbf{(A)}, CNS \textbf{(B)}, MDC \textbf{(C)} and RM \textbf{(D)} datasets. Note that we focus on a period of $12$ months to include the majority of individuals. Results are shown for $W=20$ weeks. }
\label{n_l,n_a,n_d}
\end{figure*}

\begin{figure*}[h]
\centering
\includegraphics[width=.9\textwidth]{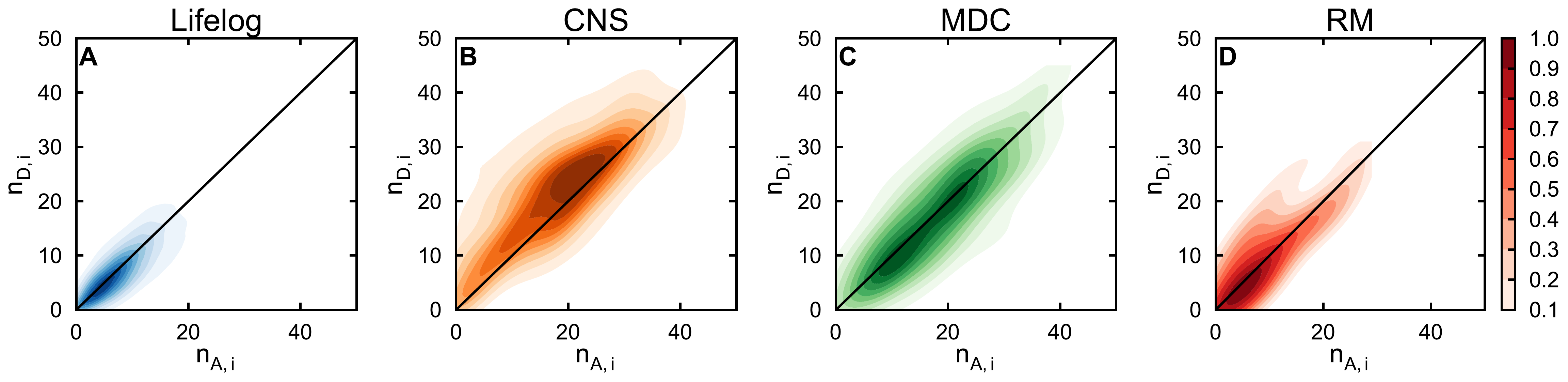}
\caption{\textbf{Correlation between number of locations adopted and dismissed in the activity set.} Heatmap showing the aggregated number of locations added $n_{A,i}$ versus the number of locations removed  $n_{D,i}$ in the activity set within a period of $12$ months. Results are shown for $W=20$ weeks. Results form the PCA indicate that for all datasets the $\sim 90\%$ of the variation can be explained by the first component in the $(0.71, 0.70)$ direction, i.e. almost the black line in the plot, $n_{A} = n_{D}$. Results are shown for the Lifelog \textbf{(A)}, CNS \textbf{(B)}, MDC \textbf{(C)} and RM \textbf{(D)} datasets. }
\label{n_a vs n_d}
\end{figure*}

\begin{figure*}[h]
\centering
\includegraphics[width=.9\textwidth]{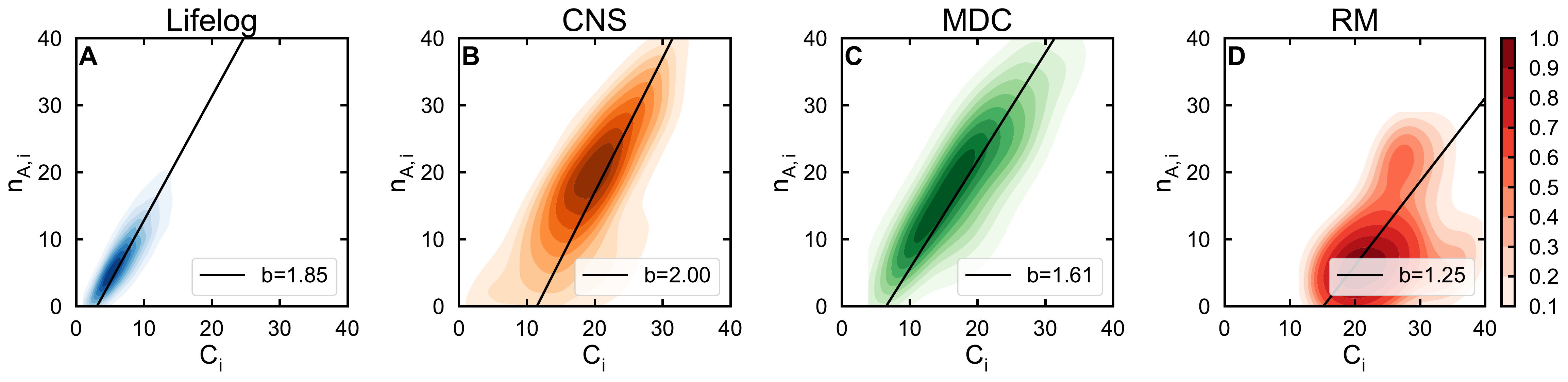}
\caption{\textbf{Correlation between the individual capacity and the total number of locations added in the activity set.} Heatmap showing the number of locations added $n_{A,i}$ during a period of $12$ months versus the individual location capacity $C_i$. Results are shown for $W=20$ weeks. The full line shows the result of a linear fit $n_{A,i} = a+b C_{i}$, where $b$ is shown in legend. Results are shown for the Lifelog \textbf{(A)}, CNS \textbf{(B)}, MDC \textbf{(C)} and RM \textbf{(D)} datasets.}
\label{n_a vs n_l}
\end{figure*}

\begin{figure*}[h]
\centering
\includegraphics[width=.5\textwidth]{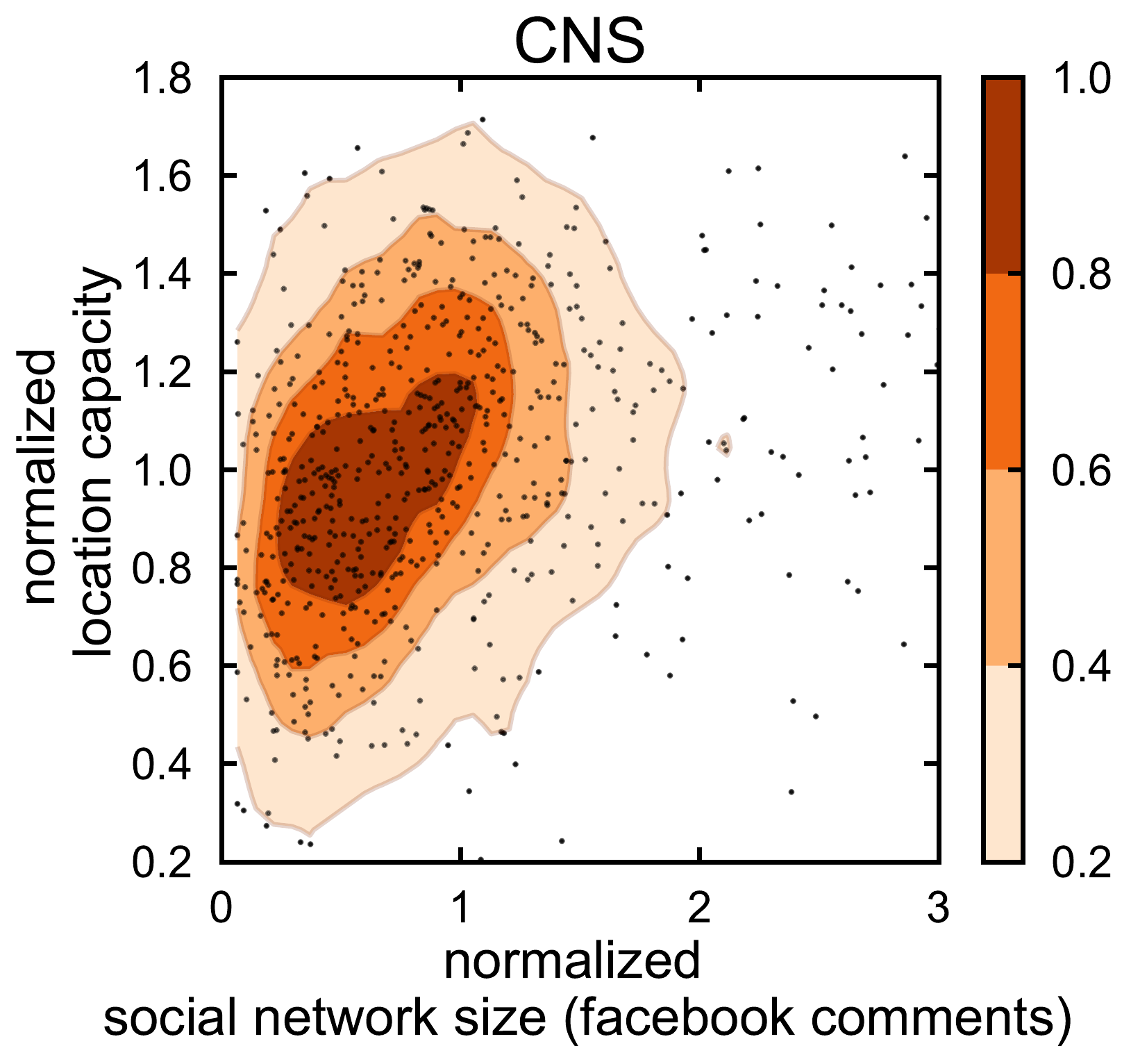}
\caption{\textbf{Correlation between location capacity and social network size measured from Facebook.} Values of individuals' average normalized location capacity vs their normalized social network size computed from Facebook (black dots). Colored filled areas correspond to cumulative probabilities estimated via Gaussian Kernel Density estimations for visualization purposes. Results are shown for the CNS dataset. The values of the Pearson correlation coefficient are 0.33 (2-tailed $p<10^{-18}$). Social network size is normalized to the population average value.}
\label{Facebook}
\end{figure*}

\clearpage
\section{Supplementary Tables}

\begin{table}[h!]
\centering
\begin{tabular}{l| ccccc}
 & N & $\delta t$ & $T$ & $\delta x$ & $TC$\\
 \hline
Lifelog & 36898 & change in motion & 19 months & 10 m & 0.57* \\
Lifelog (selected users) & 2272 & change in motion & 19 months & 10 m & 0.66* \\
CNS & 850 & 16 s & 24 months & 10 m & 0.84 \\
MDC & 185 &  60 s & 19 months &   100-200m &0.73  \\
RM & 95 & 16 s & 10 months & 100-200m & 0.93 \\
\end{tabular}\\
\vspace{0.2cm}
*computed from data including only stop-locations, after pre-processing internal at SONY Mobile
\caption{Characteristics of the datasets considered. $N$ is the number of individuals,  $\delta t$ the temporal resolution (for the Lifelog dataset, location is recorded at every change in motion), $T$ the duration of data collection, $\delta x$ the spatial resolution, $TC$ the median weekly time coverage, defined as the fraction of time an individual's location is known. Note that TC for Lifelog trajectories is computed from data including only stop-locations, where users stop for more than 10 minutes. In the other  datasets, stop-locations account on average for 80/90\% of the total TC. We also validated results considering a subset of Lifelog users with high time coverage (second row). See also Supplementary Figures~S1, S2, S3}
\label{Table_data}
\end{table}

\begin{table}
\centering
\begin{tabularx}{\linewidth}{@{}Xrrrr@{}}
\toprule

{}&  data type  & locations/week &  unique locations/week &  unique locations/week  \\
{} &   &    & &  (normalized) \\
\midrule
Lifelog &  GPS  &              25 &                      7 &                                  16 \\
CNS   & GPS+WiFi   &     28 &                     12 &                                  14 \\
MDC   & GSM   &         58 &                     11 &                                  15 \\
RM  & GSM    &          96 &                      7 &                                  13 \\
\bottomrule
\end{tabularx}
\caption{\textbf{Number of locations for different datasets.} The median number of total and unique locations visited per week. Values are reported for the 4 datasets. While noisy location data collected from GSM signals induces substantial variation in the total number of displacements, the number of unique weekly locations is comparable among the datasets, after accounting for differences in time coverage. This is the most relevant quantity for the purpose of this study.}
\label{Table0}
\end{table}

\begin{table}
\begin{tabularx}{\linewidth}{@{}ll ll ll ll l@{}}
\toprule
            data & d (m) &   W & \multicolumn{2}{c}{$H_0$} & \multicolumn{2}{c}{$H_1$} & $H_{j,k}$ \\ \cmidrule{4-5} \cmidrule{6-7}
    & &         &         $b$ & p (b)  &            $\beta$ & p ($\beta$) &  rejected \\
\midrule
 Lifelog\\(sel.user) &    50 &  10 &  $-2.69\cdot 10^{-4} \pm 3.04\cdot 10^{-3}$ &        0.94 &  - $-1.14\cdot 10^{-4} \pm 3.45\cdot 10^{-2}$ &             1.00 &       0\% \\
            Lifelog &   500 &  10 &   $3.06\cdot 10^{-3} \pm 3.48\cdot 10^{-3}$ &        0.54 &     $1.30\cdot 10^{-3} \pm 2.56\cdot 10^{-2}$ &             0.97 &       0\% \\
            Lifelog &    30 &  10 &  $-1.47\cdot 10^{-3} \pm 3.24\cdot 10^{-3}$ &        0.73 &    $-6.61\cdot 10^{-4} \pm 2.25\cdot 10^{-2}$ &             0.99 &       0\% \\
            Lifelog &    40 &  10 &  $-5.26\cdot 10^{-4} \pm 3.28\cdot 10^{-3}$ &        0.90 &    $-2.61\cdot 10^{-4} \pm 2.30\cdot 10^{-2}$ &             0.99 &       0\% \\
            Lifelog &    50 &   4 &  $-6.75\cdot 10^{-4} \pm 2.66\cdot 10^{-3}$ &        0.84 &   $-3.28\cdot 10^{-4} \pm 1.68\cdot 10^{-2}$ &             0.99 &       0\% \\
            Lifelog &    50 &   6 &  $-4.04\cdot 10^{-4} \pm 2.85\cdot 10^{-3}$ &        0.91 &   $-2.08\cdot 10^{-4} \pm 1.89\cdot 10^{-2}$ &             0.99 &       0\% \\
            Lifelog &    50 &   8 &  $-1.86\cdot 10^{-4} \pm 3.02\cdot 10^{-3}$ &        0.96 &   $-1.12\cdot 10^{-4} \pm 2.08\cdot 10^{-2}$ &             1.00 &       0\% \\
            Lifelog &    50 &  10 &  $-1.53\cdot 10^{-4} \pm 3.11\cdot 10^{-3}$ &        0.97 &    $-9.75\cdot 10^{-5} \pm 2.20\cdot 10^{-2}$ &             1.00 &       0\% \\
            Lifelog &    50 &  12 &   $2.60\cdot 10^{-4} \pm 3.33\cdot 10^{-3}$ &        0.95 &     $7.77\cdot 10^{-5} \pm 2.42\cdot 10^{-2}$ &             1.00 &       0\% \\
            Lifelog &    50 &  40 &   $2.59\cdot 10^{-3} \pm 7.05\cdot 10^{-3}$ &        0.78 &     $1.10\cdot 10^{-3} \pm 5.77\cdot 10^{-2}$ &             0.99 &       0\% \\
            Lifelog &    50 &  20 &   $2.13\cdot 10^{-3} \pm 3.91\cdot 10^{-3}$ &        0.68 &     $8.76\cdot 10^{-4} \pm 3.21\cdot 10^{-2}$ &             0.98 &       0\% \\
             CNS &     2 &  10 &  $-3.74\cdot 10^{-3} \pm 3.42\cdot 10^{-3}$ &        0.47 &  $-1.50\cdot 10^{-3} \pm 4.21\cdot 10^{-2}$ &             0.98 &       0\% \\
             CNS &     5 &   4 &  $-2.06\cdot 10^{-3} \pm 3.66\cdot 10^{-3}$ &        0.67 &   $-1.04\cdot 10^{-3} \pm 3.39\cdot 10^{-2}$ &             0.98 &       0\% \\
             CNS &     5 &   6 &  $-1.81\cdot 10^{-3} \pm 3.57\cdot 10^{-3}$ &        0.70 &  $-8.37\cdot 10^{-4} \pm 3.71\cdot 10^{-2}$ &             0.99 &       0\% \\
             CNS &     5 &   8 &  $-2.92\cdot 10^{-3} \pm 3.50\cdot 10^{-3}$ &        0.56 &   $-1.16\cdot 10^{-3} \pm 3.94\cdot 10^{-2}$ &             0.98 &       0\% \\
             CNS &     5 &  10 &  $-3.84\cdot 10^{-3} \pm 3.43\cdot 10^{-3}$ &        0.46 &   $-1.54\cdot 10^{-3} \pm 4.10\cdot 10^{-2}$ &             0.98 &       0\% \\
             CNS &     5 &  12 &  $-4.09\cdot 10^{-3} \pm 3.33\cdot 10^{-3}$ &        0.43 &   $-1.66\cdot 10^{-3} \pm 4.18\cdot 10^{-2}$ &             0.97 &       0\% \\
             CNS &     5 &  20 &  $-4.41\cdot 10^{-3} \pm 4.05\cdot 10^{-3}$ &           0.00 &  $-1.83\cdot 10^{-3} \pm 6.19\cdot 10^{-2}$ &             0.98 &       0\% \\
             CNS &     5 &  40 &  $-1.77\cdot 10^{-3} \pm 8.92\cdot 10^{-3}$ &        0.87 &   $-8.57\cdot 10^{-4} \pm 1.41\cdot 10^{-1}$ &             1.00 &       0\% \\
             CNS &     5 &  50 &  $-2.76\cdot 10^{-3} \pm 1.78\cdot 10^{-2}$ &        0.90 &  $-1.19\cdot 10^{-3} \pm 2.86\cdot 10^{-1}$ &             1.00 &       0\% \\
             CNS &    10 &  10 &  $-3.39\cdot 10^{-3} \pm 3.39\cdot 10^{-3}$ &        0.50 &   $-1.37\cdot 10^{-3} \pm 4.04\cdot 10^{-2}$ &             0.98 &       0\% \\
             MDC &     0 &   4 &  $-1.08\cdot 10^{-3} \pm 2.70\cdot 10^{-3}$ &        0.76 &   $-5.12\cdot 10^{-4} \pm 2.74\cdot 10^{-2}$ &             0.99 &       0\% \\
             MDC &     0 &   6 &  $-9.54\cdot 10^{-4} \pm 2.75\cdot 10^{-3}$ &        0.79 &   $-4.70\cdot 10^{-4} \pm 3.11\cdot 10^{-2}$ &             0.99 &       0\% \\
             MDC &     0 &   8 &  $-7.25\cdot 10^{-4} \pm 2.82\cdot 10^{-3}$ &        0.84 &   $-3.54\cdot 10^{-4} \pm 3.41\cdot 10^{-2}$ &             0.99 &       0\% \\
             MDC &     0 &  10 &  $-5.98\cdot 10^{-4} \pm 2.88\cdot 10^{-3}$ &        0.87 &  $-2.98\cdot 10^{-4} \pm 3.64\cdot 10^{-2}$ &             0.99 &       0\% \\
             MDC &     0 &  12 &  $-4.52\cdot 10^{-4} \pm 2.95\cdot 10^{-3}$ &        0.90 &   $-2.39\cdot 10^{-4} \pm 3.83\cdot 10^{-2}$ &             1.00 &       0\% \\
             MDC &     0 &  40 &   $1.74\cdot 10^{-3} \pm 5.13\cdot 10^{-3}$ &        0.79 &     $7.45\cdot 10^{-4} \pm 7.85\cdot 10^{-2}$ &             0.99 &       0\% \\
             MDC &     0 &  50 &   $3.77\cdot 10^{-3} \pm 7.52\cdot 10^{-3}$ &        0.70 &     $1.60\cdot 10^{-3} \pm 1.19\cdot 10^{-1}$ &             0.99 &       0\% \\
             MDC &     0 &  20 &  $-5.93\cdot 10^{-4} \pm 3.22\cdot 10^{-3}$ &       0.00 &  $-2.53\cdot 10^{-4} \pm 4.72\cdot 10^{-2}$ &             1.00 &       0\% \\
              RM &     0 &   4 &   $4.73\cdot 10^{-3} \pm 7.05\cdot 10^{-3}$ &        0.62 &     $1.15\cdot 10^{-3} \pm 7.76\cdot 10^{-2}$ &             0.99 &       0\% \\
              RM &     0 &   6 &   $3.77\cdot 10^{-3} \pm 8.47\cdot 10^{-3}$ &        0.73 &   $8.58\cdot 10^{-4} \pm 1.08\cdot 10^{-1}$ &             0.99 &       0\% \\
              RM &     0 &   8 &   $4.31\cdot 10^{-3} \pm 8.87\cdot 10^{-3}$ &        0.71 &    $9.40\cdot 10^{-4} \pm 1.23\cdot 10^{-1}$ &             1.00 &       0\% \\
              RM &     0 &  10 &   $2.16\cdot 10^{-3} \pm 9.46\cdot 10^{-3}$ &        0.86 &   $3.87E-06 \pm 1.38\cdot 10^{-1}$ &             1.00 &       0\% \\
              RM &     0 &  12 &  $-3.52\cdot 10^{-4} \pm 1.05\cdot 10^{-2}$ &        0.98 &   $-9.48\cdot 10^{-4} \pm 1.60\cdot 10^{-1}$ &             1.00 &       0\% \\
              RM &     0 &  20 &   $6.01\cdot 10^{-3}  \pm 1.97\cdot 10^{-2} $ &          0.10 &   $1.85\cdot 10^{-3}  \pm 3.49\cdot 10^{-1} $ &             1.00 &       0\% \\

\bottomrule
\end{tabularx}

\caption{\textbf{Conservation of capacity: evidence 1}. The results of hypotheses testing $H_0$, $H_1$ and $H_{j,k}$ (see section \emph{Robustness Tests}) for different values of the threshold used to define locations $d$, and sliding window size $W$. For $H_0$, we report the value of the linear fit coefficient $b$ and the p-value. $H_0: b = 0$ is rejected for $p<0.05$. For $H_1$, we report the value of the power-law fit coefficient $\beta$ and the corresponding p-value. $H_1: \beta = 0$ is rejected for $p<0.05$. For $H_{j,k}$, we report the percentage of rejected hypotheses $H_{j,k}: C_j = C_k$, with $j$ and $k$ two different time-intervals.}
\label{conservation_cap_1}
\end{table}
\clearpage
\begin{table}
\begin{tabularx}{.3\linewidth}{@{}Xrr rr rr@{}}
\toprule
 data & d (m) &   W & $|G_i|<\sigma_{G_i}$ \\
\midrule
 Lifelog &    500 &  10 &           98\%  \\
 Lifelog &    30 &  10 &           98\%  \\
 Lifelog &    40 &  10 &     98\%\\
 Lifelog &    50 &   4 &       99\%\\
 Lifelog &    50 &   6 &      99\%  \\
 Lifelog &    50 &   8 &       98\% \\
 Lifelog &    50 &  10 &       98\%  \\
 Lifelog &    50 &  12 &          98\%  \\
 Lifelog &    50 &  40 &     27\%  \\
 Lifelog &    50 &  20 &         89\%  \\
  CNS &     2 &  10 &        98\%  \\
  CNS &     5 &   4 &       98\%  \\
  CNS &     5 &   6 &               98\% \\
  CNS &     5 &   8 &              97\%  \\
  CNS &     5 &  10 &              98\%  \\
  CNS &     5 &  12 &            98\% \\
  CNS &     5 &  40 &                95\% \\
  CNS &     5 &  50 &              94\% \\
  CNS &    10 &  10 &                98\%  \\
  MDC &     0 &   4 &                98\%  \\
  MDC &     0 &   6 &          95\% \\
  MDC &     0 &   8 &           97\% \\
  MDC &     0 &  10 &                   99\% \\
  MDC &     0 &  12 &             99\%\\
  MDC &     0 &  40 &                94\%  \\
  MDC &     0 &  50 &                  83\%  \\
  MDC &     0 &  20 &                  95\%  \\
   RM &     0 &   4 &         93\%  \\
   RM &     0 &   6 &           90\%  \\
   RM &     0 &   8 &        87\%  \\
   RM &     0 &  10 &            84\%  \\
   RM &     0 &  12 &               85\%\\
   RM &     0 &  20 &                85\%  \\
\bottomrule
\end{tabularx}
\caption{\textbf{Conservation of capacity: evidence 2.} For different values of the threshold used to define locations $d$, and sliding window size $W$, the percentage of individuals such that $|G_{i}|<\sigma_{G_i}$ (see section Robustness Tests). }
\label{conservation_cap_2}
\end{table}
\clearpage
\begin{centering}
\begin{table}
\begin{tabularx}{\linewidth}{@{}X rrrr@{}}
\toprule
 data & KS statistics (local) & p-value (local) & KS statistics (global) & p-value (global)\\
 \midrule
Lifelog & 0.21 & 0 &   &  \\
CNS & 0.29 & 0.0 & 0.94 & 0.0\\
MDC & 0.36 & 0.0 & 0.99 & 0.0 \\
RM & 0.35 & 0.0 & 0.99 & 0.0 \\
\bottomrule
\end{tabularx}
\caption{\textbf{Discrepancy with the randomized case.} The Kolomogorov-Smirnov (KS) test statistics measuring the discrepancy between the capacity in the real and randomized case, with the corresponding p-values. Since $p<0.05$ we can reject the hypothesis that the distributions underlying the two samples are the same under a 2-tailed test. Results are shown for the local and global randomization, for different datasets. }
\label{ks_statistics}
\end{table}
\end{centering}

\begin{centering}
\begin{table}
\begin{tabularx}{\linewidth}{@{}X rr rrrrrr @{}}
\toprule
 data & d (m) &   W & $\Delta T$=10-30min & $\Delta T=$ 30-60 min& $\Delta T$=1-6 h & $\Delta T$=6-12h & $\Delta T=$12-48 h & $\Delta T>$48 h  \\
  & &    &                p (b) &                p (b) &                 p (b) &                  p (b) &                   p (b) &                     p (b) \\
\midrule
 Lifelog &    500 &  10 &                       0.99 &                       1.00 &                        0.99 &                         0.99 &                          1.00 &                            1.00 \\
 Lifelog &    30 &  10 &                       1.00 &                       1.00 &                        0.98 &                         0.99 &                          1.00 &                            1.00 \\
 Lifelog &    40 &  10 &                       0.96 &                       1.00 &                        1.00 &                         1.00 &                          1.00 &                            1.00 \\
 Lifelog &    50 &   4 &                       1.00 &                       1.00 &                        0.99 &                         0.99 &                          1.00 &                            1.00 \\
 Lifelog &    50 &   6 &                       0.99 &                       1.00 &                        1.00 &                         0.99 &                          1.00 &                            1.00 \\
 Lifelog &    50 &   8 &                       0.92 &                       0.98 &                        0.97 &                         0.99 &                          1.00 &                            1.00 \\
 Lifelog &    50 &  10 &                       0.99 &                       1.00 &                        0.98 &                         0.99 &                          1.00 &                            1.00 \\
 Lifelog &    50 &  12 &                       0.99 &                       0.99 &                        0.98 &                         0.98 &                          1.00 &                            1.00 \\
 Lifelog &    50 &  40 &                       0.75 &                       0.96 &                        0.78 &                         0.99 &                          0.98 &                            1.00 \\
 Lifelog &    50 &  20 &                       0.83 &                       0.99 &                        1.00 &                         0.98 &                          0.99 &                            1.00 \\
  CNS &     2 &  10 &                       0.94 &                       0.98 &                        1.00 &                         0.99 &                          0.99 &                            1.00 \\
  CNS &     5 &   4 &                       0.97 &                       0.99 &                        0.99 &                         1.00 &                          0.99 &                            1.00 \\
  CNS &     5 &   6 &                       0.97 &                       0.99 &                        0.99 &                         1.00 &                          0.99 &                            1.00 \\
  CNS &     5 &   8 &                       0.96 &                       0.98 &                        0.99 &                         1.00 &                          1.00 &                            1.00 \\
  CNS &     5 &  10 &                       0.94 &                       0.98 &                        0.99 &                         0.99 &                          1.00 &                            1.00 \\
  CNS &     5 &  12 &                       0.93 &                       0.98 &                        0.97 &                         1.00 &                          0.99 &                            1.00 \\
  CNS &     5 &  40 &                       0.94 &                       0.99 &                        0.94 &                         0.99 &                          0.99 &                            1.00 \\
  CNS &     5 &  50 &                       0.92 &                       0.98 &                        0.92 &                         0.99 &                          0.99 &                            0.99 \\
  CNS &    10 &  10 &                       0.95 &                       0.98 &                        0.99 &                         0.99 &                          0.99 &                            0.99 \\
  MDC &     0 &   4 &                       0.96 &                       0.99 &                        0.97 &                         0.97 &                          0.96 &                            1.00 \\
  MDC &     0 &   6 &                       0.97 &                       0.99 &                        0.99 &                         0.98 &                          0.97 &                            1.00 \\
  MDC &     0 &   8 &                       0.98 &                       1.00 &                        0.99 &                         0.97 &                          0.96 &                            1.00 \\
  MDC &     0 &  10 &                       0.96 &                       0.99 &                        0.99 &                         0.97 &                          0.96 &                            1.00 \\
  MDC &     0 &  12 &                       0.95 &                       0.99 &                        0.98 &                         0.99 &                          0.96 &                            0.99 \\
  MDC &     0 &  40 &                       0.91 &                       0.96 &                        0.98 &                         0.95 &                          1.00 &                            0.99 \\
  MDC &     0 &  50 &                       0.95 &                       0.91 &                        0.90 &                         0.95 &                          0.94 &                            0.99 \\
  MDC &     0 &  20 &                       0.90 &                       0.98 &                        0.97 &                         0.99 &                          0.97 &                            0.99 \\
   RM &     0 &   4 &                       0.97 &                       0.95 &                        0.93 &                         0.91 &                          0.96 &                            0.99 \\
   RM &     0 &   6 &                       0.93 &                       0.93 &                        1.00 &                         0.93 &                          0.98 &                            0.99 \\
   RM &     0 &   8 &                       0.97 &                       0.84 &                        0.99 &                         0.97 &                          0.98 &                            0.99 \\
   RM &     0 &  10 &                       0.99 &                       0.89 &                        0.95 &                         0.94 &                          0.95 &                            0.99 \\
   RM &     0 &  12 &                       0.94 &                       0.87 &                        0.92 &                         0.93 &                          0.93 &                            0.99 \\
   RM &     0 &  20 &                       0.80 &                       0.86 &                        0.89 &                         0.96 &                          0.87 &                            1.00 \\
\bottomrule
\end{tabularx}
\caption{\textbf{Conservation of time allocation.}
The results of hypotheses testing $H_0$ for different classes of locations $\Delta T$. Results are shown for different values of the threshold used to define locations $d$, and sliding window size $W$. We report the p-value, testing the hypothesis $H_0: b = 0$ is rejected for $p<0.05$.}
\label{table_time_alloc}
\end{table}
\end{centering}

\begin{center}
\begin{table}
\begin{tabularx}{\linewidth}{@{}Xrrrr@{}}
\toprule
     $\delta T$      &  Lifelog &   CNS &   MDC &                                                 RM \\
\midrule
10 - 30 min    &     0.13 &  0.09 &  0.09 &                                          0.076 \\
30  - 60  min  &     0.15 &  0.08 &  0.07 &                                          0.013 \\
1 - 6 h   &     0.37 &  0.21 &  0.19 &                                           0.13 \\
6 - 12 h   &     0.65 &  0.31 &  0.18 &                                          0.09 \\
12 - 48 h  &     0.83 &  0.54 &  0.47 &                                          0.02 \\
$>$48 h &     0.99 &  0.86 &  0.77 & 1 \\
\bottomrule
\end{tabularx}
\caption{\textbf{Evolution of various classes of locations.} The average Jaccard similarity between the subsets of the activity set $AS_i(t)^{\Delta T}$ and $AS_i(t+w)^{\Delta T}$. Results are shown for several classes $\Delta T$, considering $W=20$ weeks. } 
\label{subset_evolution}
\end{table}
\end{center}

\end{document}